\documentclass{article}

\usepackage{arxiv}

\usepackage[utf8]{inputenc} 
\usepackage[T1]{fontenc}    
\usepackage{hyperref}       
\usepackage{url}            
\usepackage{booktabs}       
\usepackage{amsfonts}       
\usepackage{nicefrac}       
\usepackage{microtype}      
\usepackage{graphicx}
\usepackage{natbib}
\usepackage{doi}

\usepackage{amsmath,amsfonts}

\title{A phase defect framework\\  for the analysis of cardiac arrhythmia patterns} 

\author{Louise Arno \\
    KULeuven Campus KULAK \\ Department of Mathematics\\ 8500 Kortrijk, Belgium \\
    \And
	Jan Quan \\
	KULeuven Campus KULAK\\ 8500 Kortrijk, Belgium \\
	\AND
	Nhan T. Nguyen \\
	KULeuven Campus KULAK\\ 8500 Kortrijk, Belgium \\ 
	\And
	Maarten Vanmarcke \\
	KULeuven Campus KULAK\\ 8500 Kortrijk, Belgium \\ 
	\AND
	Elena G. Tolkacheva \\
     University of Minnesota\\ Biomedical Engineering Department\\ Minneapolis, MN 55455, Minnesota, USA \\
     \And
	Hans Dierckx \\
	KULeuven Campus KULAK\\ Department of Mathematics \\ 8500 Kortrijk, Belgium \\
	\texttt{h.dierckx@kuleuven.be} 
}

\renewcommand{\headeright}{}
\renewcommand{\undertitle}{}
\renewcommand{\shorttitle}{Phase defects in cardiac arrhythmias}
\date{\today} 

\hypersetup{
pdftitle={A template for the arxiv style},
pdfauthor={Louise Arno, Jan Quan, Nhan T. Nguyen, Elena G. Tolkacheva, Hans Dierckx},
pdfkeywords={},
}

\newcommand{\phiact}{\ensuremath{\phi_{\rm act}}}
\newcommand{\phiarr}{\ensuremath{\phi_{\rm arr}}}
\newcommand{\tlap}{\ensuremath{t_{\rm elapsed}}}
\newcommand{\tact}{\ensuremath{t_{\rm activation}}}

\newcommand{\VS}{V_*}
\newcommand{\RS}{R_*}
\newcommand{\uu}{\mathbf{u}}
\newcommand{\dd}{\partial}

\newcommand{\figbreakup}{2.5cm}
\newcommand{\figrevisit}{3.2cm}
\newcommand{\figspiralphiarr}{3.5cm}

\graphicspath{{figures/}}

\begin{document}
\maketitle

\begin{abstract}
During cardiac arrhythmias, dynamical patterns of electrical activation form and evolve, which are of interest to understand and cure heart rhythm disorders. The analysis of these patterns is commonly performed by calculating the local activation phase and searching for phase singularities (PSs), i.e. points around which all phases are present. 
Here we propose an alternative framework, which focuses on phase defect lines (PDLs) and surfaces (PDSs) as more general mechanisms, which include PSs as a specific case. The proposed framework enables two conceptual unifications: between the local activation time and phase description, and between conduction block lines and the central regions of linear-core rotors. A simple PDL detection method is proposed and applied to data from simulations and optical mapping experiments. Our analysis of ventricular tachycardia in rabbit hearts $(n=6)$ shows that nearly all detected PSs were found on PDLs, but the PDLs had a significantly longer lifespan than the detected PSs. Since the proposed framework revisits basic building blocks of cardiac activation patterns, it can become a useful tool for further theory development and experimental analysis.
\end{abstract}

\section{Introduction}\label{sec:intro}

About once per second, a wave of electrical depolarization travels through the heart, coordinating its mechanical contraction. The heart is a prime example of a dynamical system that can self-organize across different scales: changes in the flow of ions through the cell membrane affect the dynamics of the emergent pattern and may lead to life-threatening heart rhythm disorders. For more than a century, great efforts have been allocated to understand the different dynamical mechanisms of arrhythmia initiation and maintenance in the heart. It was found that electrical activity during cardiac arrhythmias may travel in a closed circuit within the cardiac wall, and thus re-excite the heart \citep{Mines:1913}. Later on, Allessie et al. demonstrated with electrode recordings that re-excitation of the tissue by the wave itself was also possible without a central obstacle \cite{Allessie:1973}. These rotating vortices, also called rotors, were later demonstrated using optical mapping techniques, both in ventricular tachycardia \cite{Gray:1995b} and ventricular fibrillation \cite{Gray:1998}. 

\begin{figure}
    \centering
   \raisebox{3cm}{\textbf{A}} \includegraphics[trim={7cm 0.3cm 6cm 1.2cm},clip, height = 3.2cm]{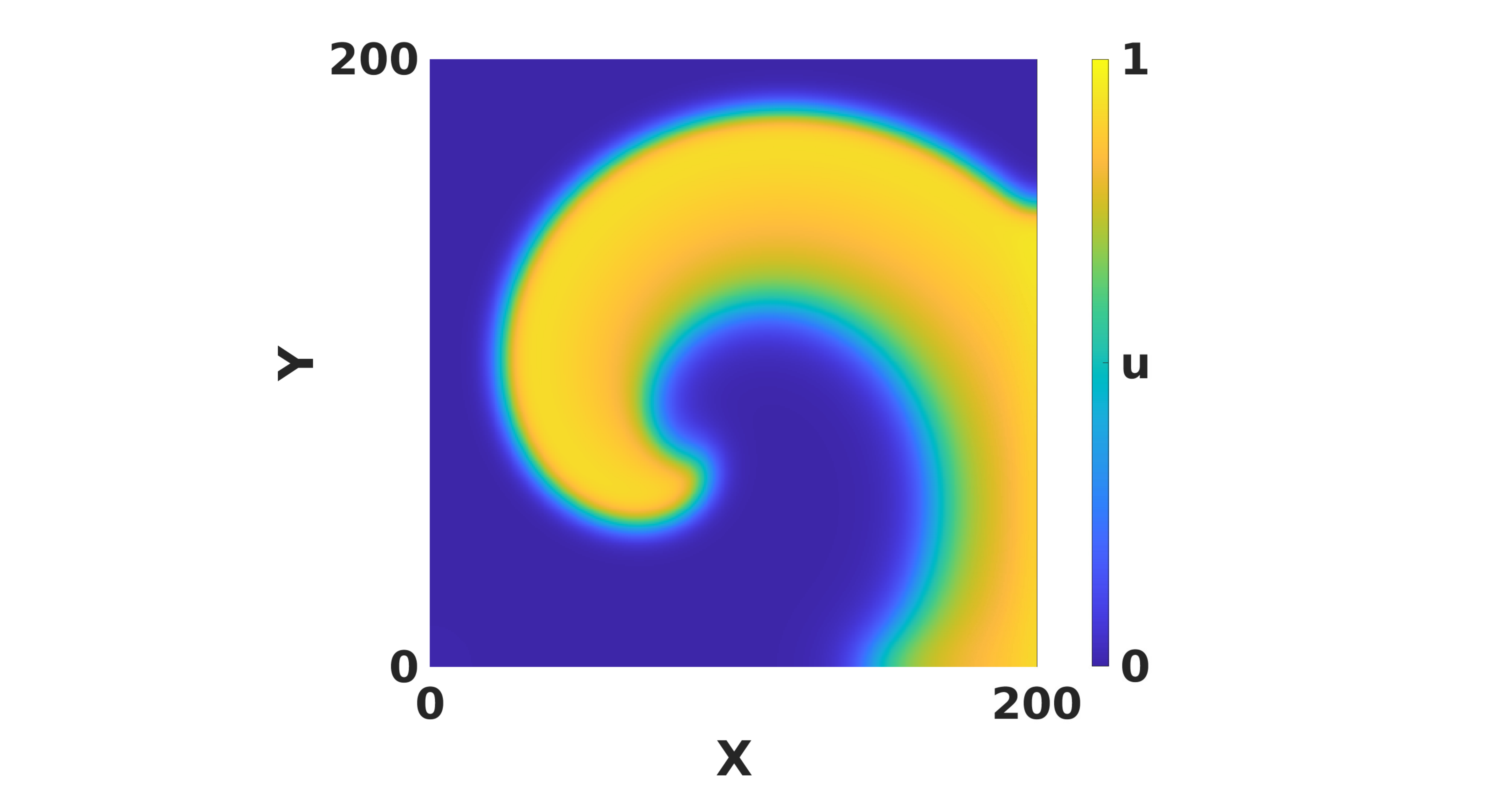}
   \raisebox{3cm}{\textbf{B}} \includegraphics[trim={1cm 0.3cm 0cm 1.3cm},clip, height = 3.cm]{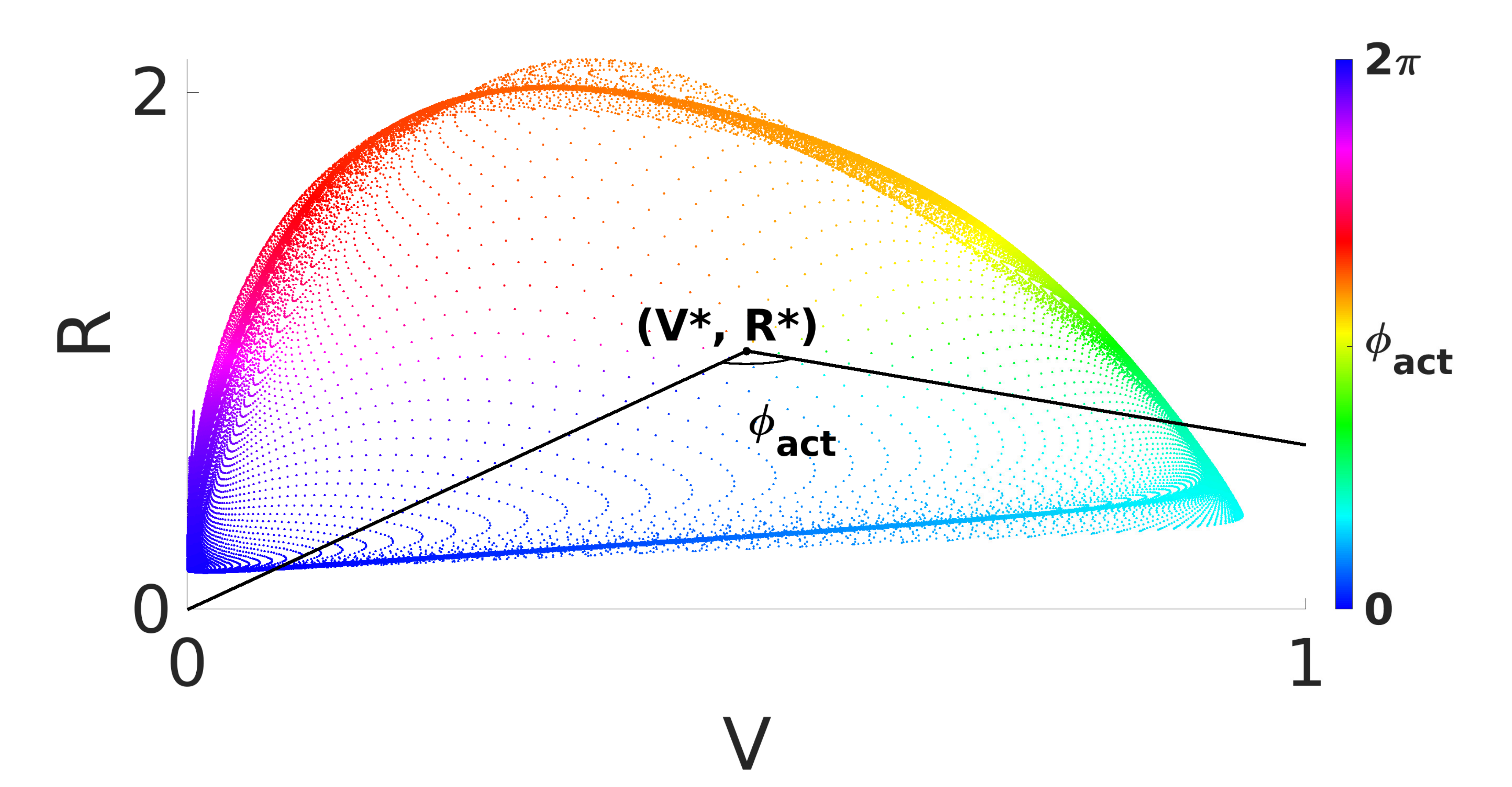}
     \raisebox{3cm}{\textbf{C}} \includegraphics[trim={8cm 0.3cm 8cm 1.2cm},clip, height = 3.2cm]{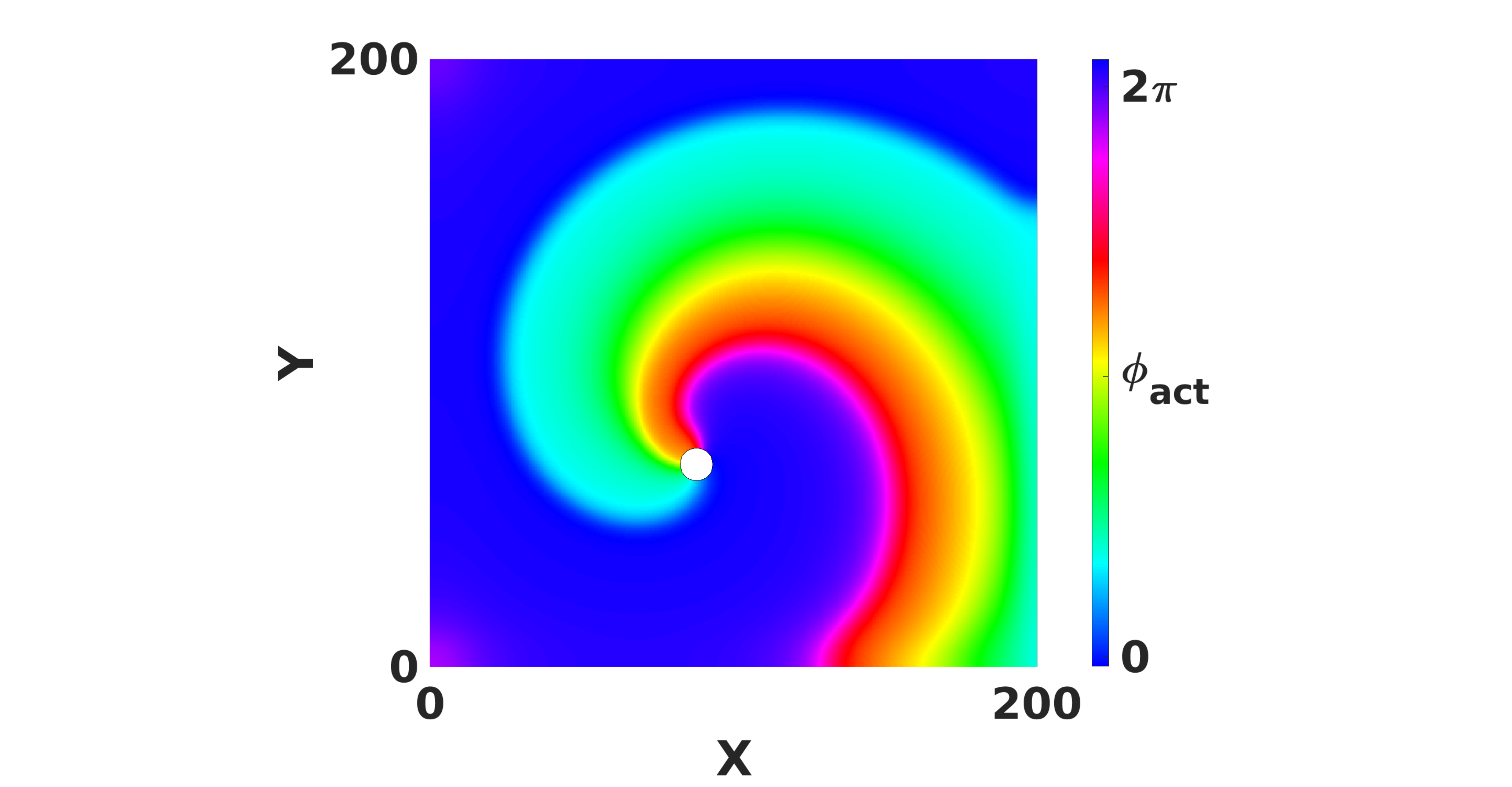} \\
    \caption{Classical phase analysis of cardiac rotors. (A) Circular-core rotor with Aliev-Panfilov kinetics \citep{Aliev:1996}, where in each point an activation variable $u$ and a recovery variable $v$ are defined. (B) Two observables of the system, $V=u$ and $R=v$ plotted against each other reveal a cycle corresponding to the action potential. The polar angle with respect to a point $(\VS, \RS)$ situated within the cycle serves as a definition of activation phase. (C) Coloring the rotor using phase and a periodic colormap reveals a special point where all phases meet, the phase singularity (PS), see white dot. This point has $V=\VS$, $R=\RS$. }
    \label{fig:classic}
\end{figure}

\cite{Gray:1998} also developed the concept of phase analysis to identify structures in the excitation patterns in the heart. 
They considered two fields $V(\vec{r},t)$ and $R(\vec{r},t)$ of the system and calculated the phase of a point of the medium as the polar angle in the $(V,R)$-plane, see Fig. \ref{fig:classic}:
\begin{align}
   \phiact(V,R) = \mathrm{atan2}( R-\RS, V-\VS) + b.  \label{phiact}
\end{align}
An additive constant in Eq. \eqref{phiact} is included here to fix the absolute phase, e.g. such that resting state corresponds to $\phiact =0$. To discriminate from an alternative phase definition below, we call this classical phase the `activation phase' and denote it as $\phiact$. 

The application of phase analysis to cardiac electrical signals during arrhythmia showed that there are few points in the medium where all possible phases meet each other, such that the point itself has no well-defined phase, see \citep{Gray:1998} and Fig. \ref{fig:classic}C. Such point is now generally accepted to be a phase singularity (PS), and since then PSs have been widely used in the analysis of cardiac excitation patterns. 
In two dimensions (2D), PSs are associated with the rotor core, i.e. the regions around which electrical waves revolve tachycardia or fibrillation. 

In three dimensions (3D), the set of PSs in the medium extend to a dynamical curve, the rotor filament. The rotor filament has been used in many modeling and theoretical studies, as it allows for easy visualization of the rotor dynamics \citep{Clayton:2005,Wellner:2002, Verschelde:2007,Dierckx:2012}. 

\begin{figure}
 \begin{tabular}{ c c}
\raisebox{3.1cm}{\textbf{A}}  &  \includegraphics[ trim={8cm 0cm 10.5cm 1.2cm},clip,width=0.24\textwidth]{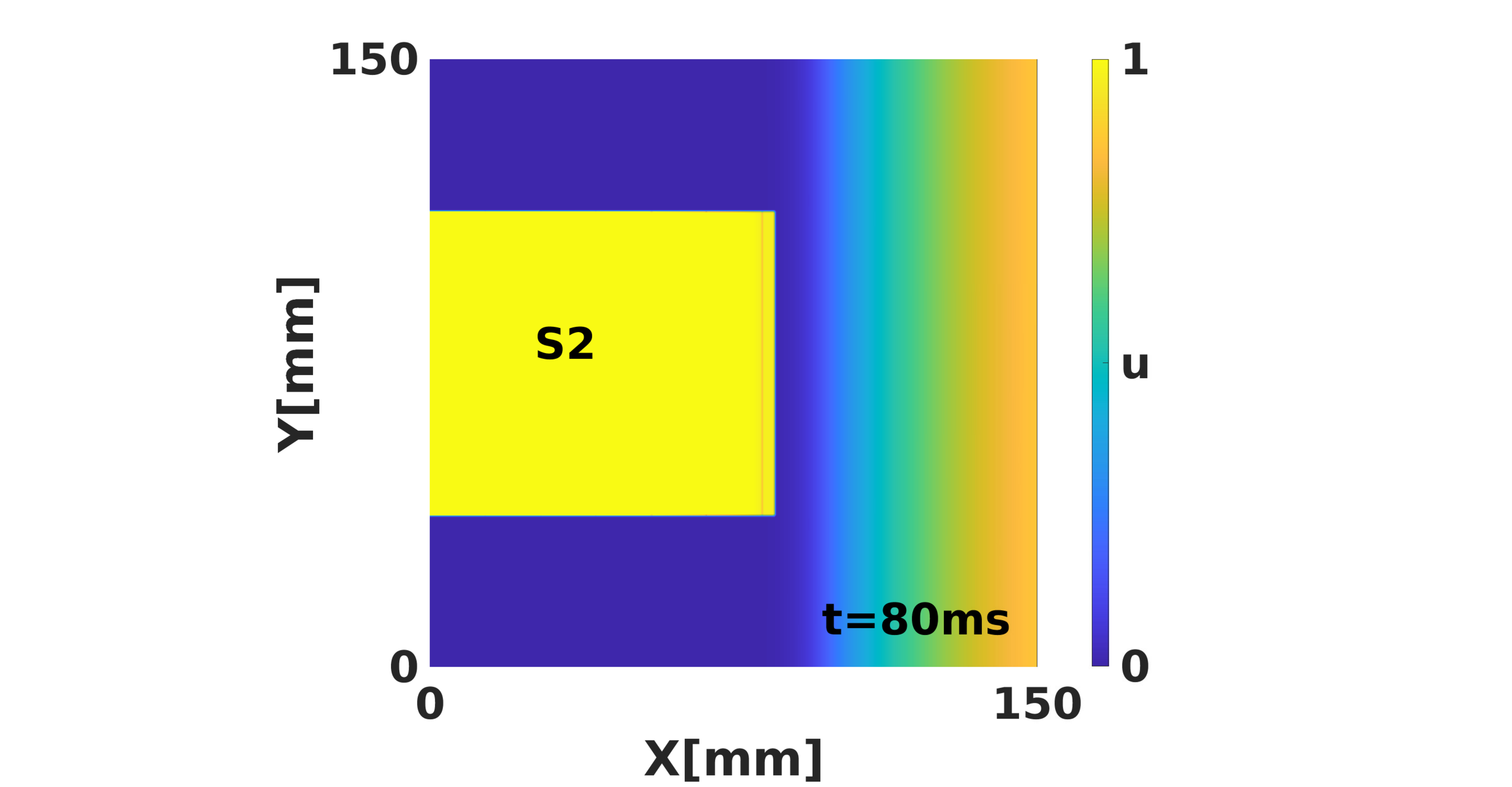}
 \includegraphics[ trim={8cm 0cm 10.5cm 1.2cm},clip,width=0.24\textwidth]{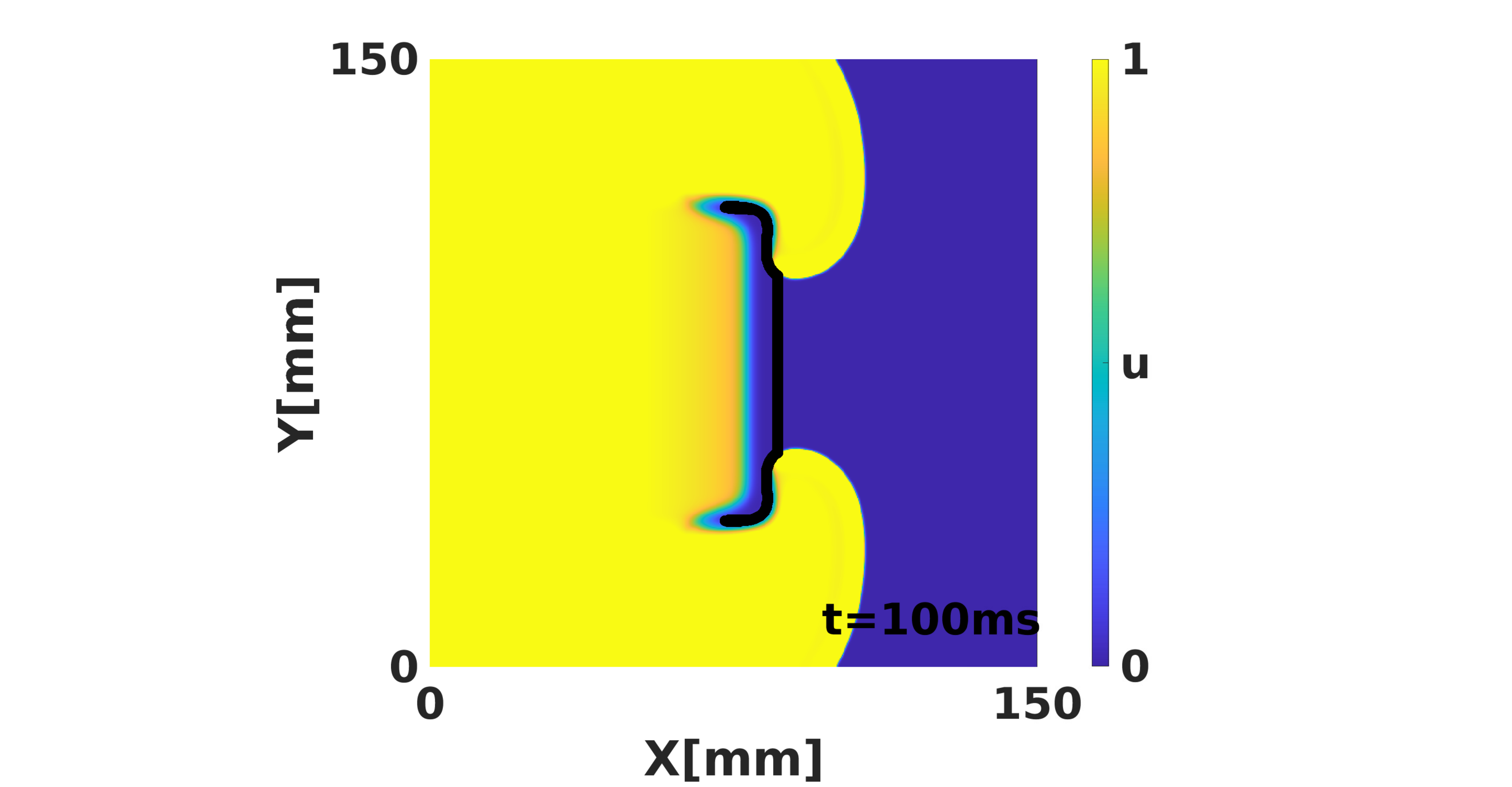}
     \includegraphics[trim={8cm 0cm 10cm 1.2cm},clip, width=0.24\textwidth]{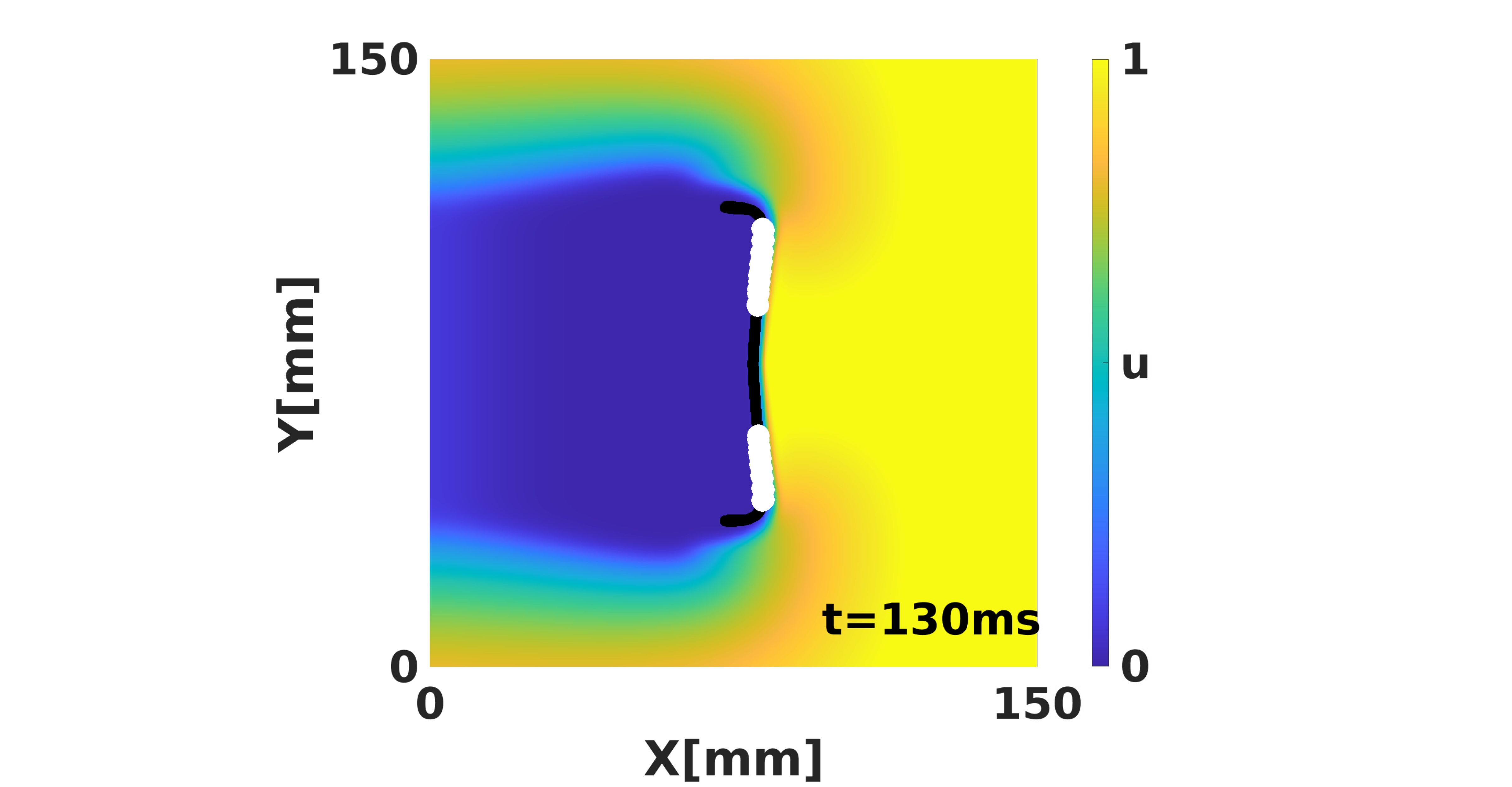}
    \includegraphics[trim={8cm 0cm 10cm 1.2cm},clip, width=0.24\textwidth]{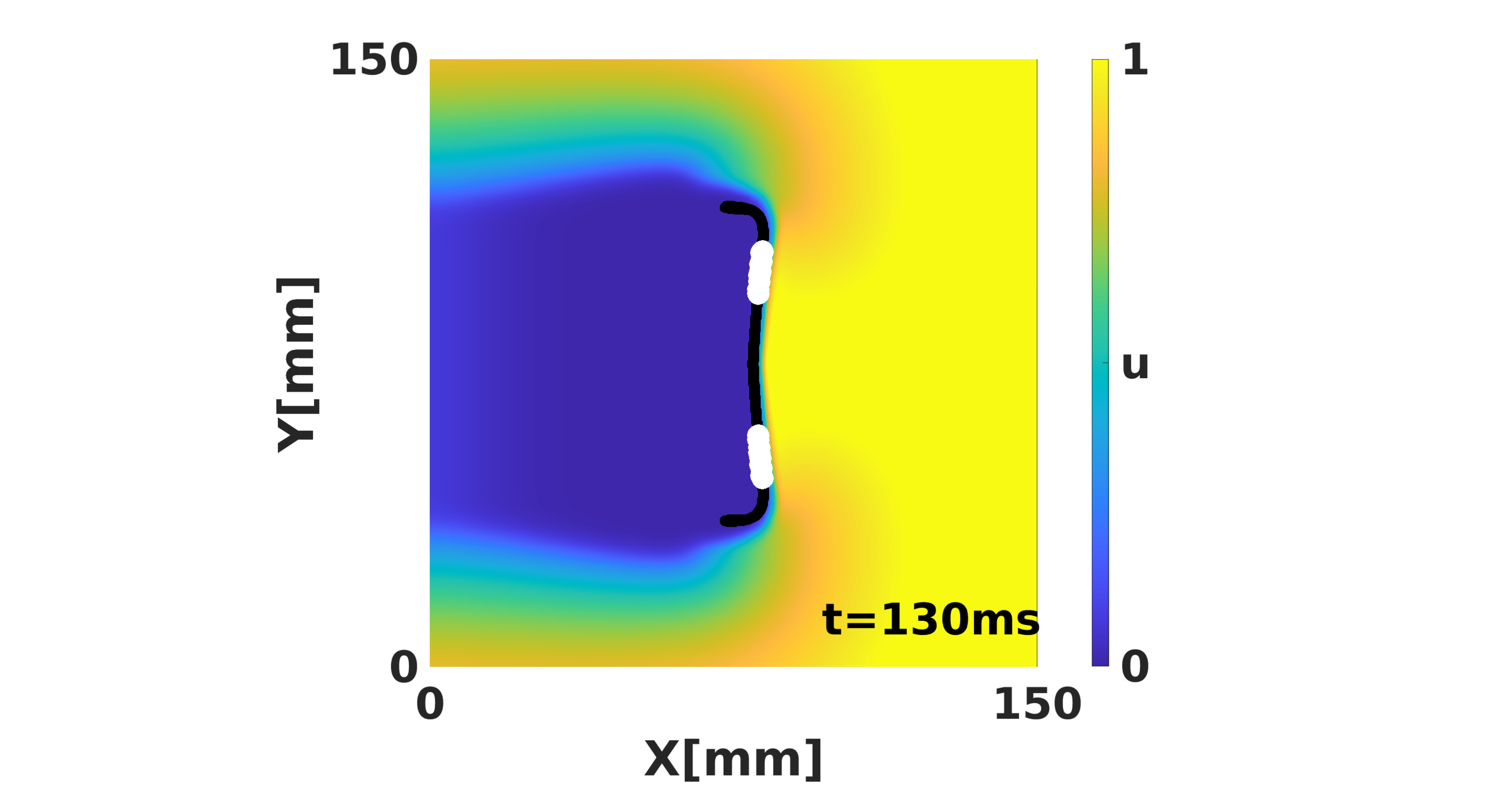} 
 \\
\raisebox{4.1cm}{\textbf{B}}  &  \includegraphics[trim={8cm 0cm 8cm 1.2cm},clip, width=0.33\textwidth]{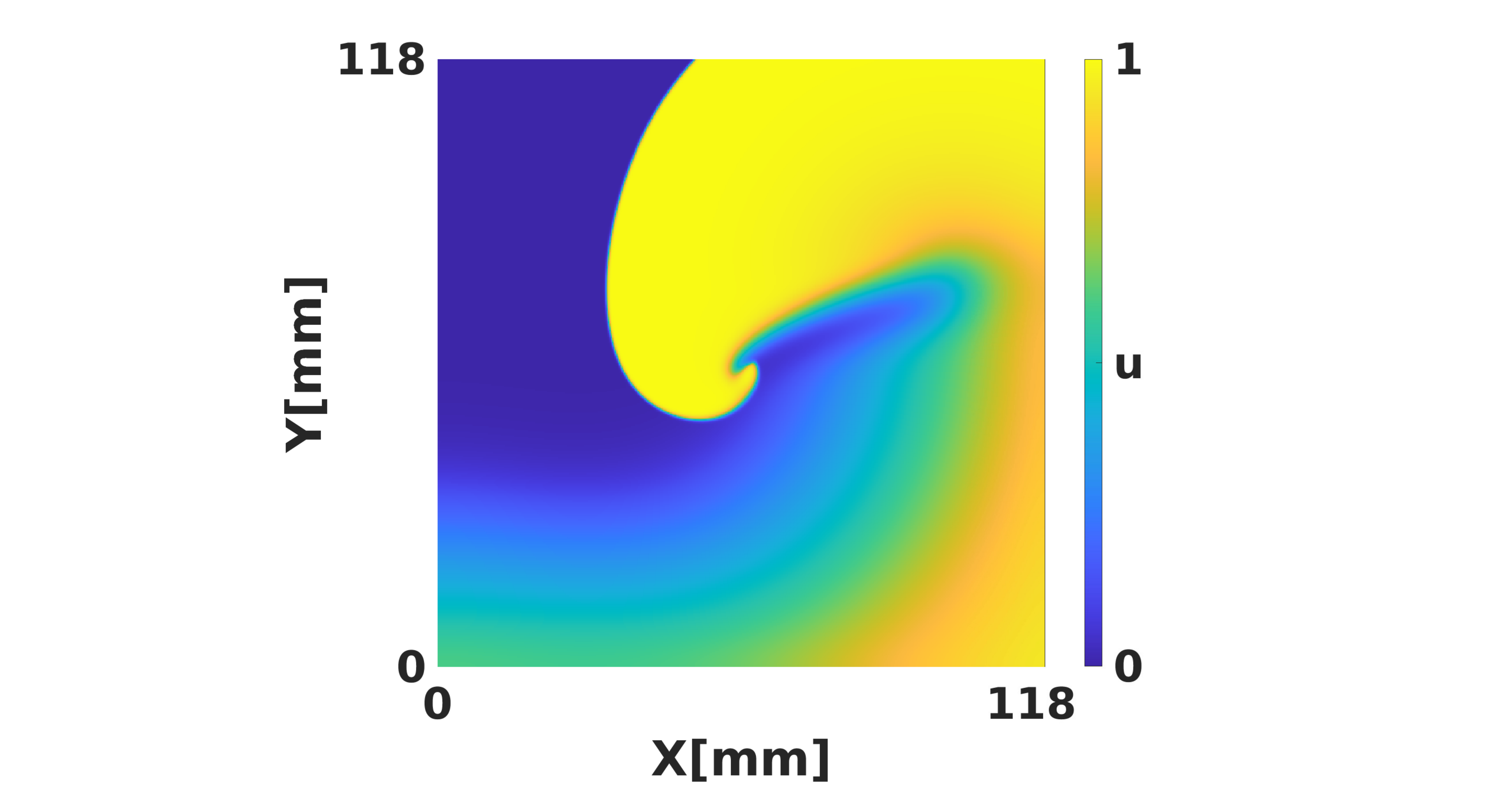}
   \includegraphics[trim={8cm 0cm 13.5cm 1.2cm},clip,width=0.32\textwidth]{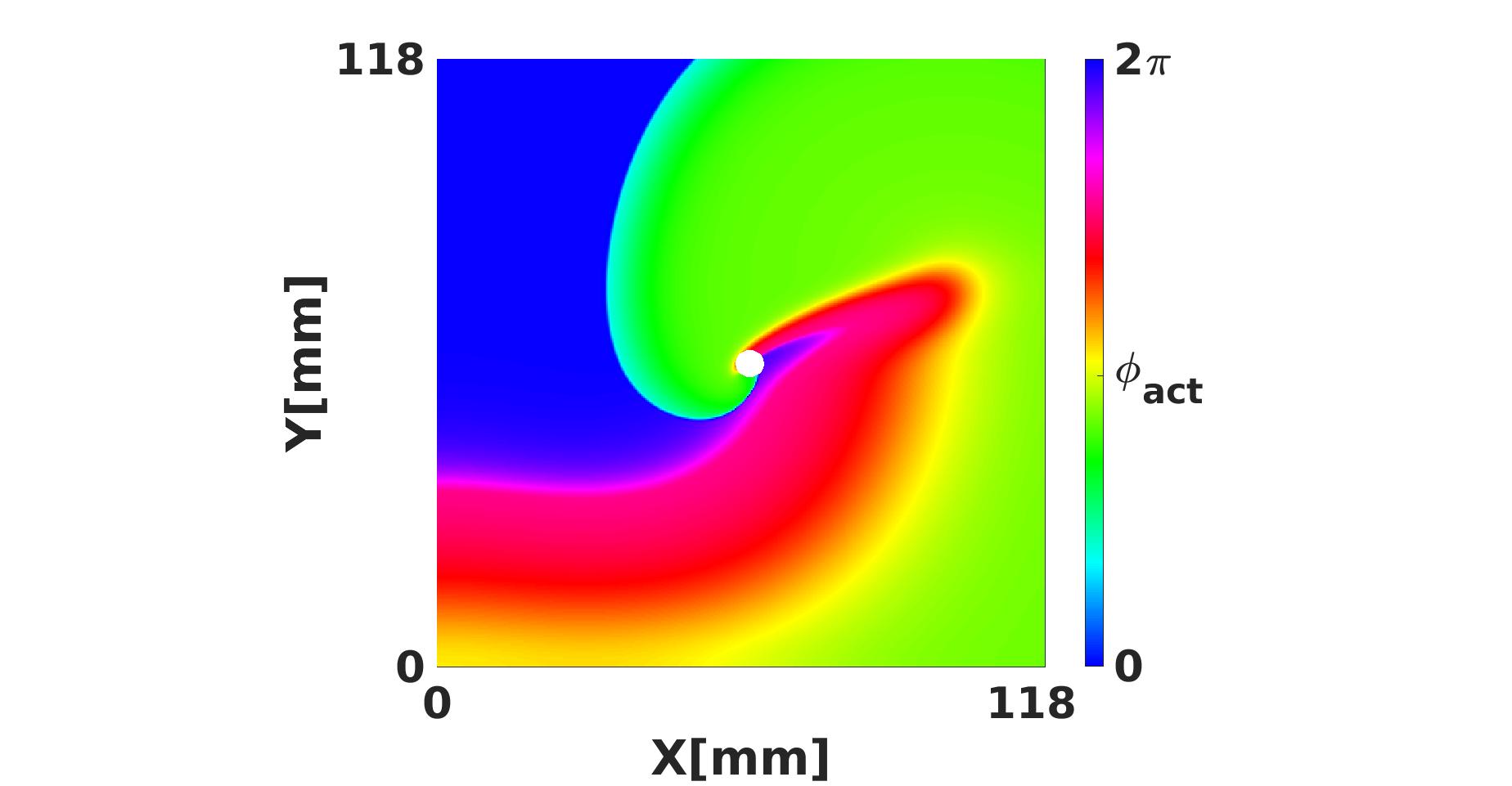}
    \includegraphics[trim={8cm 0cm 9.5cm 1.2cm},clip,width=0.31\textwidth]{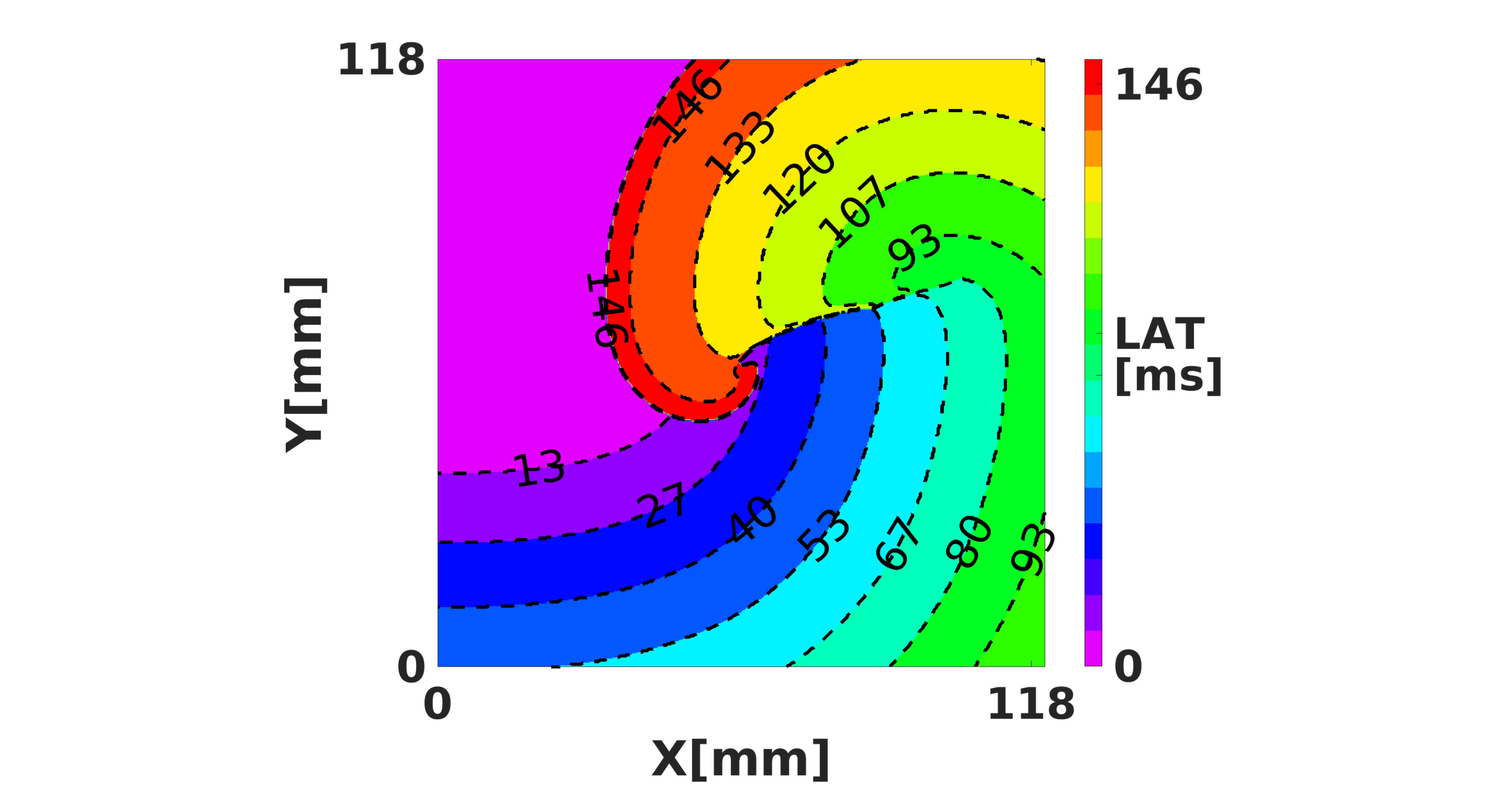}\\
    \raisebox{5cm}{\textbf{C}}&
    \includegraphics[trim={4cm 9.5cm 3.9cm 1cm}, clip, width=0.97\textwidth]{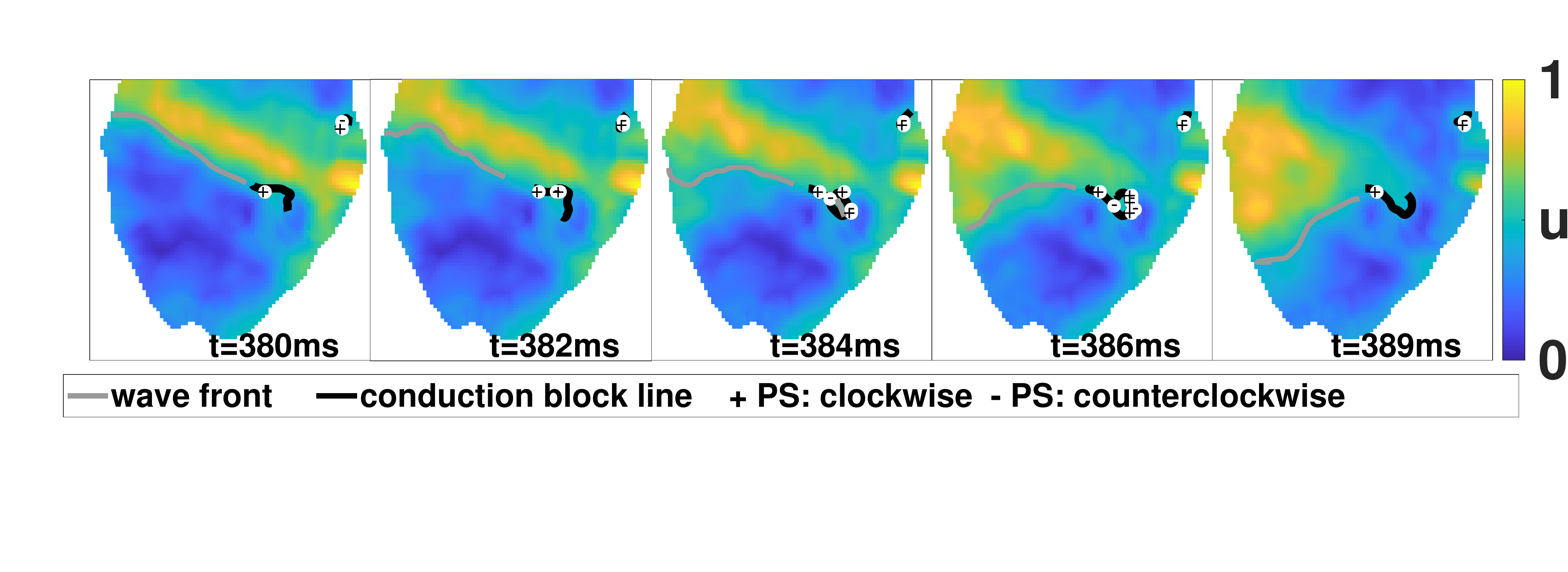}
 
\end{tabular}
    \caption{Limitations of current PS detection algorhithms. (A) Application of the S1-S2 stimulation protocol in the BOCF \citep{BuenoOrovio:2008} model to initiate a rotor. White dots denoted detected PSs with either a $2\times2$ ring (third panel) or $2\times2+4\times4$ ring (fourth panel) using the method of \cite{Kuklik:2017}. Both methods identify multiple PS on the CBL (black line). %
     (B) A linear-core rotor in the Fenton Karma model, from 3 perspectives: transmembrane voltage (left), activation phase (middle panel) with PS indicated (white) and LAT (right). (C) Optical mapping of rabbit hearts during ventricular tachycardia showing that detected PSs are all located on CBLs.}
    \label{fig:problems}
\end{figure}

To identify PSs in datasets, different methods have been developed \citep{Gray:1998,Fenton:1998,Bray:2003,Clayton:2005,Kuklik:2015,Kuklik:2017}. However, these methods that look for PSs have limitations when performing detections near conduction block lines, as illustrated in Fig. \ref{fig:problems}. 
A first limitation of the classical PS concept is non-robustness of PS detection methods. Fig. \ref{fig:problems}A shows an example of a simulation of 2D sheet of cardiac tissue using the S1-S2 stimulation protocol. The second (S2) pulse undergoes unidirectional block in the wave of the S1 pulse, leading to the formation of a conduction block line (CBL), rendered in black. The region excited by the S2 stimulus grows, turning around the CBL endpoints, but no PSs are detected at its end. Later on, when the initial S2 region has repolarized, a large number of PSs is found at the CBL, even though there is no rotor core. Moreover, the precise number of PSs found depends on the algorithm used, as shown in the third and fourth panel of \ref{fig:problems}A. In literature, other false positives of PS identification have been reported \citep{Podziemski:2018}.

A second limitation of classical PS detection algorithms is related to rotors with so-called linear cores, i.e. rotors which have a CBL in the central region (core) around which they rotate, see Fig. \ref{fig:problems}B. If one looks at the point where the wave front ends on the CBL, this is in the classical picture a PS, see Fig. \ref{fig:problems}B. However, looking closer, only 3 phases are found in the neighborhood of this point: excitable tissue (ahead of the front, blue), recently excited tissue (behind the front, green), and refractory tissue (across the CBL, red). A point touching three phases is not necessarily a mathematical PS, as the latter should touch all phases (see figure \ref{fig:phasediagram}).

If we represent the same linear core rotor in terms of the local activation time (LAT), as indicated in the rightmost panel of Fig. \ref{fig:problems}B, an extended CBL is clearly seen, where isochrones of activation coincide. However, this extended CBL is not captured by the classical PS analysis.  

A third limitation of the classical PS detection methods concerns the coincidence of PSs and CBLs. \cite{Podziemski:2018} and our own optical mapping experiments of ventricular tachycardia in rabbits indicate that the PSs are generally localized at a CBL, see Fig. \ref{fig:problems}C. When comparing different time frames, we moreover find that the PSs tend to jump between different locations on the CBL, while the CBL itself seems to persist longer in time. 

The goal of this paper is to present a new framework for phase analysis, which will lead to identification of novel structures in the regions of conduction block: phase defect lines (PDLs) and phase defect surfaces (PDSs) instead of PSs and filaments.
In the context of cardiac arrhythmia patterns, interpreting linear rotor cores as a phase defect was only done very recently by \cite{Tomii:2021} and in our preprint to this manuscript \citep{Arno:2021arxiv}. Within this paper, we additionally introduce a phase based on local activation time, which bridges the gap between the phase viewpoint and LAT viewpoint. We propose a simple PDL detection method and apply it to data from simulations (in 2D and 3D) and optical mapping experiments. We draw conclusions from the analysis of our experiments and outline how future theory development can be initiated from the phase defect concept. 


\section{Methods} \label{sec:methods}

\subsection{Methods for data generation and PS and CBL detection}

\subsubsection{Numerical simulations}

To illustrate our theoretical concepts, numerical simulations were performed in a cardiac monodomain model:
\begin{align}
    \dd_t \uu =  \Delta \mathbf{P} \uu +  \mathbf{F}(\uu). \label{RDE}
\end{align}
This partial differential equation states how a column matrix $\uu(\vec{r},t)$ of $m$ state variables evolves in time. The constant matrix $\mathbf{P} = \rm{diag}(P_{11},0,...0)$ makes sure that only the transmembrane potential $u_1$ undergoes diffusion. 
Note that our analysis methods are applicable to excitation patterns in general and are not restricted to reaction-diffusion systems only. 

In this work, we use the reaction-kinetics functions $\mathbf{F}(\uu)$ from \cite{Aliev:1996} (AP model, $m=2$), \cite{Fenton:1998} (FK model, $m=3$) and \cite{BuenoOrovio:2008} (BOCF model, $m=4$). The equations are numerically integrated using the Euler method with the spatial resolution $dx$ and time step $dt$ as specified in Tab. \ref{tab:sims}. Only in the example with a biventricular geometry, uniaxial anisotropy was included: the Laplacian in \eqref{RDE} was replaced by $\dd_i (D^{ij} \dd_j)$ with $D^{ij} = D_1 \delta^{ij} + (D_1-D_2) e_f^i e^f_j$, with $D_1=1, D_2=1/5$ and $\vec{e}_f$ the local fiber orientation as measured by \cite{Hren:1995}. 

The first and second state variables in these different models were used to calculate $\phiact$, see Tab. \ref{tab:sims} for definitions and threshold values. 

\begin{table}[b]
    \centering
    \begin{tabular}{c|ccc}
    kinetics & AP & FK & BOCF \\ \hline
    reference & \small{\citep{Aliev:1996}} & \small{\citep{Fenton:1998}} & \tiny{\citep{BuenoOrovio:2008}} \\
    parameter set & standard & MLRI(2D) and MBR(3D) & epicardial (EPI) \\
    diffusion coefficient $P_{11}$ & 1 & 0.1 mm$^2/$ms & 0.11 mm$^2/$ms  \\
    dx (2D) & 0.5  & 0.262 mm & 0.25 mm \\ 
    dx (3D)  & - & 0.31 mm & 0.5 mm \\ 
    dt & 0.0029 & 0.16 ms(2D) and 0.1 ms(3D) & 0.1 ms \\
   domain size (2D) & 200 $\times$ 200 &  118\,mm $\times$ 118\,mm & 150\,mm $\times$ 150\,mm \\
  domain size (3D) & -  & \tiny{186\,mm $\times$ 186\,mm $\times $6.2\,mm} & \tiny{168\,mm $\times$ 208\,mm $\times $231mm} \\
    variable used as V ($\VS$) & u (0.5) & u (0.5) &u (0.5)  \\
    variable used as R ($\RS$) & v (1) & 1-v (0.8) & 1-v (0.2) \\
    $\Delta t$ for LAT & 1 & 1.6 ms & 5 ms\\
    $\Delta t_c$ for CBL detection & 10 & 16 ms & 45 ms\\
    $\Delta\phi_{\rm arr, crit}$ for CBL detection & 0.7 rad & 2 rad  & 0.5 rad \\
   $\tau$ in $\phiarr$& 20 & 70 ms & 300\,ms \\
    simulation used in &  Figs. \ref{fig:classic}, \ref{fig:complex}, \ref{fig:AP_spiral} & Figs. \ref{fig:problems}, \ref{fig:phiarr}, \ref{fig:complex}, \ref{fig:breakup} & Figs. \ref{fig:problems}, \ref{fig:S1S2revisited}, \ref{fig:S1S2revisited_phiarr}, \ref{fig:CBL_Riemann}, \ref{fig:3D} \\ 
    \end{tabular}
    \caption{Overview of mathematical models of cardiac excitation and parameters used in simulations throughout this work. }
    \label{tab:sims}
\end{table}

\subsubsection{Optical mapping experiments} \label{sec:opticalmappingmethods} 

We applied our new framework to experimentally recorded movies of rotors in isolated Langendorff-perfused rabbit hearts (n=6), as described by \cite{Kulkarni:2018}. Optical movies corresponding
to the fluorescence signal were recorded from
the epicardial surfaces of the left or right ventricular surface of the heart (LV
and RV) at 1000 frames per second, with 14-bit,
80 x 80-pixel resolution cameras (Little Joe, RedShirt
Imaging, SciMeasure) after a period of stabilization of
approximately 30 minutes.  Data processing was performed using a custom-made program in Matlab \cite{matlab:2020}. The background was removed by thresholding, and in each pixel the intensity was normalized against the minimal and maximal value of optical intensity attained in that pixel over the full recording. The characteristic timescale of activation was found as the inverse of the  dominant frequency of activation, and used to evaluate the Hilbert transform to obtain the activation phase $\phiact$ \citep{Bray:2002}. 

\subsubsection{Methods for PS and CBL detection} \label{sec:PDLmethods} 

\begin{figure}
    \centering
    \includegraphics[width=0.25\textwidth]{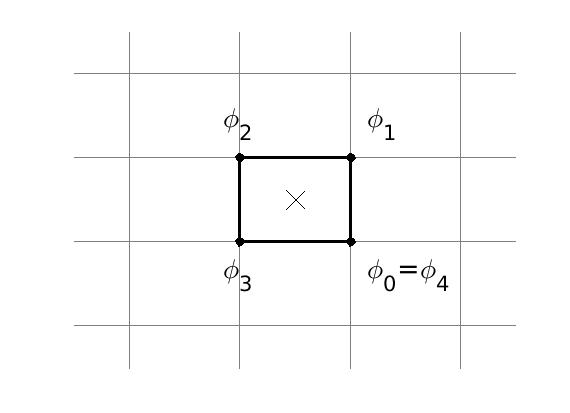}
    \includegraphics[width=0.25\textwidth]{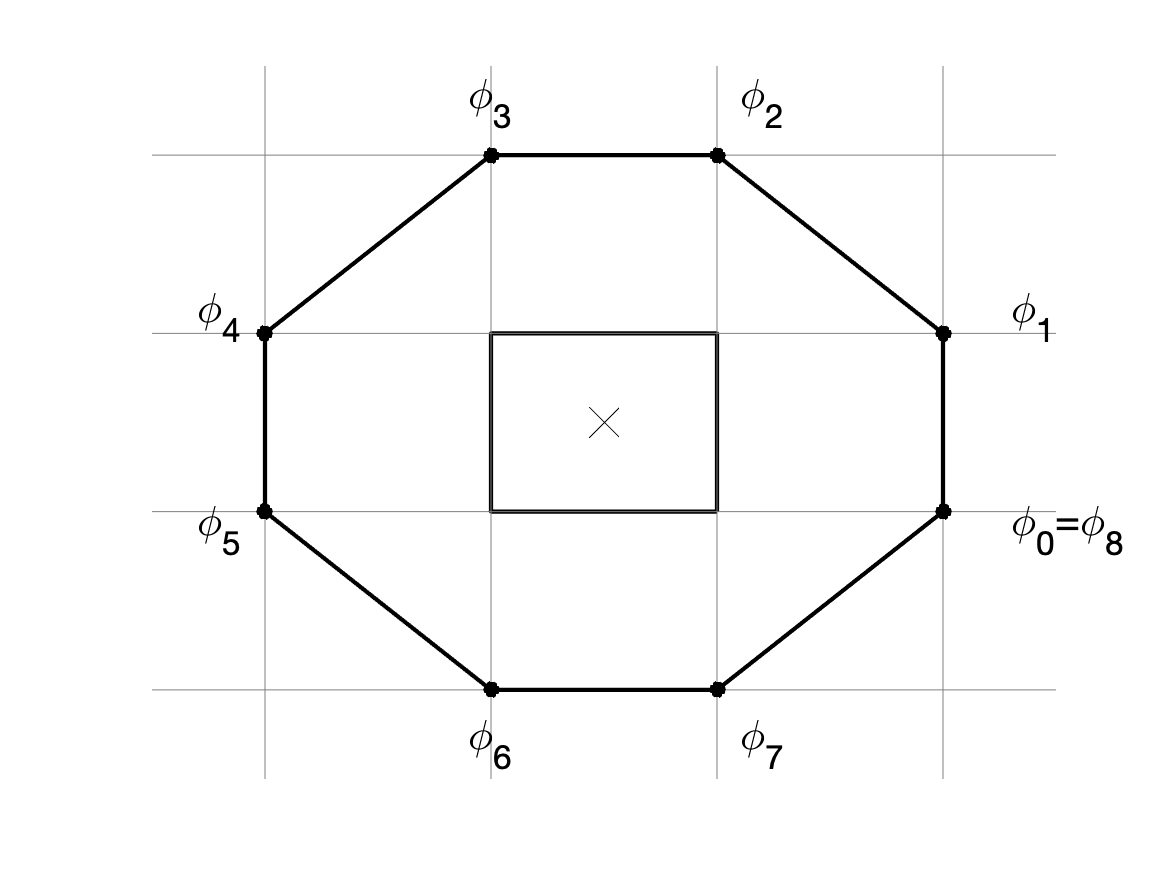}
    \caption{Method of PS detection by \cite{Kuklik:2017} as used in this paper. Phase differences are considered between pairs of points on a ring of 4 points (left, $2\times 2$ method) or 8 points (right, $4\times 4$ method), to assess whether a PS is present at the central location. }
    \label{fig:PSdetection}
\end{figure}

PSs were detected using the method of \cite{Kuklik:2017}, see Fig. \ref{fig:PSdetection}. A ring of N=4 or N=8 pixels was considered around each grid cell of the medium, and the phase difference was computed between adjacent points on the ring. If exactly one of these phase differences was larger than $\pi$ in absolute value, a PS was assigned to that grid element:
\begin{align}
PS\ \rm{at}&\  \vec{r}_0    &\Leftrightarrow   \# \{ k |\   \left| \phi_k(\vec{r}_0) - \phi_{k-1}(\vec{r}_0) \right| > \pi \} &=1. 
\end{align}
In the $2\times 2$ method, only the 4 points around each grid cell are used; in the $2\times 2 + 4 \times 4$ method, a second ring of 8 pixels was used and a PS is assigned if both the small ring and large ring have one phase difference larger than $\pi$, see Fig. \ref{fig:PSdetection}.  

To localize CBLs, frames were recorded every $\Delta t$ time units, and if in the new frame, $V$ rose above $\VS$, this time was locally saved as the newest LAT. If an edge connecting two neighboring points of the grid had two LAT that differed by more than $\Delta t_c \approx dx/c$ (with $c$ the plane wave velocity in the medium), the middle of that edge was considered to be part of a CBL: 
\begin{align}
\left. \begin{array}{l} 
| \tact(\vec{r}_1) - \tact(\vec{r}_2)) | > \Delta t_c  \\
\mathrm{and} \ \tact(\vec{r}_1) > 0 \\
\mathrm{and} \ \tact(\vec{r}_2) > 0 
\end{array} \right\}
\Rightarrow \frac{\vec{r_1} + \vec{r_2}}{2} \in CBL. \label{CBLdetection}
\end{align}
Here, the first condition selects the union of WF and CBL. The threshold values $\Delta t_c$ that we used with the different model kinetics are given in Tab. \ref{tab:sims}. The second and third condition are imposed to obtain CBLs only. 


\subsection{Theory}\label{sec:theory}

\subsubsection{Introduction of phase defects}

In many physical sciences, interfaces are found between regions of different phase, where they are known as domain walls or phase defects. In this sub-section, we demonstrate the existence of phase defects. 

\begin{figure}
    \centering
    \raisebox{3cm}{\textbf{A}} 
    \includegraphics[trim={12cm 0cm 18.5cm 0.5cm},clip, height= \figrevisit]{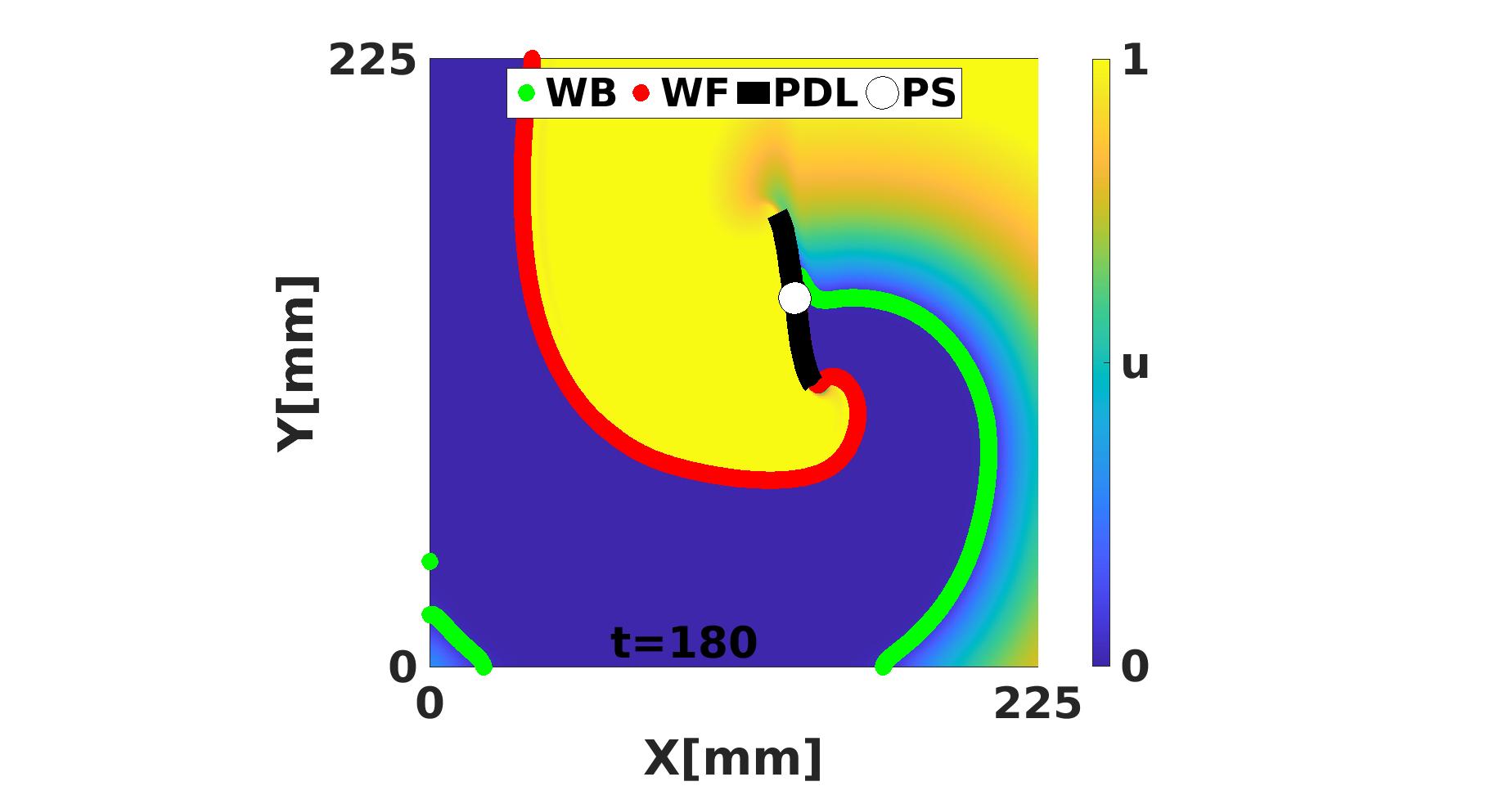}
    \includegraphics[trim={12cm 0cm 18.5cm 0.5cm},clip, height= \figrevisit]{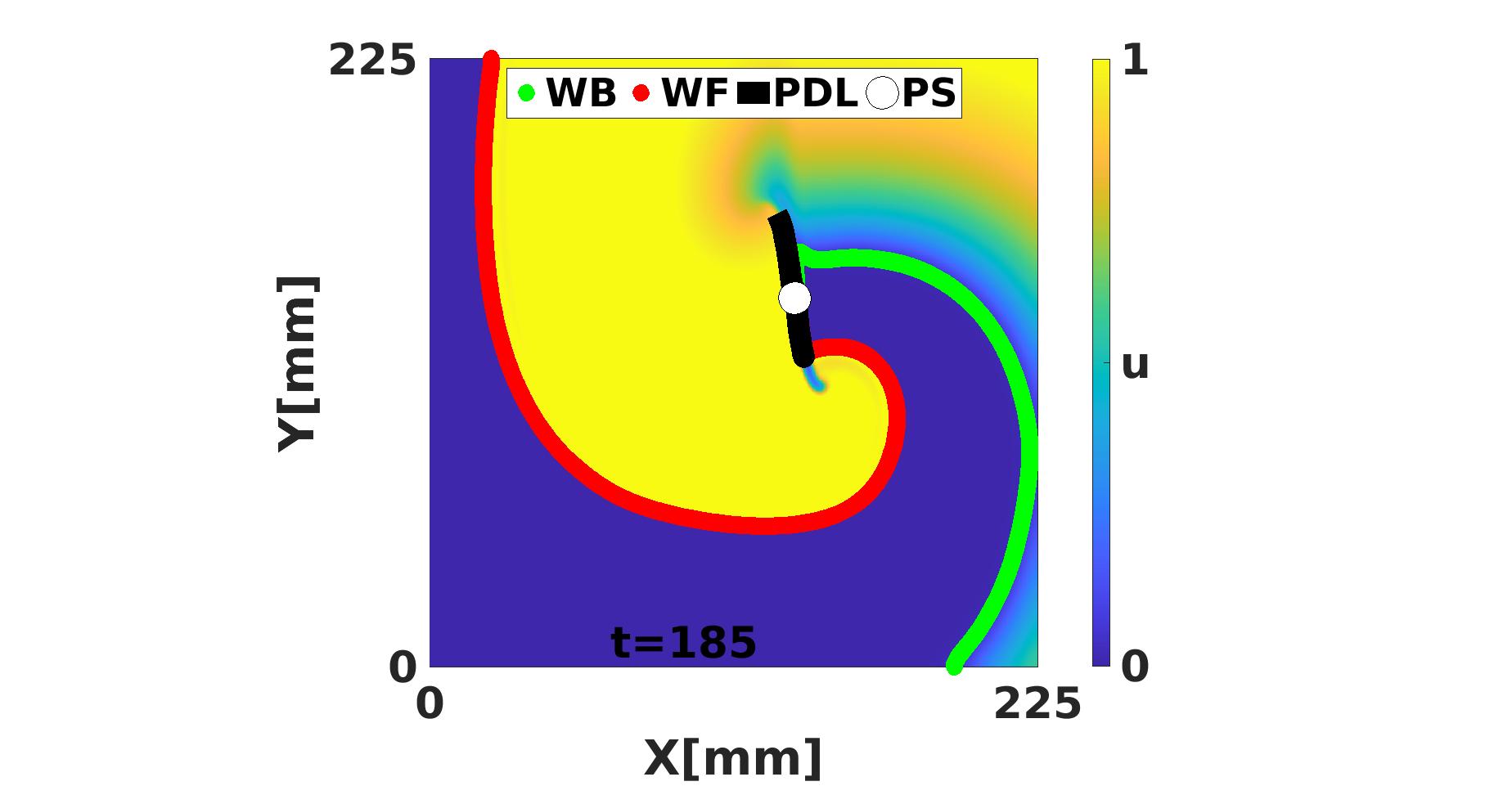}
    \includegraphics[trim={12cm 0cm 18.5cm 0.5cm},clip, height= \figrevisit]{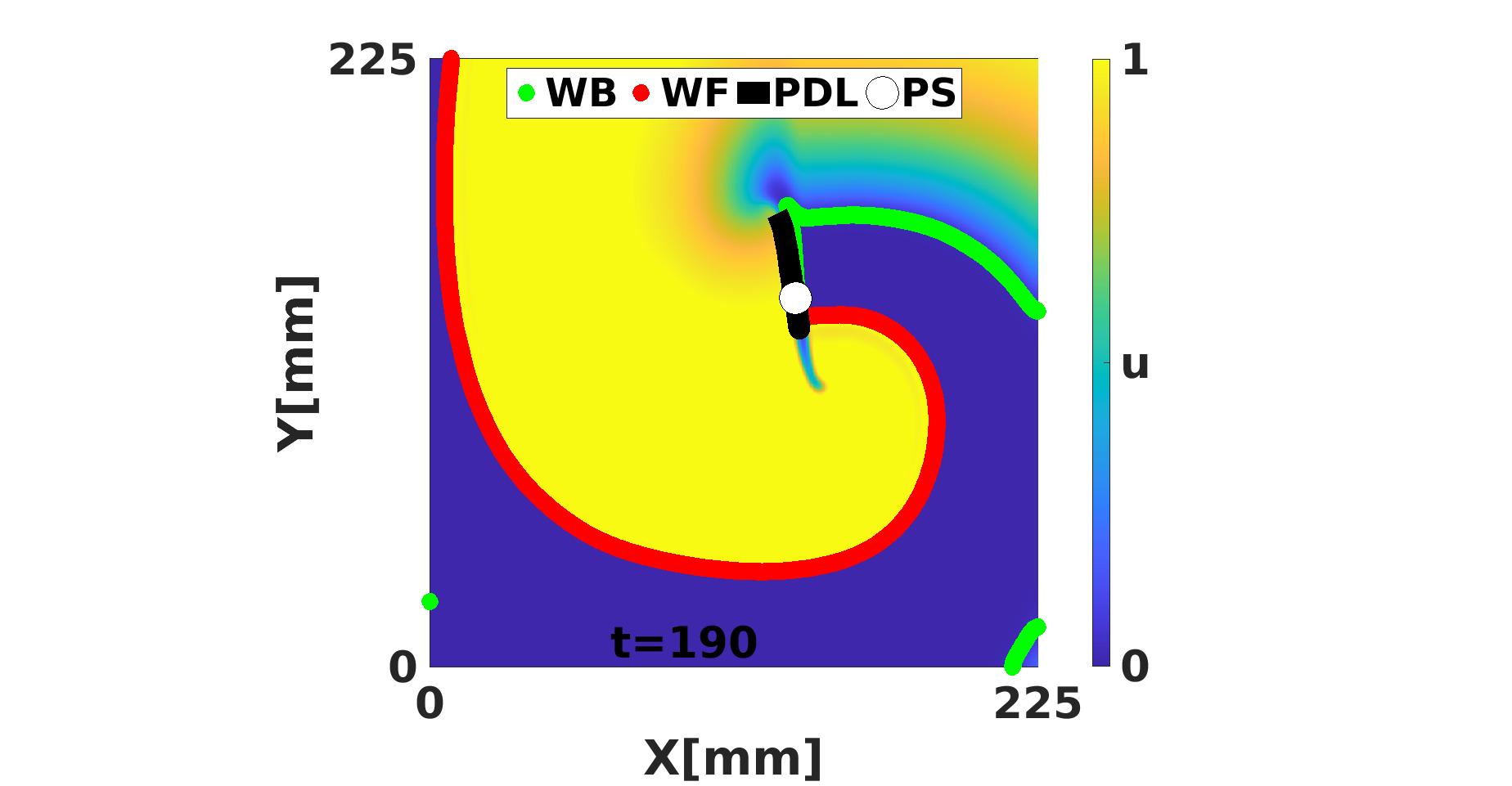}
    \includegraphics[trim={12cm 0cm 8cm 0.5cm},clip, height= \figrevisit]{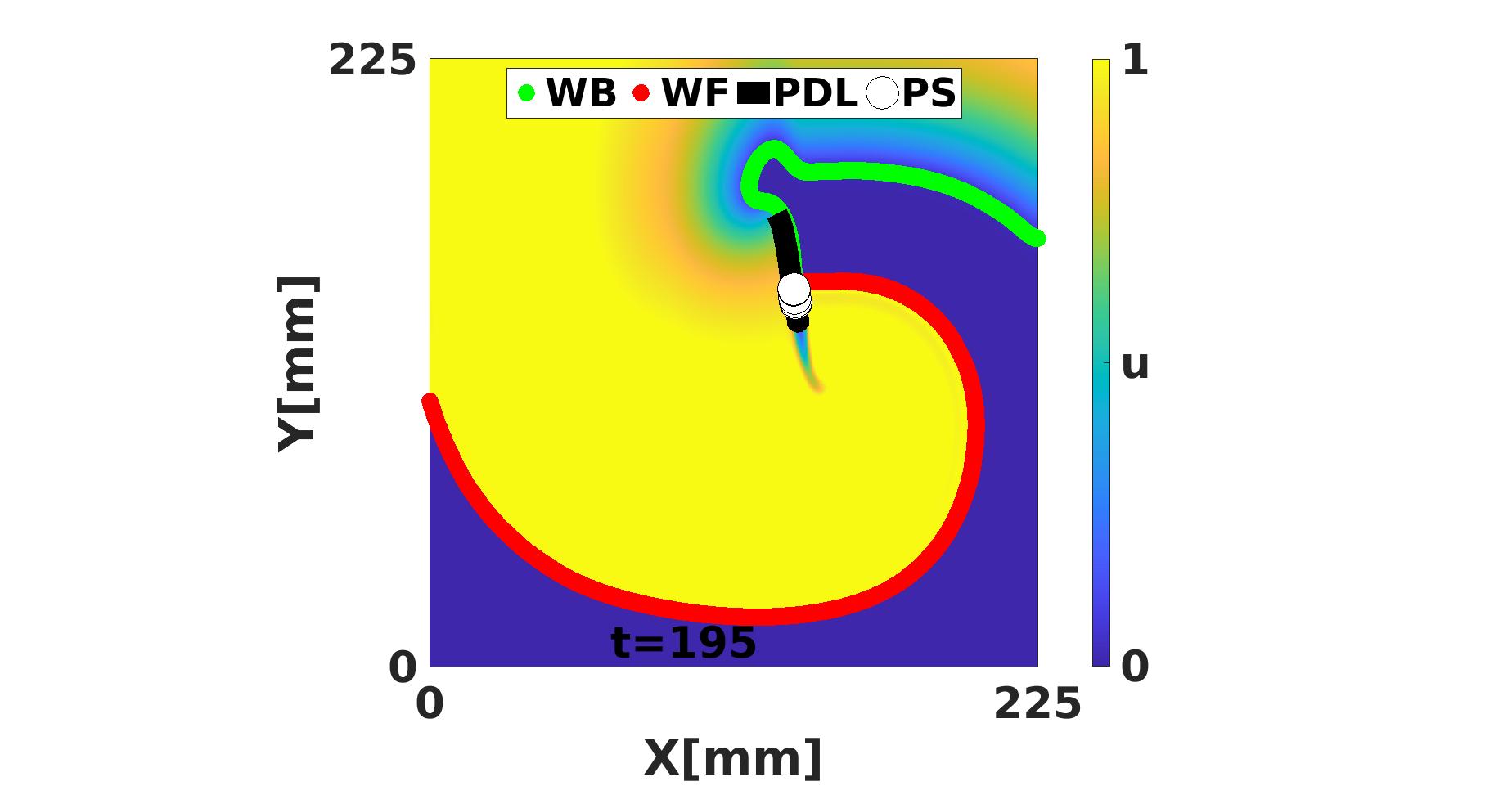} \\
    
         \raisebox{3cm}{\textbf{B}} 
       \includegraphics[trim={12cm 0cm 18.5cm 0.5cm},clip, height= \figrevisit]{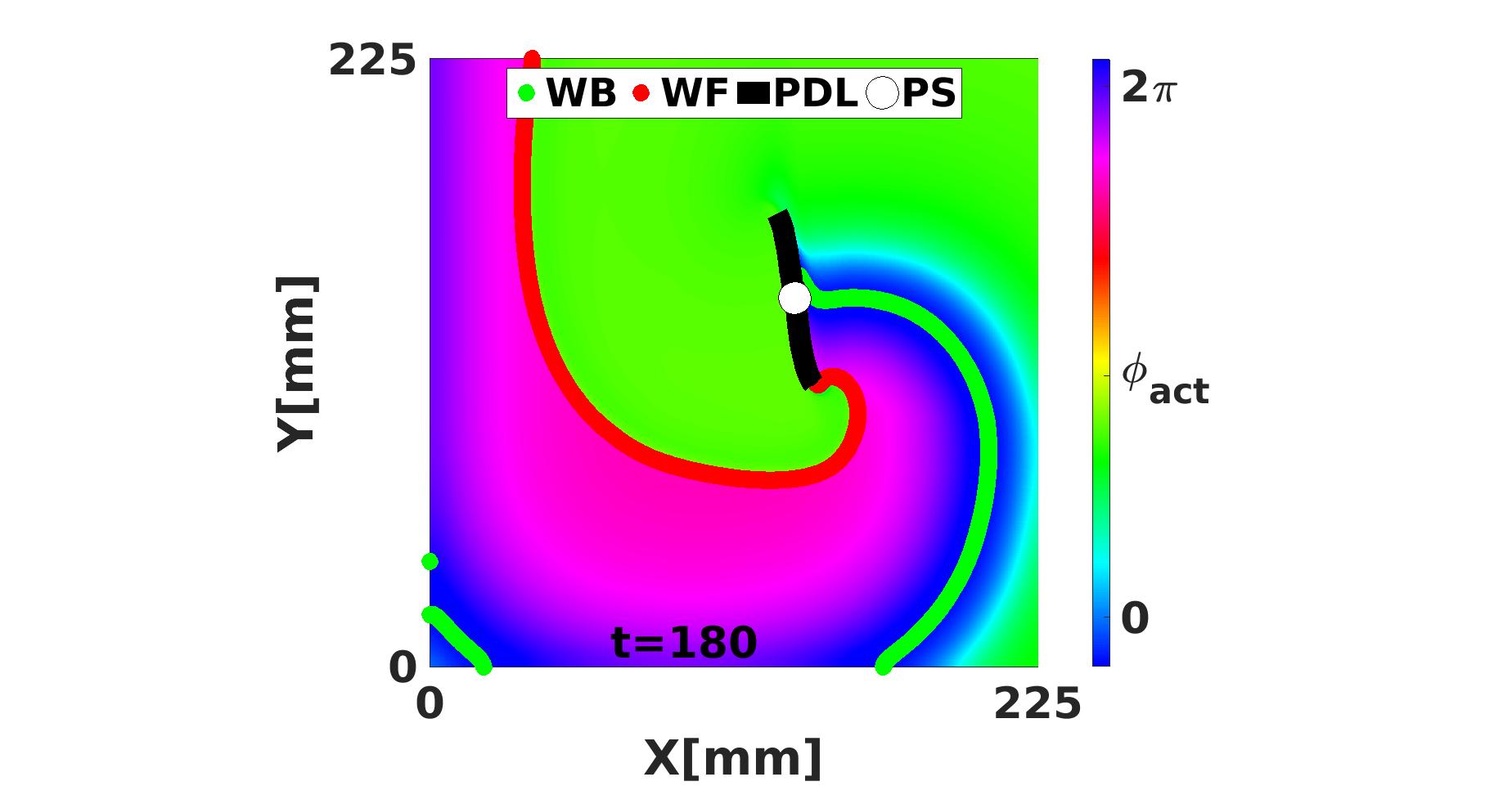}
    \includegraphics[trim={12cm 0cm 18.5cm 0.5cm},clip, height= \figrevisit]{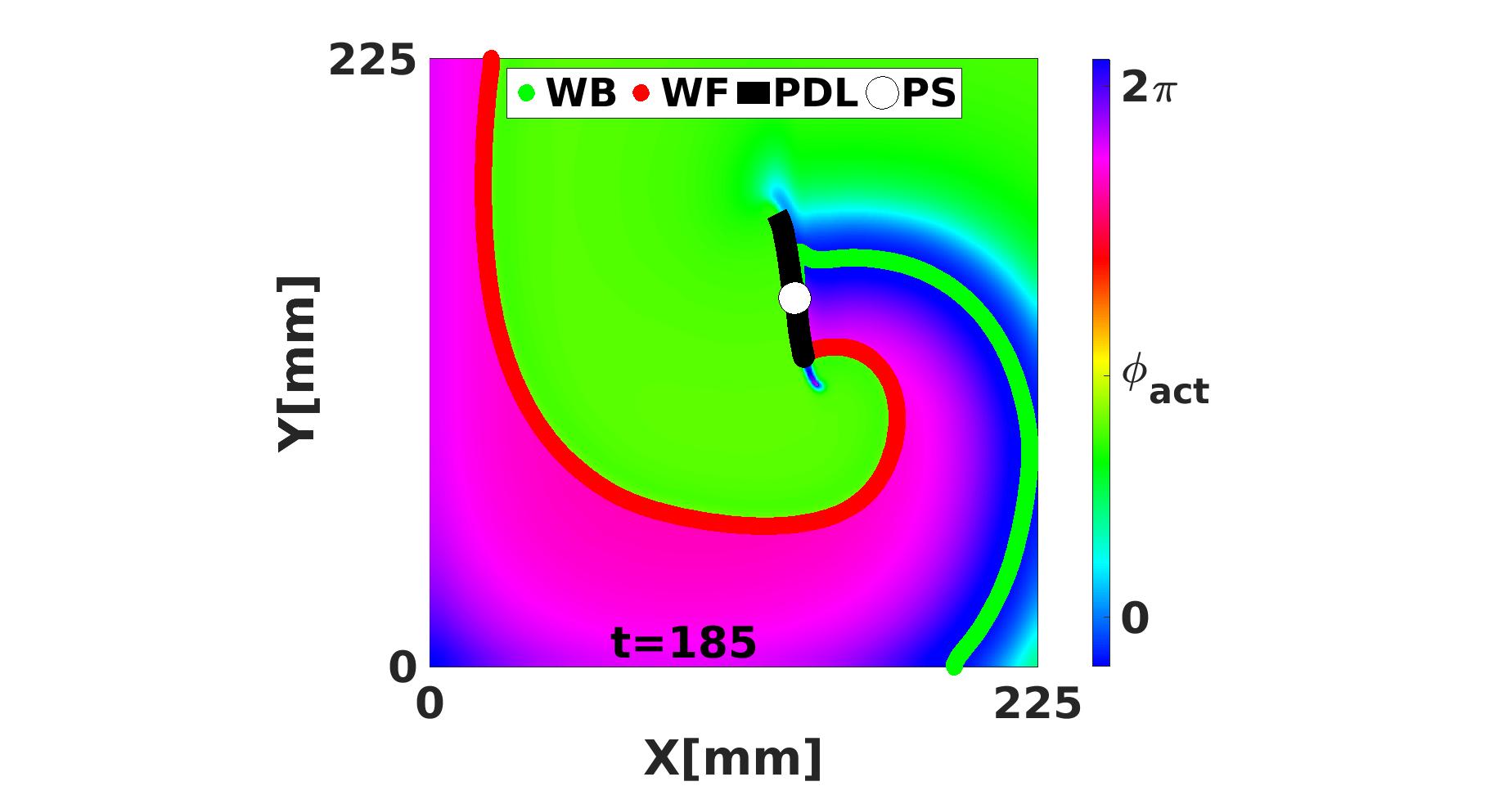}
    \includegraphics[trim={12cm 0cm 18.5cm 0.5cm},clip,height= \figrevisit]{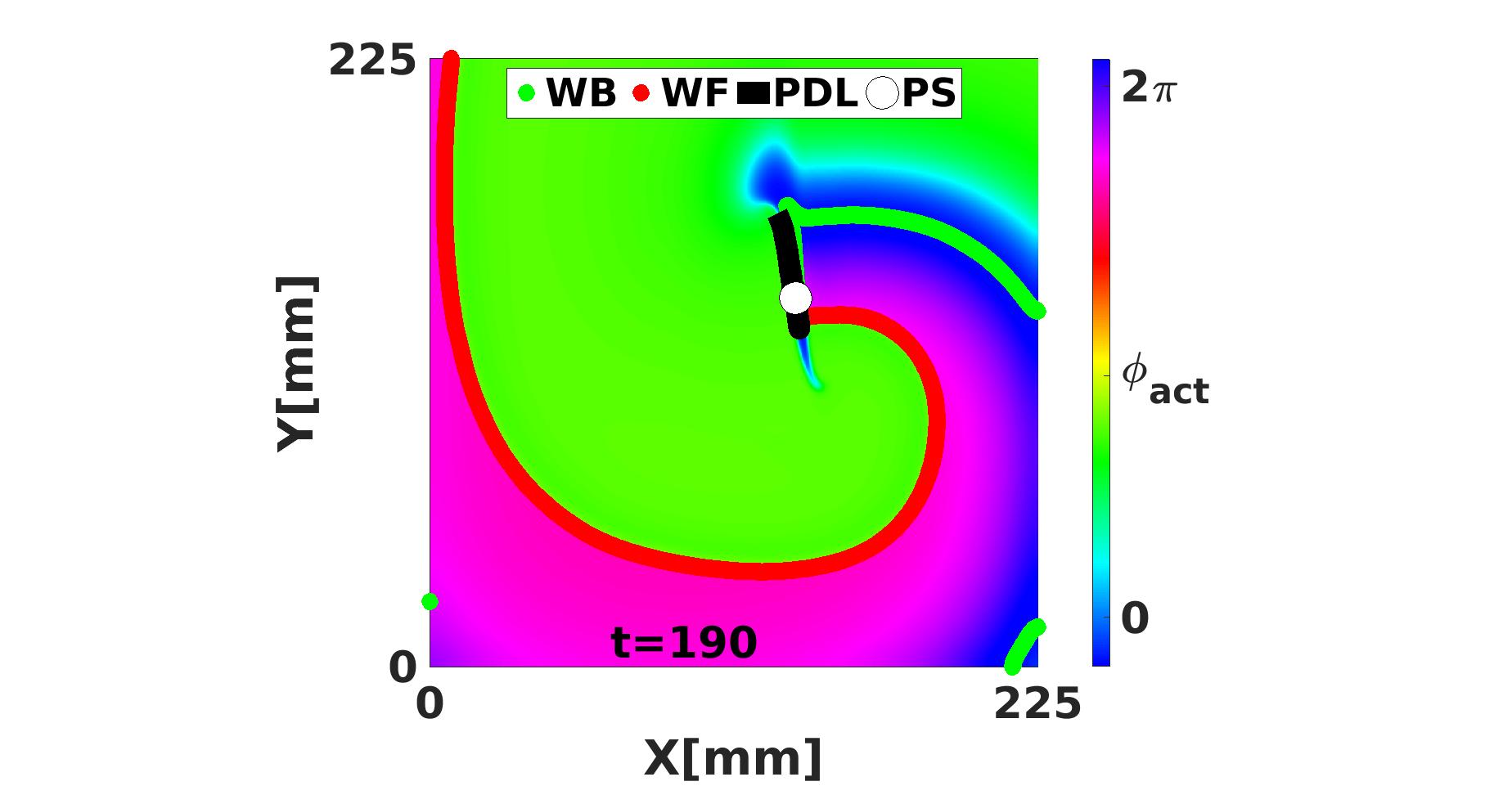}
    \includegraphics[trim={12cm 0cm 8cm 0.5cm},clip, height= \figrevisit]{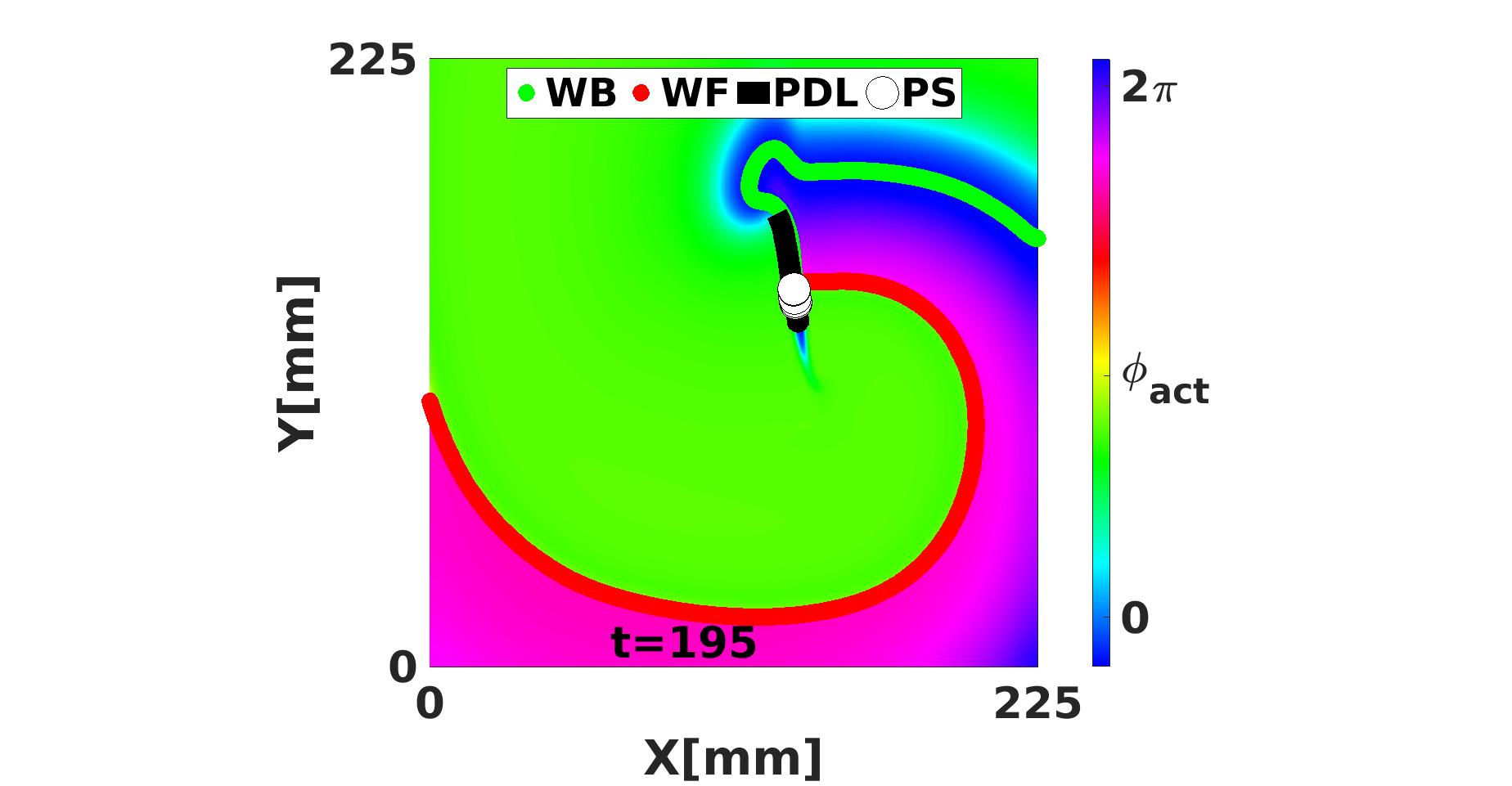} \\
    
    \raisebox{3cm}{\textbf{C}} 
    \includegraphics[trim={12cm 0cm 19cm 0cm},clip, height= \figrevisit]{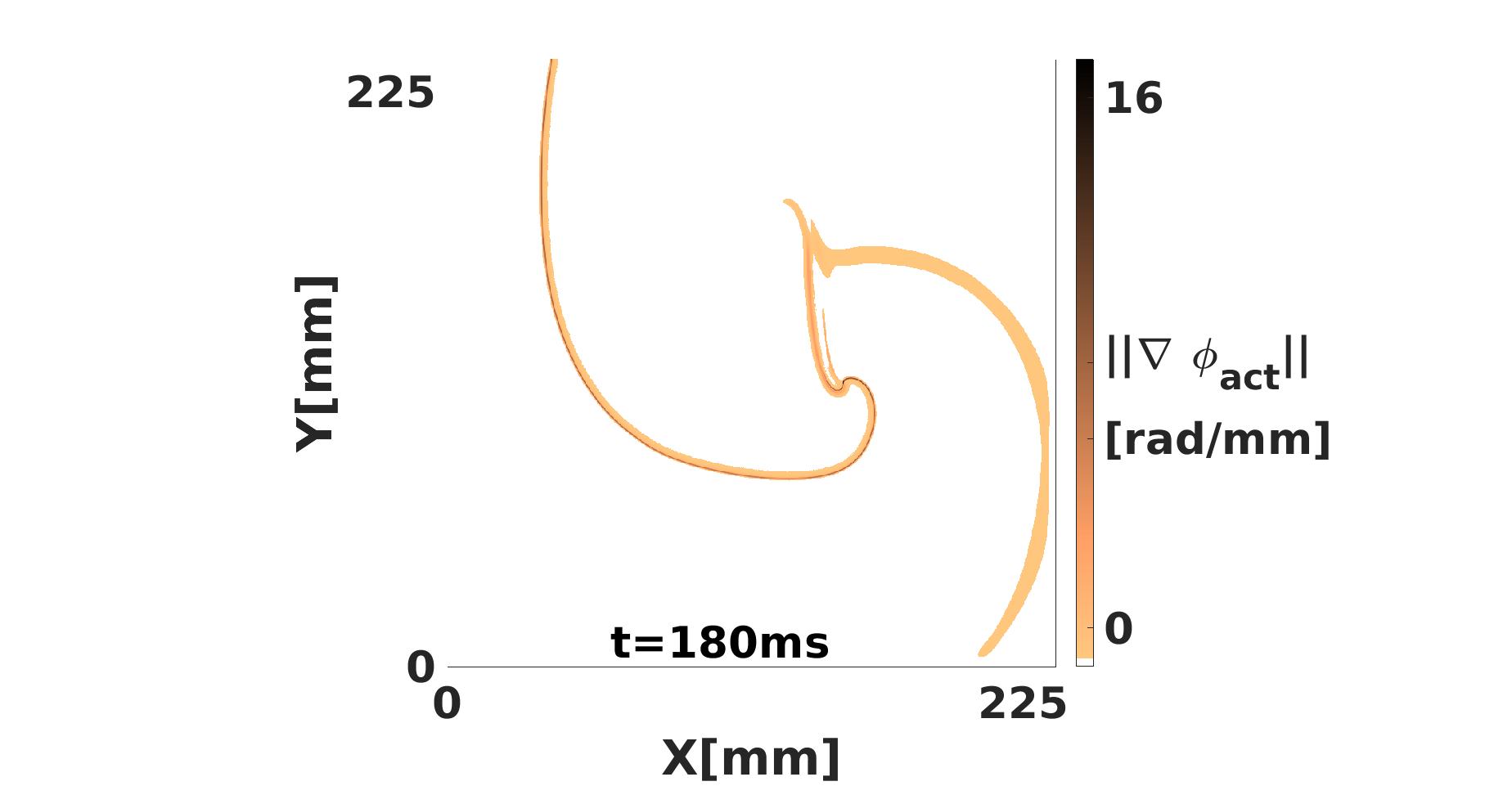}
    \includegraphics[trim={12cm 0cm 19cm 0.5cm},clip, height= \figrevisit]{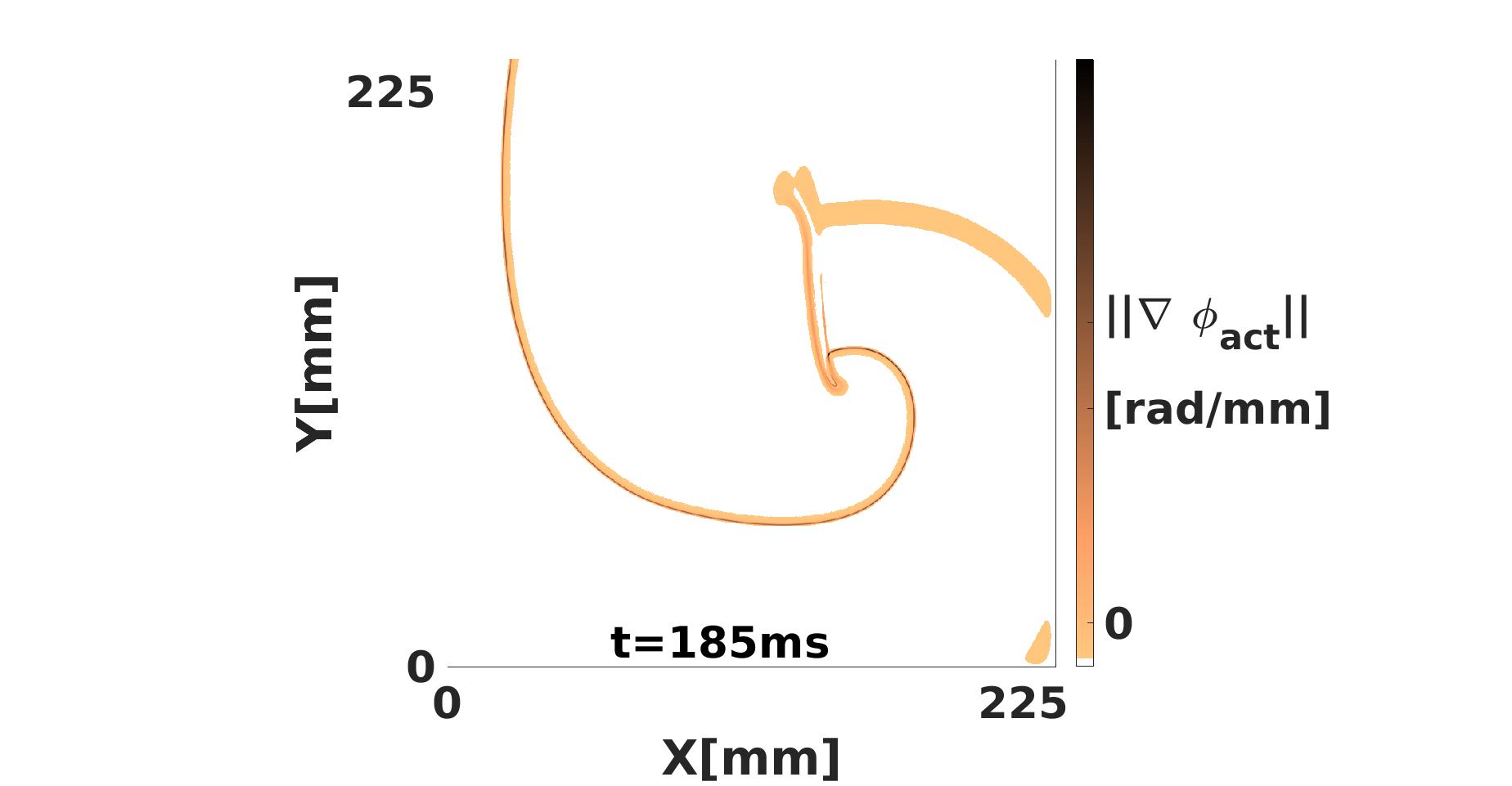}
    \includegraphics[trim={12cm 0cm 19.2cm 0cm},clip, height=\figrevisit]{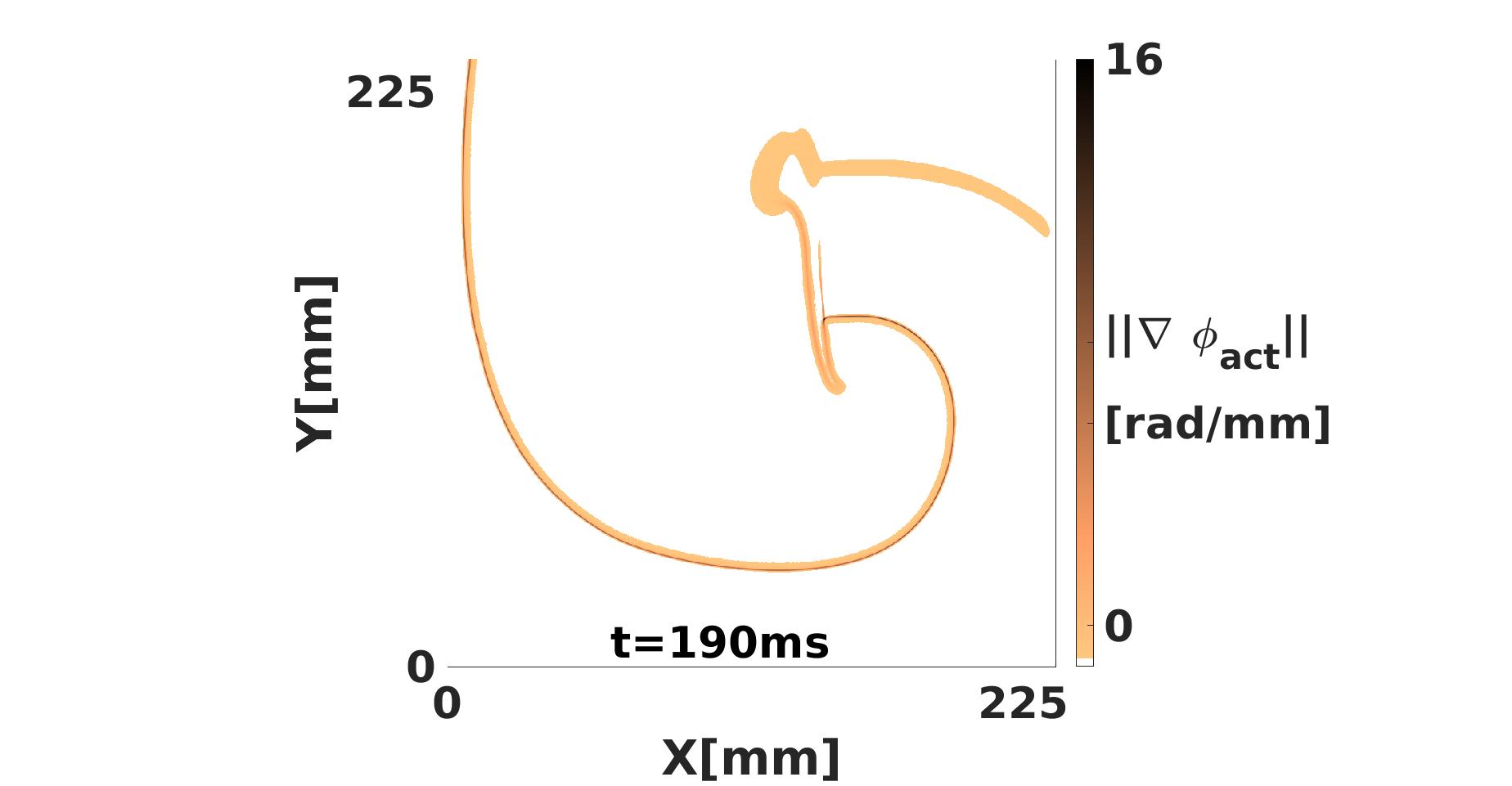}
    \includegraphics[trim={12cm 0cm 7.9cm 0.5cm},clip, height= \figrevisit]{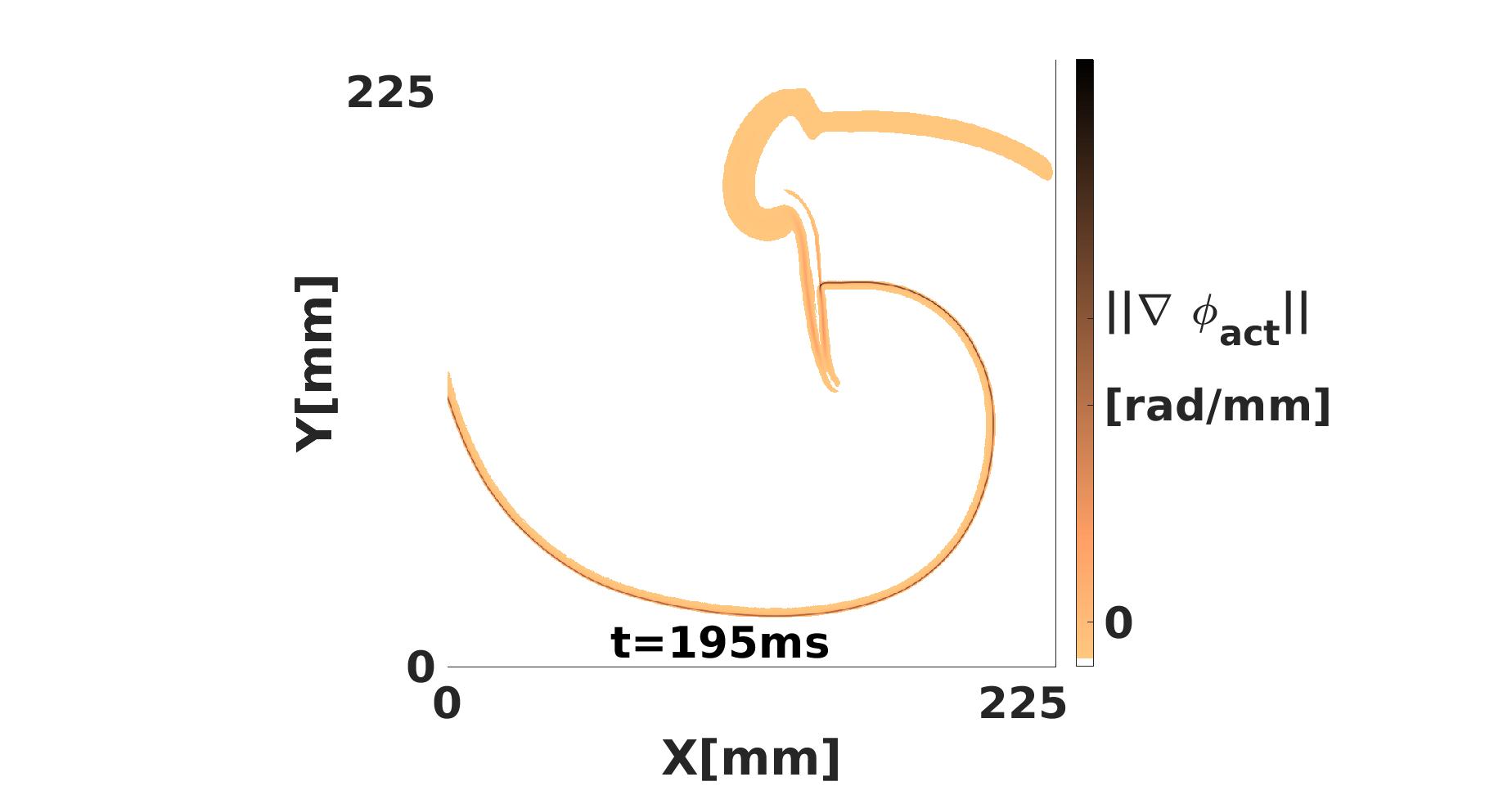}\\
        \caption{Analysis of a simulated linear-core rotor (BOCF model) using activation phase, showing the WF, WB, classical PS and CBL/PDL at different times. The WF and WB were computed as points with $V=\VS$ with positive or negative $dV/dt$. The CBL/PDL was  computed with Eq. \eqref{CBLdetection}.
        (A) Transmembrane potential $u$. The classical PS is located near the position where the WF joins the PDL. At $t=195$, several PSs are found near this intersection.  
        (B) Same frames colored with the classical activation phase, showing sudden transitions at WF, WB and PDL, see gradient of $\phiact$ in (C). }
    \label{fig:S1S2revisited}
\end{figure}

Fig. \ref{fig:S1S2revisited} shows in each row the time evolution of a simulated linear-core rotor (at four different times). When applying classical phase analysis with $\phiact$, see panels B, we note that there are several interfaces that connect excited (panel A, yellow) to non-excited areas (panel A, blue). These interfaces can be shown by plotting the spatial gradient of $\phiact$, see Fig. \ref{fig:S1S2revisited}C. When taking the spatial gradient, it is important to diregard phase differences of $2\pi$. We compute spatial derivatives of phase in the grid point with discrete position $(i,j)$ as: 
\begin{align} 
(\partial_x \phi)_{ij} \approx \ U(\phi_{i+1,j} - \phi_{i,j})/dx
\end{align}
with $U$ the unwrap function, which adds an integer multiple of $2\pi$ to its argument, to bring the result closest to zero:
 \begin{align}
     U(x) = min_{k \in \mathbb{Z}} [ x + 2k \pi ] =  mod( x + \pi, 2\pi) - \pi.
 \end{align}

In the linear-core rotor, we note different steep transitions between zones of approximately equal phase, and discriminate between them as follows. 
We denote by $\phi_1$ a phase in the middle of the upstroke (depolarization), and $\phi_2$ a phase in the middle of the down stroke (repolarization). 
These values are found as
\begin{align}
    \phi_1 &= mean \{ \phi(\vec{r},t) | V(\vec{r},t) = \VS \ \mathrm{and}\ \frac{dV}{dt}(\vec{r},t) >0\}, \nonumber \\
    \phi_2 &= mean \{ \phi(\vec{r},t) | V(\vec{r},t) = \VS \ \mathrm{and}\ \frac{dV}{dt}(\vec{r},t) < 0\}. \label{defphi12}
\end{align}
Thereafter, the wave front (WF) and wave back (WB) can be defined as (with $\heartsuit$ depicting the domain or cardiac tissue): 
\begin{align}
     WF (t)  = \{\vec{r} \in \heartsuit | \phiact(\vec{r},t) = \phi_1 \}, \label{WF} \\
     WB (t)  = \{\vec{r} \in \heartsuit | \phiact(\vec{r},t) = \phi_2 \}. \nonumber
\end{align}
In the classical view, the only way to connect WF and WB is in a PS, where different phases $\phi_1$ and $\phi_2$ can meet. 

\begin{figure}
     \centering
     \includegraphics[width=0.3\textwidth]{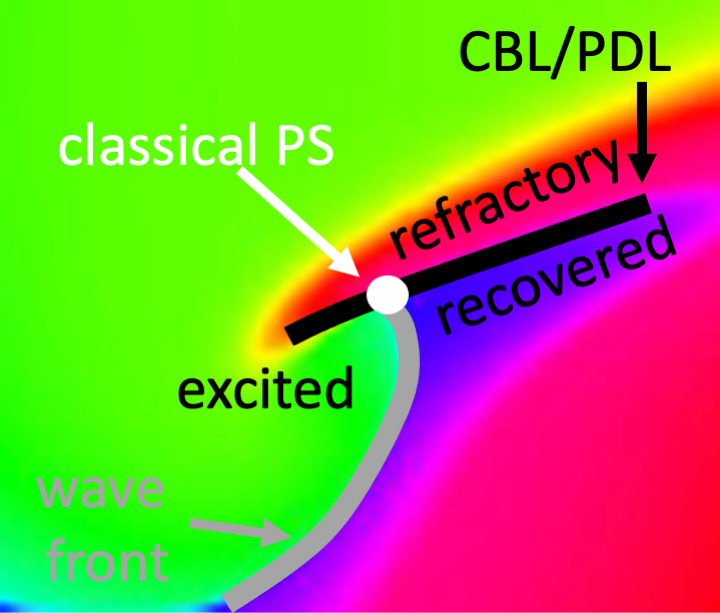}
     \caption{Closer look at a linear-core rotor. At the location where classical methods detect a PS, three distinct phases come together: recovered, excited and refractory tissue. At either side of the CBL, two distinct phases are present: refractory vs. either excitable or recovered. Therefore, the CBL is a phase defect line (PDL).}
     \label{fig:phasediagram}
 \end{figure}

 However, if a wave hits unrecovered (refractory) tissue, unidirectional conduction block occurs and a CBL is formed. In that case, a third kind of interface is seen, in addition to WF and WB, see Fig. \ref{fig:phasediagram}. Since the tissue at either side of a CBL was activated at a different time, it also has different phase, see different colors across the black line in Fig. \ref{fig:phasediagram}. This observation implies that not only the end point of a wave front (the classical PS) is a special point, but that all points on a CBL have a non-trivial phase structure: they are situated on a region where phase suddenly transitions from one value to another, which we call a phase defect line (PDL). 
 
 Note that the PDL shown here persists in time until either both sides have fully recovered, or when one side of the line is re-excited again, while the other is not. 



Fig. \ref{fig:S1S2revisited}C shows the norm of $\vec \nabla \phiact$, which visualizes the domain walls between the different zones of similar phase; note that without further filtering, the transition zones consist of the union of WF, WB and PDL.  

\subsubsection{An alternative phase definition based on LAT}


We find it useful to discriminate WF and WB from CBLs, since the former are `natural' behaviour in an excitable system, while conduction block can lead to the formation of abnormal patterns and arrhythmias and therefore, deserves to be detected separately. Therefore, we now introduce another definition of phase, which is only sensitive to CBLs.

As we described before, the classical definition of phase is not unique, as one could make various choices for the observables $V$ and $R$, and their threshold values $\VS$, $\RS$ can be also chosen at will, as long as they are located within the cycle in the $(V,R)$-plane. 

The most important feature of the phase is its periodicity. For convenience, we choose the phase of the resting state to be phase equal to $0$. Apart from that, there is need to let the phase correspond to the polar angle in one choice of $(V,R)$-coordinates: we can define new equivalent phases $\phi$ by a continuous mapping of $\phiact$: 
\begin{align}
    \phi &= h(\phiact), & h(0) &= 0, & h(2\pi) &=2\pi, & h'(\phiact) &>0.
\end{align}
The last condition ensures that the transformation between the phases is invertible. In a mathematical sense, fixing a phase is nothing but introducing coordinates on the inertial manifold (dynamical attractor) on which the dynamics takes place. As with coordinates in the plane or on a surface, many choices are possible, and depending on the circumstances, some may be more appropriate than others. 

\begin{figure}
    \centering
\raisebox{5.5cm}{\textbf{A}}    \includegraphics[trim={8cm 0cm 7.5cm 1cm},clip,width=0.45\textwidth]{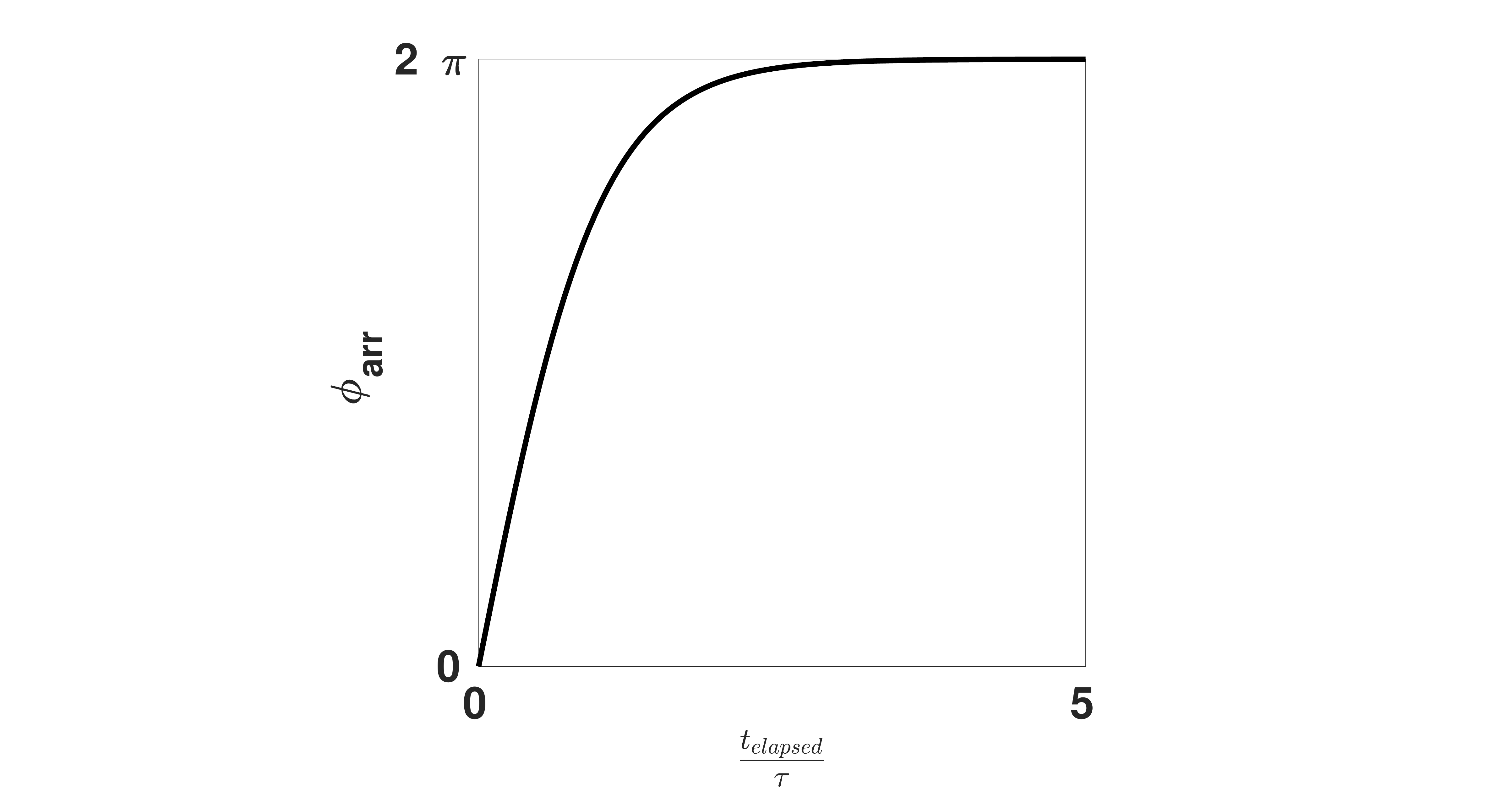}
\raisebox{5.5cm}{\textbf{B}}    \includegraphics[trim={8cm 0cm 7.5cm 1.5cm},clip,width=0.45\textwidth]{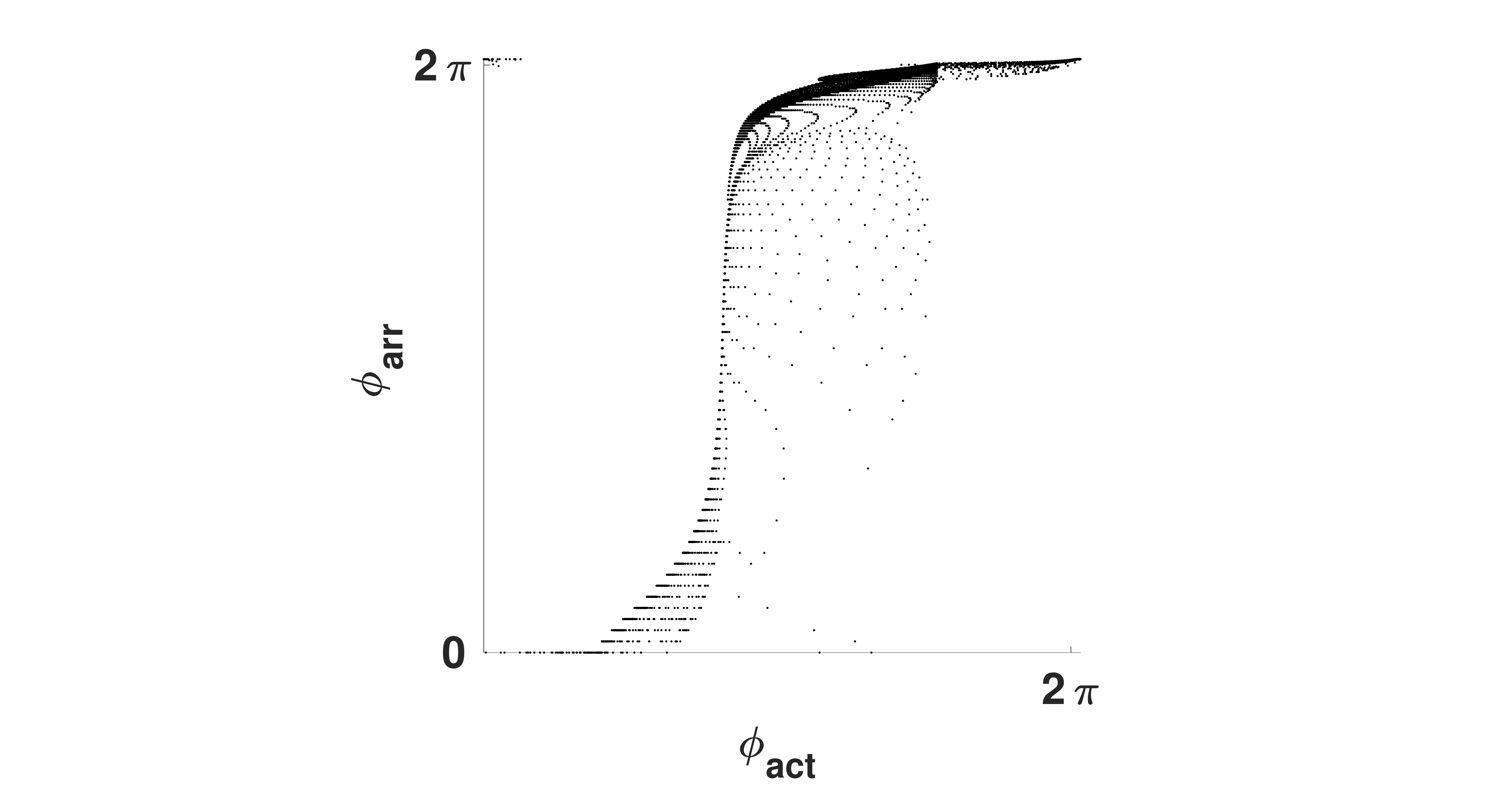} \\
\raisebox{5.5cm}{\textbf{C}}    \includegraphics[trim={8cm 0cm 7.5cm 1.2cm},clip,width=0.45\textwidth]{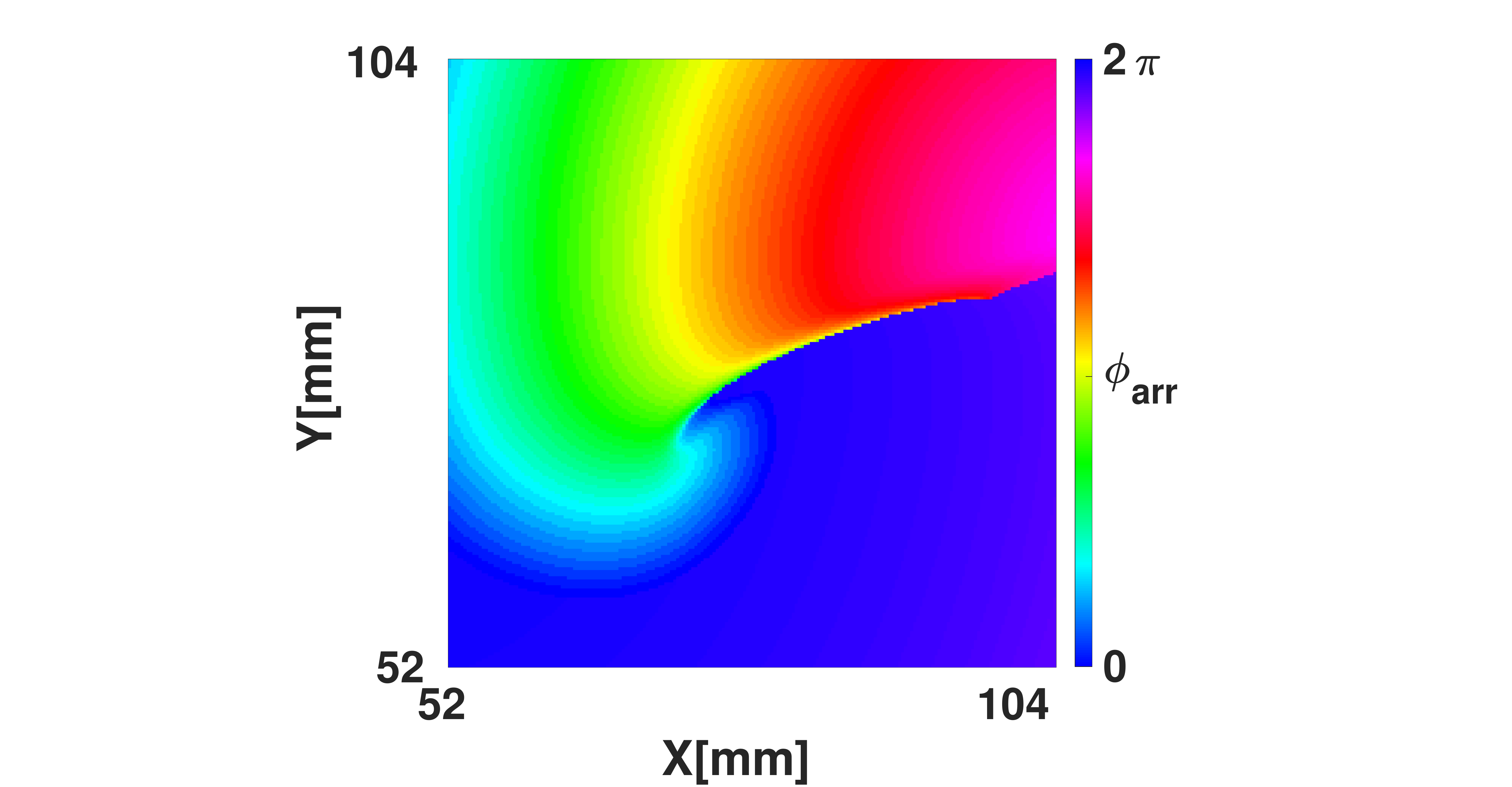}
\raisebox{5.5cm}{\textbf{D}}    \includegraphics[trim={8cm 0cm 7cm 1.5cm},clip,width=0.45\textwidth]{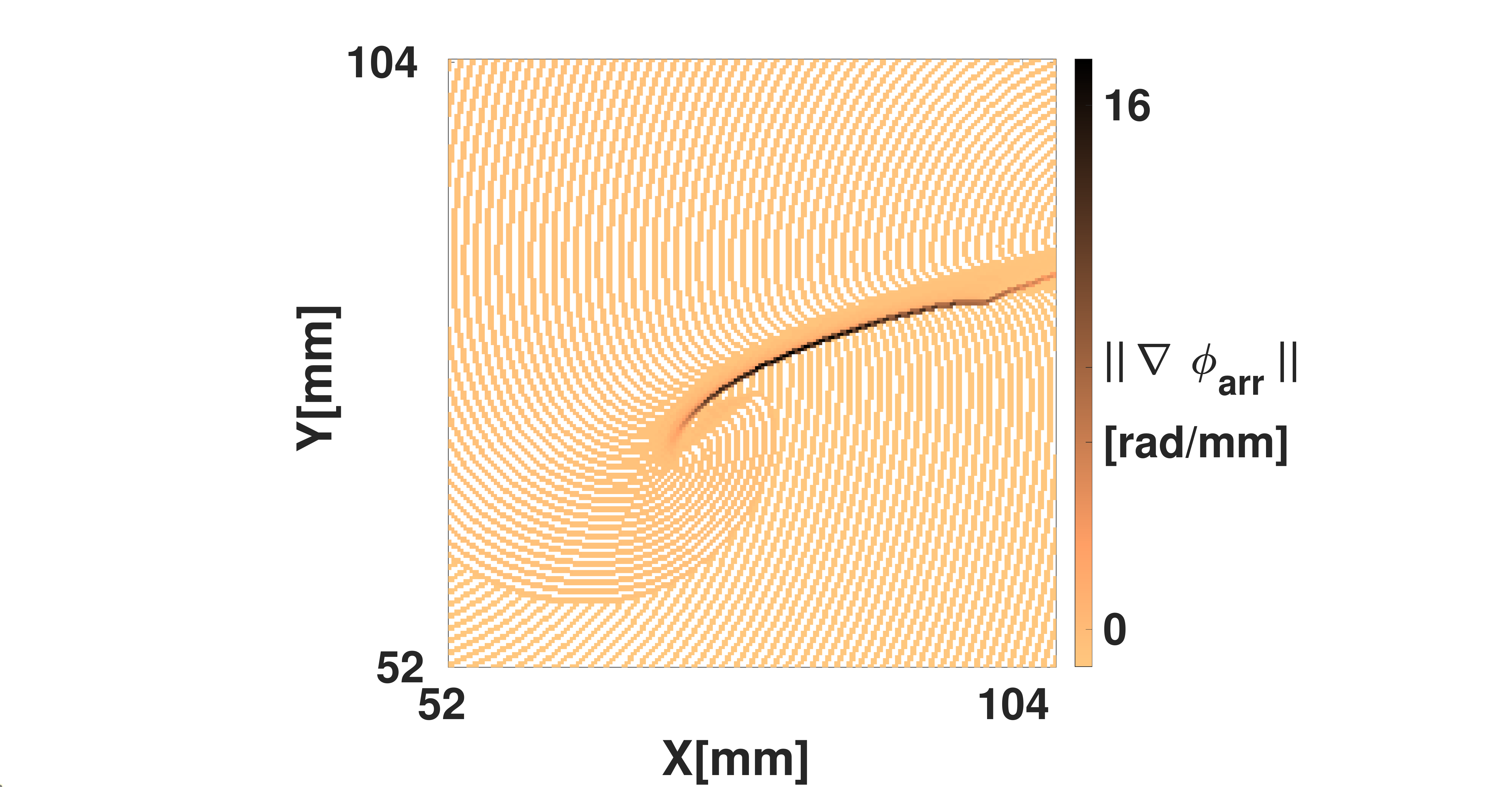}
     \caption{Phase defects shown using the LAT-based phase $\phiarr$. (A) The function $\phiarr(\tlap)$ from Eq. \eqref{phiarr}. (B) Scatter plot of $\phiarr$ vs. $\phiact$, showing that one is a reparameterization of the other on the interval $[0, 2\pi]$. (C) Same linear-core rotor in the FK model as in Fig. \ref{fig:problems}B, now shown with $\phiarr$. Note that WF and WB are no longer showing abrupt phase variations, these only happen at the PDL, i.e. the points where conduction block happened. (D) Magnitude of the gradient of $\phiarr$. Due to the discrete sampling of LAT, a staircase artefact in the gradient is seen (no smoothing was applied here). 
    }
    \label{fig:phiarr}
\end{figure}

To link the concept of LAT with phase, we propose to take the elapsed time
\begin{align}
\tlap(\vec{r}) = t - \tact(\vec{r})
\end{align}
as the new phase, since in leading order (neglecting inhomogeneity, electrotonic effects and long-term memory), the state of a point will depend on the elapsed time since previous activation. However, $\tlap \in \mathbb{R}^+$, which is an unbound interval, such that $\tlap$ itself is not a suitable replacement for $\phiact$. To this purpose, we apply a sigmoidal function to $\tlap$, in order to define an `arrival time phase': 
\begin{align}
\phiarr(\vec{r},t) = 2 \pi \tanh (3 \tlap(\vec{r},t) / \tau). \label{phiarr}
\end{align}
Here, $\tau$ is a characteristic time constant of the medium, e.g. the mean action potential duration. Fig. \ref{fig:phiarr}A-B show how $\phiarr$ depends on both $t$ and $\phiact$ for FK reaction kinetics. From this, it can be seen that, to a good approximation, $\phiarr$ is a reparameterization of $\phiact$. The new phase $\phiarr$ is not constant in time, but gets update at the places where the wave front excites tissue, just like the updates in LAT. It relies on two parameters, which we chose manually for now: the timescale $\tau$ and the critical threshold $\VS$ to consider tissue as excited, see definition \eqref{WF}. 

Importantly, $\phiarr$ unifies the classical phase description (since $\phiact$ is a phase) with the LAT description (since it is defined based on LAT). When a linear-core rotor is plotted using $\phiarr$, we see that the WF and WB are not strongly emphasized anymore, since both occur near $\phiarr= 0$ (omitting phase differences of $2\pi$). Still visible as a spatial phase transition is the region of conduction block, or the PDL, see Fig. \ref{fig:phiarr}C. 

The `arrival time phase' $\phiarr$ is thus a kind of hybrid between classical phase and the LAT viewpoint. For, the lines of equal phase will now correspond to the isochrones that commonly appear in medical literature and our Fig. \ref{fig:problems}B. At a CBL, LAT changes discontinuously across the conduction block site, and this phase defect region now clearly exhibits a discontinuous phase. So, depending on the definition of phase used ($\phiact$ or $\phiarr$ or another one) the phase defect will be either a steep but continuous phase transition, or a true discontinuity in phase. 

\begin{figure}
      \raisebox{2.5cm}{\textbf{A}} 
     \includegraphics[trim={12cm 0cm 18.5cm 0.5cm},clip, height=\figspiralphiarr ]{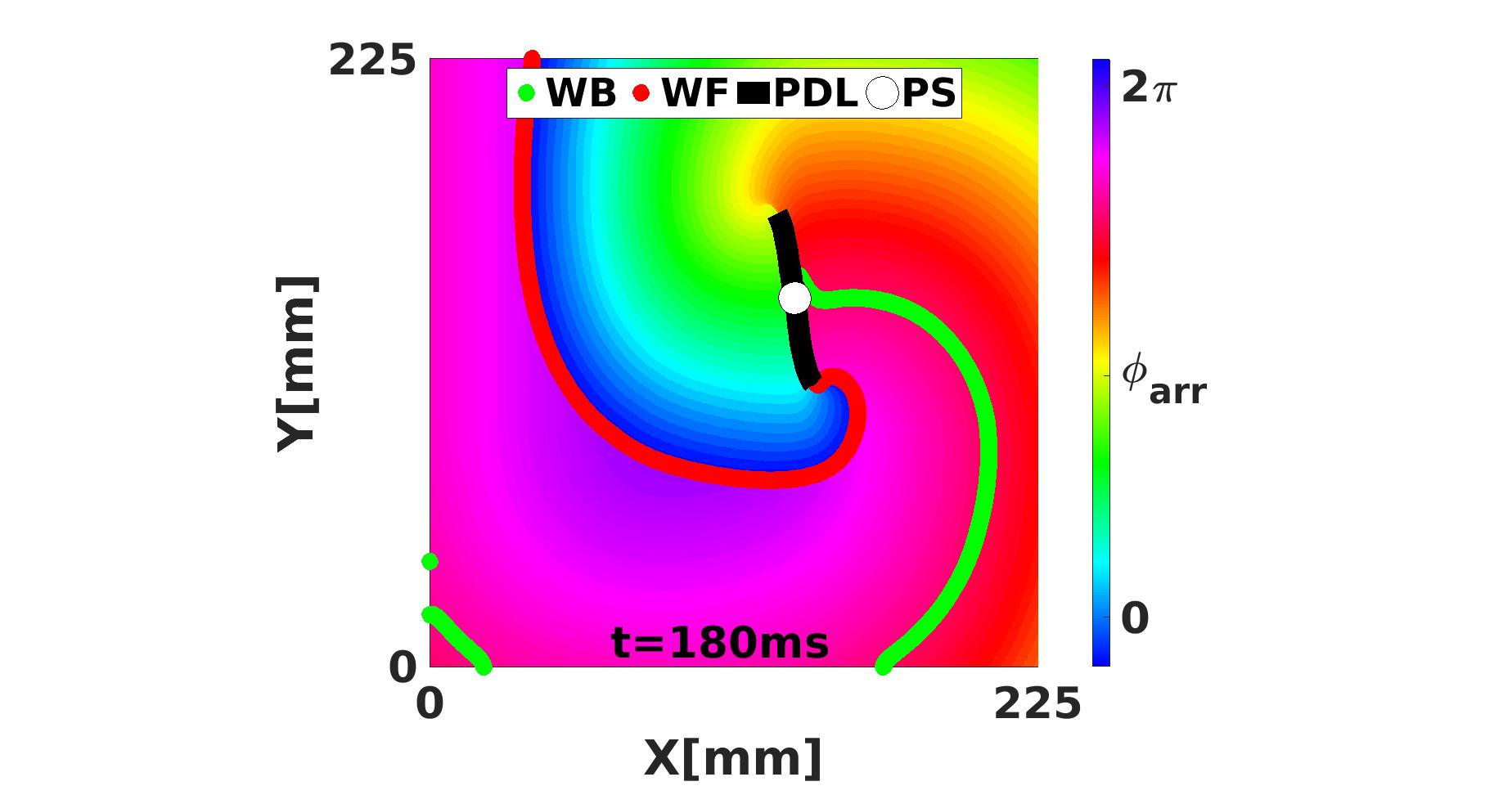}
    \includegraphics[trim={12cm 0cm 18.5cm 0.5cm},clip,height=\figspiralphiarr]{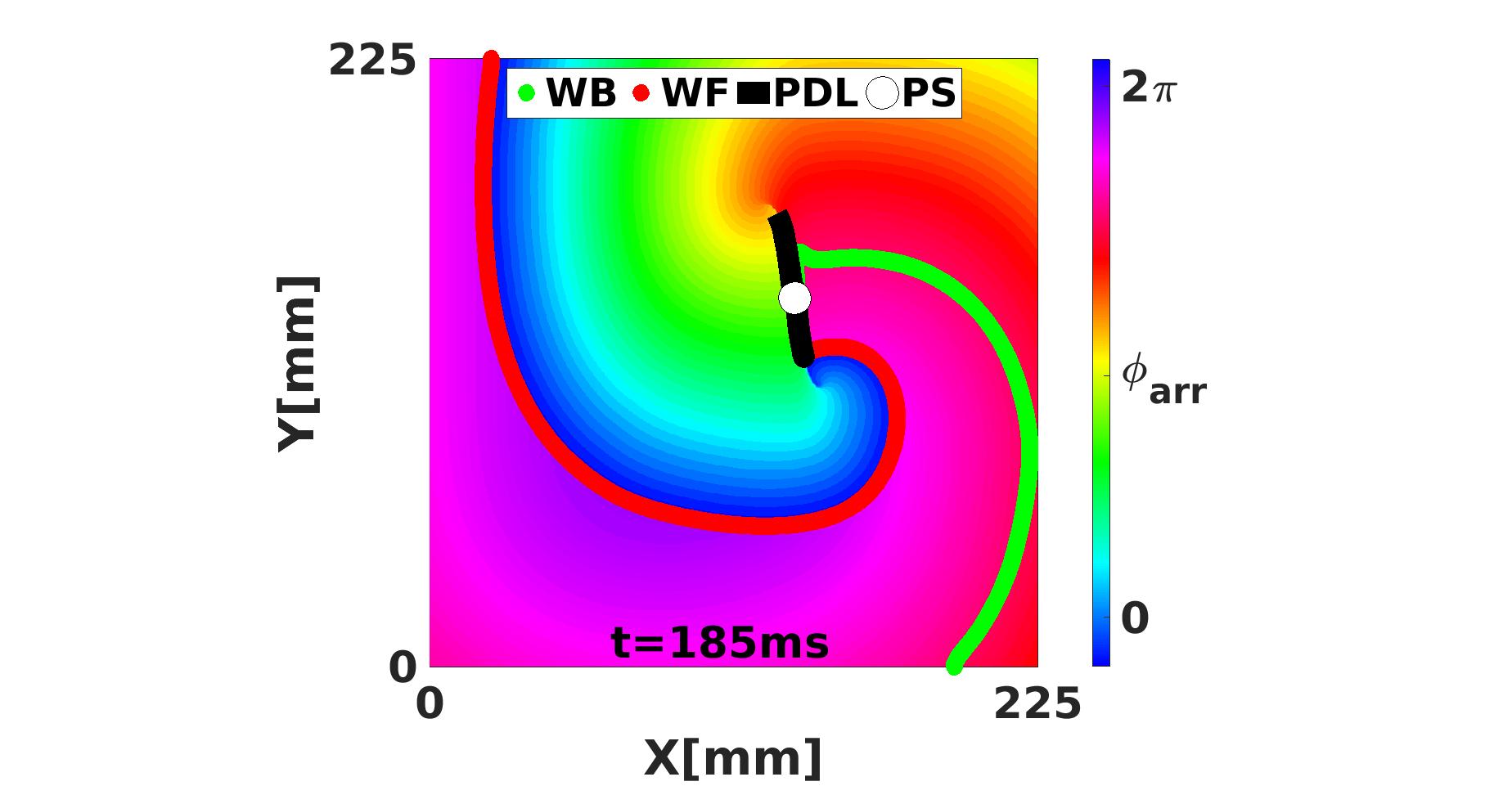}
    \includegraphics[trim={12cm 0cm 18.5cm 0.5cm},clip, height= \figspiralphiarr]{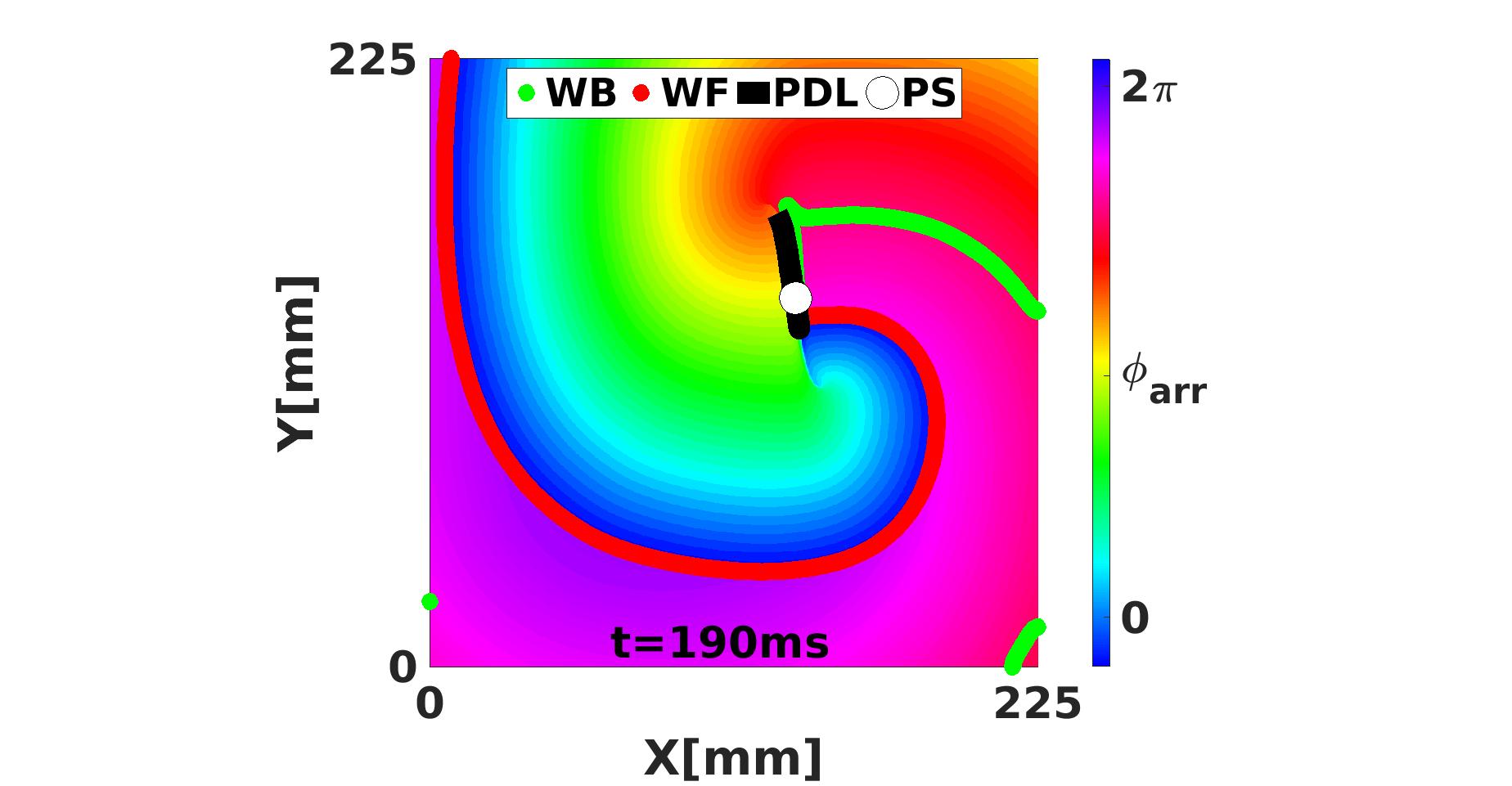}
    \includegraphics[trim={12cm 0cm 8cm 0.5cm},clip, height= \figspiralphiarr]{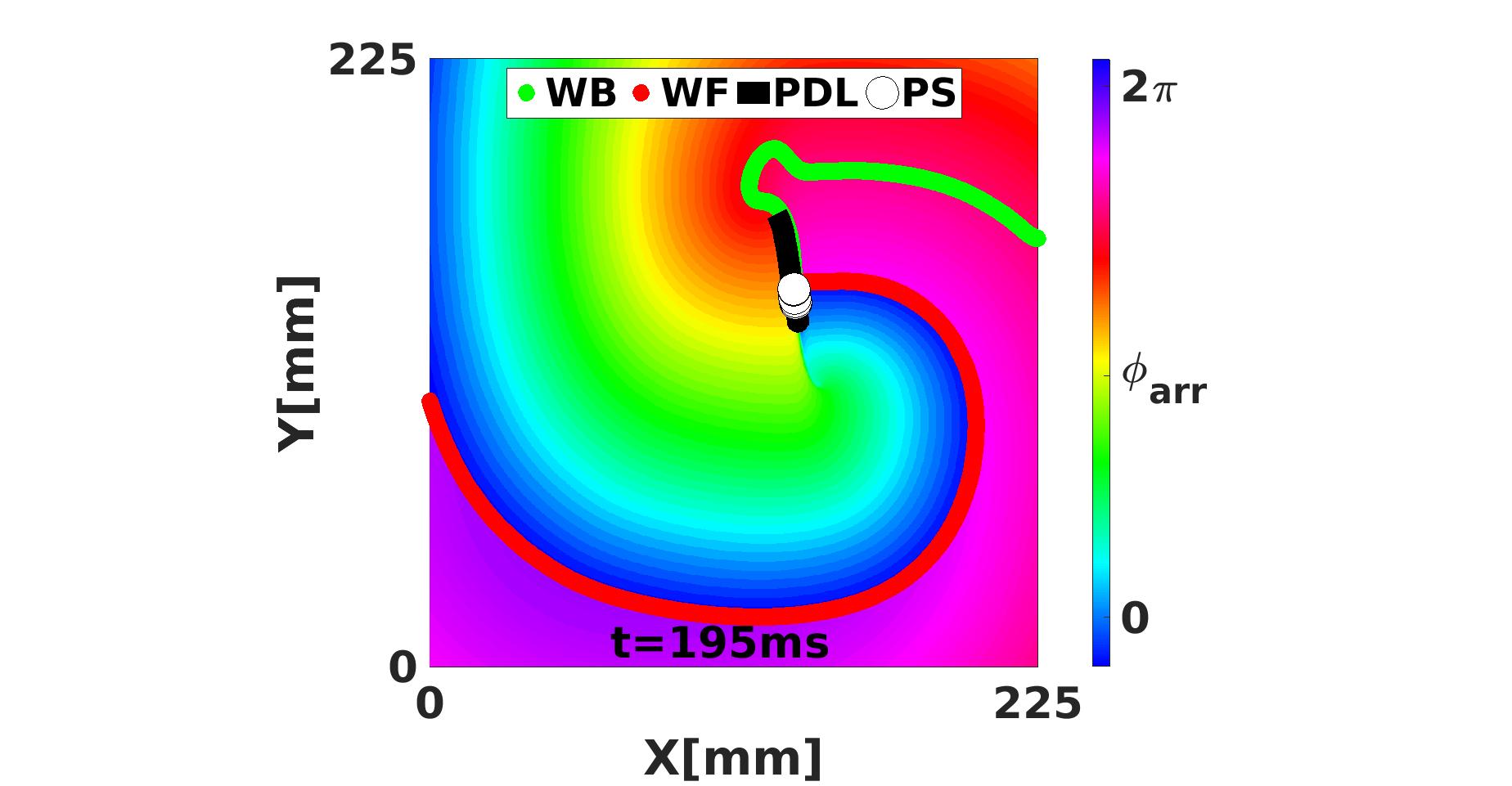} \\
    
    \raisebox{2.5cm}{\textbf{B}} 
    \includegraphics[trim={12cm 0cm 19cm 0.5cm},clip, height= \figspiralphiarr]{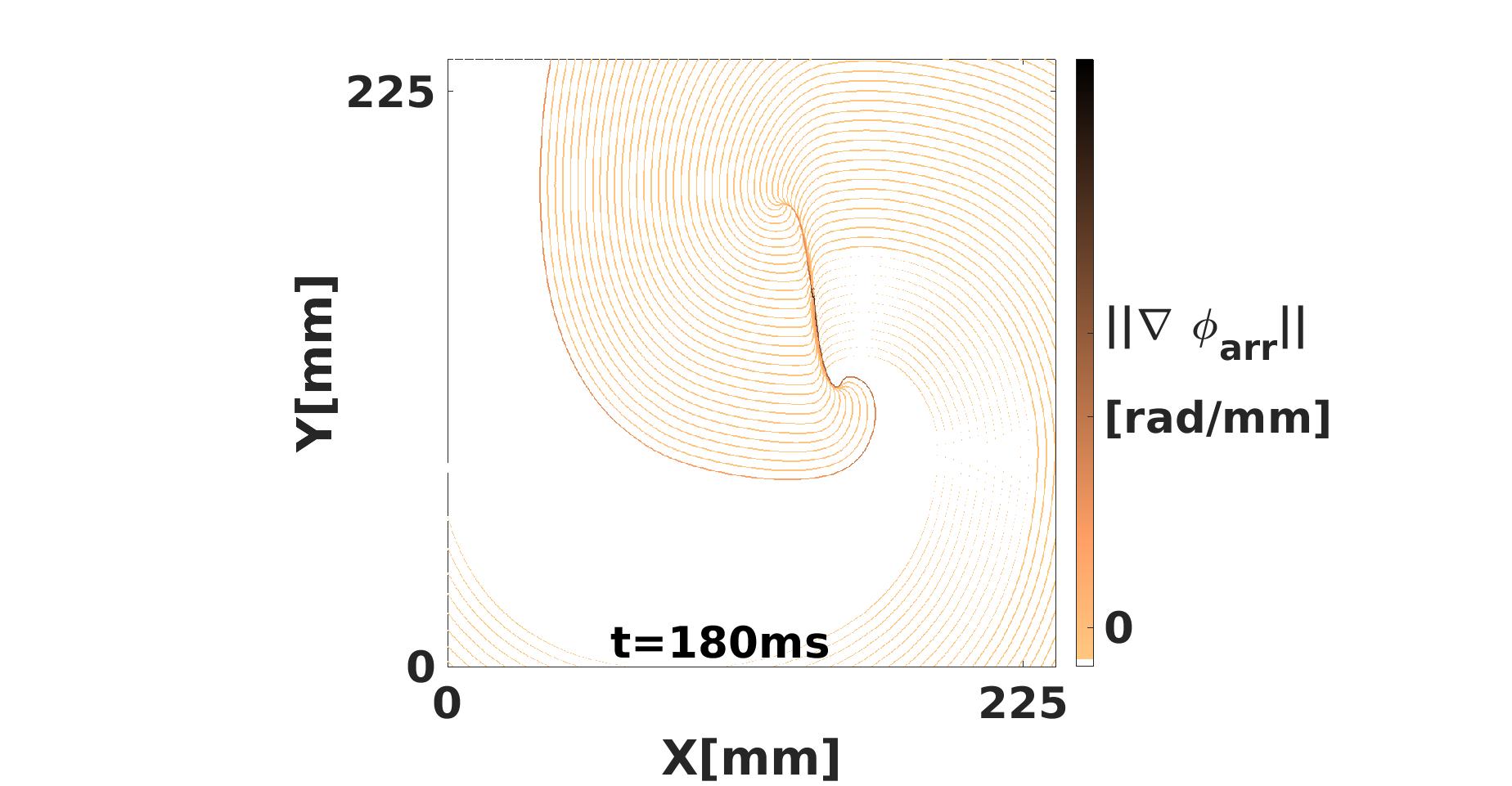}
    \includegraphics[trim={12cm 0cm 19cm 0.5cm},clip, height= \figspiralphiarr]{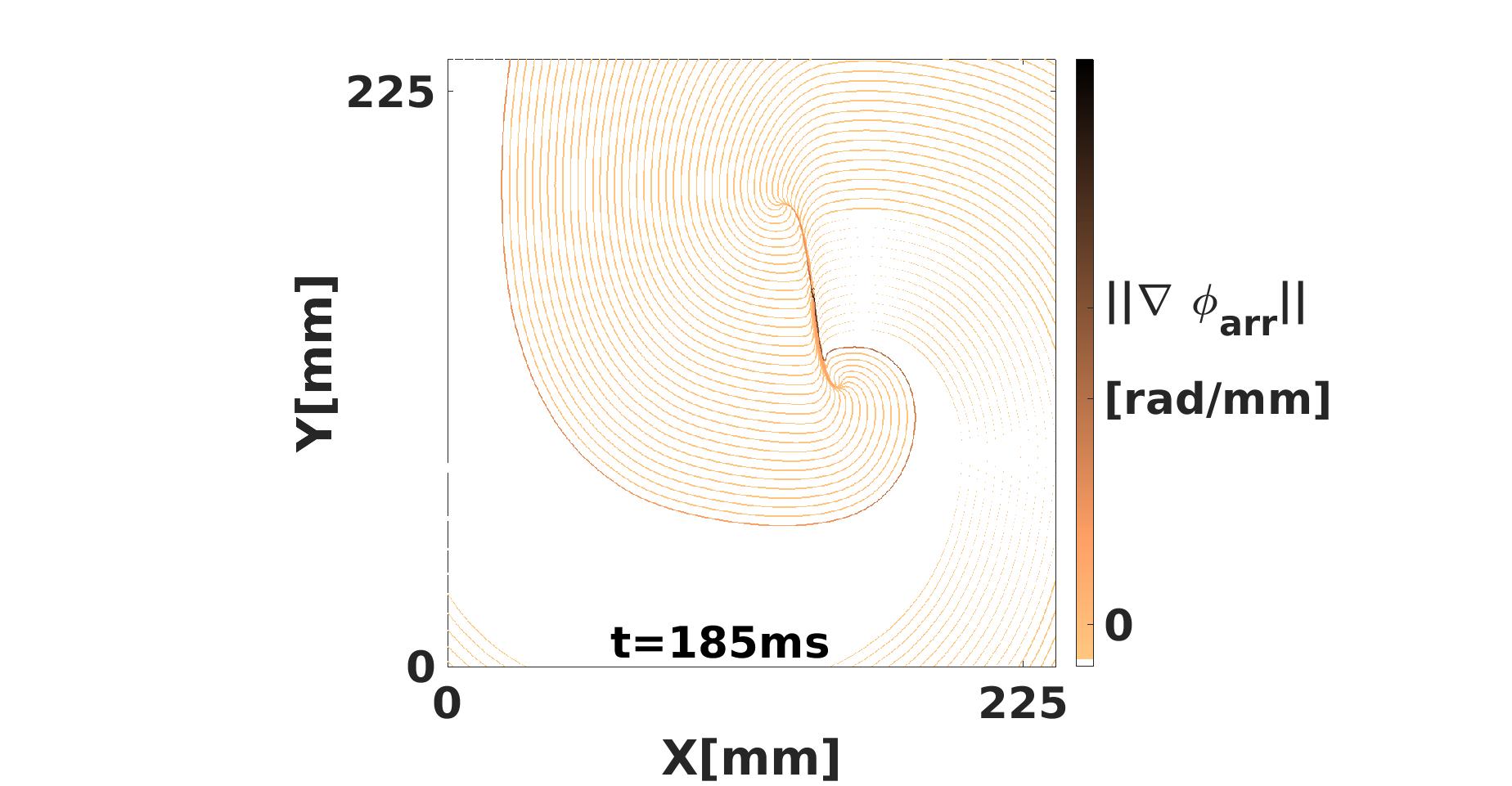}
    \includegraphics[trim={12cm 0cm 19cm 0.5cm},clip, height= \figspiralphiarr]{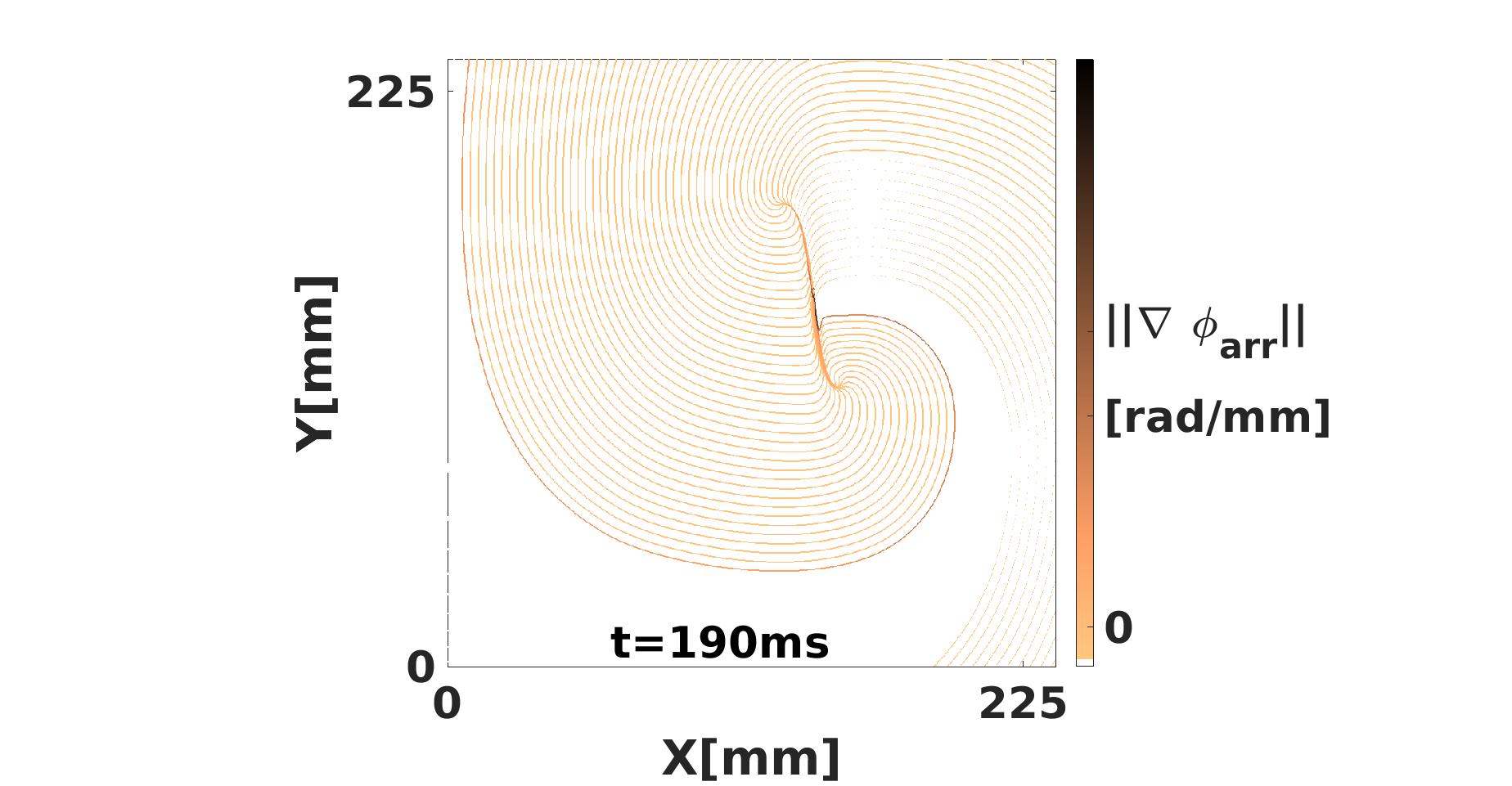}
    \includegraphics[trim={12cm 0cm 8.cm 0.5cm},clip, height= \figspiralphiarr]{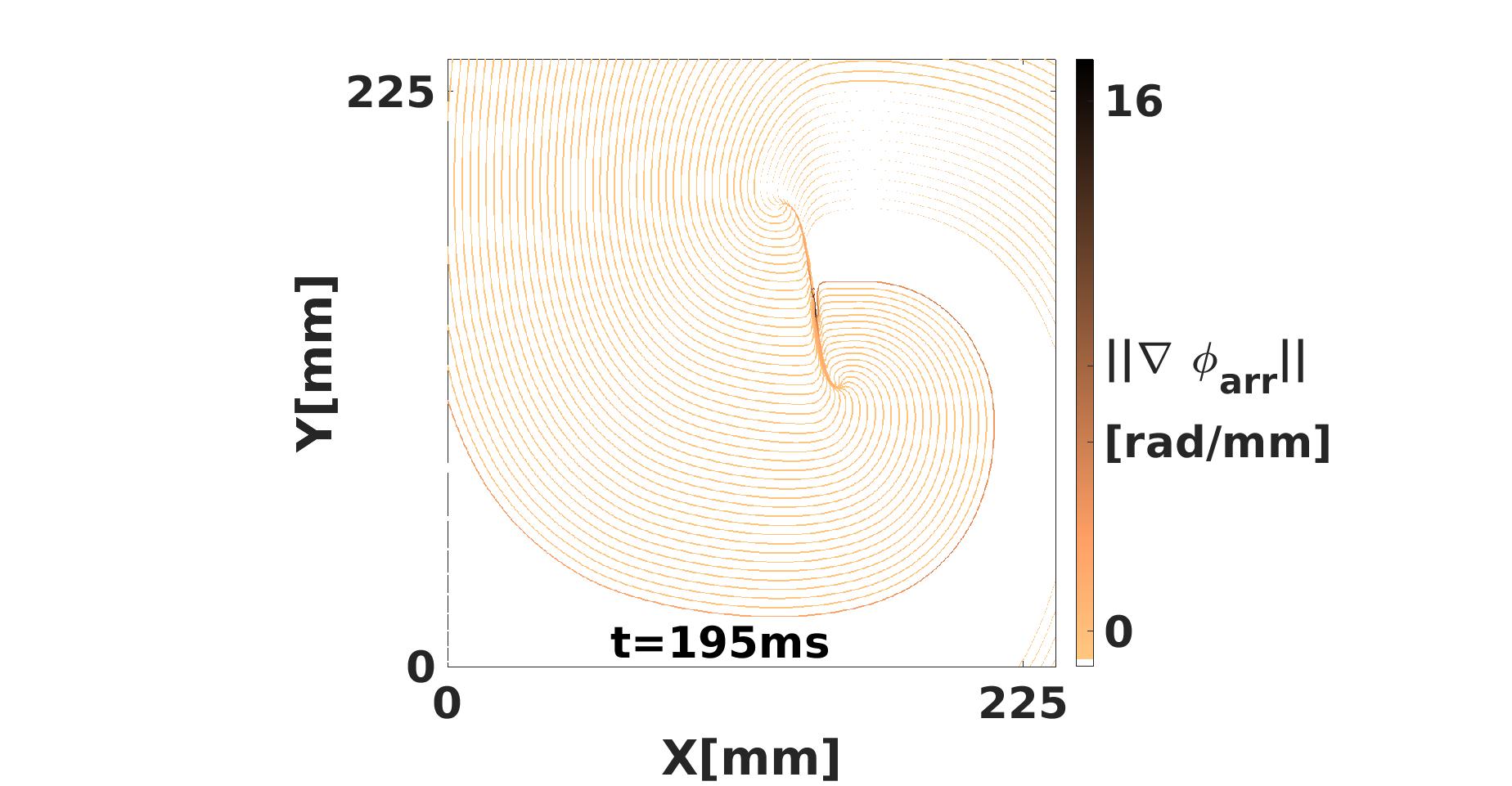}
        \caption{Analysis of the simulated linear-core rotor from Fig. \ref{fig:S1S2revisited} using $\phiarr$. (A) With $\phiarr$, there are no sudden transitions in phase along the WF and WB, such that only CBLs are shown as phase defects, with large local phase gradient (B).  }
    \label{fig:S1S2revisited_phiarr}
\end{figure}

\subsubsection{Interpretation of phase defects as branch cuts from complex analysis}


Here, we demonstrate how phase defects can be interpreted as a branch cut from complex analysis \citep{Arfken}. For any non-zero complex number of the form $z = x+yi$ with $i^2=-1$, the phase can be calculated using
\begin{align}
    \arg(z) = \mathrm{atan2}(y,x).
\end{align}
Next, one can consider functions of a complex variable: $w= f(z)$ with both $w$ and $z$ complex numbers, e.g. $w=z^2$. To make a visual representation of the function value, one can draw colormaps of the phase $\arg(w)$, or show $\arg(w)$ as an XY-dependent elevation in a so-called Riemann surface; see Fig. \ref{fig:complex}. In case of polynomial or rational functions, such phase map will reveal point singularities similarly to PSs found in circular-core spiral waves, see Fig. \ref{fig:complex}A. When the phase of the simple function $w=f(z) = z$ is shown as a graph of $(x,y)$ the phase surface resembles a helicoid. If one walks around the PS once in the XY-plane, the phase will change by $2\pi$, but since phase is measured disregarding terms of $2\pi$, one ends in the same state. This effect is more easily seen using a cyclic colormap in 2D, see Fig. \ref{fig:complex}A. Close to the central axis of the helicoid, the height (phase) is undefined, which is typical for a PS. 

In contrast to polynomial and rational functions, functions $f$ which include roots, exhibit a discontinuity in their phase, also called `branch cut' in mathematics. In particular, for the function $w= f(z) = \sqrt{z^2-1}$, there is a line of discontinuous phase for $(x,y)$ between $(-1,0) $ and $(1,0)$. This line is easily noticed in the representation as a color map as a sudden transition in color, see Fig. \ref{fig:complex}. This situation is in our opinion very similar to the phase discontinuity in the linear-core rotor throughout this paper. 

For, if one regards two neighboring points on both sides of a conduction block line at the rotor core, they will have distinct phases. There need not be a single point where all phases spatially converge (that is, a PS), as on both sides the phases can gradually change to take the same value at the points where the branch cut stops. 

From this paragraph and Fig. \ref{fig:complex} it is apparent that the PSs and PDLs are different topological structures. We argue in this paper that we and \cite{Tomii:2021} are the first to note this difference in a cardiac electrophysiology context, and that discriminating between these different structures can be useful in theory development and applications. 

\newcommand{\fs}{0.2}
\begin{figure}[t] 
\begin{tabular}{cc|cc}
\multicolumn{2}{c}{\textbf{A} PSs} & 
\multicolumn{2}{c}{\textbf{B} PDLs / branch cuts} \\
AP model (circular core) & $\arg{z}$
& FK model (linear core) & $\arg(\sqrt{z^2-1})$ \\
\includegraphics[width=\fs\textwidth]{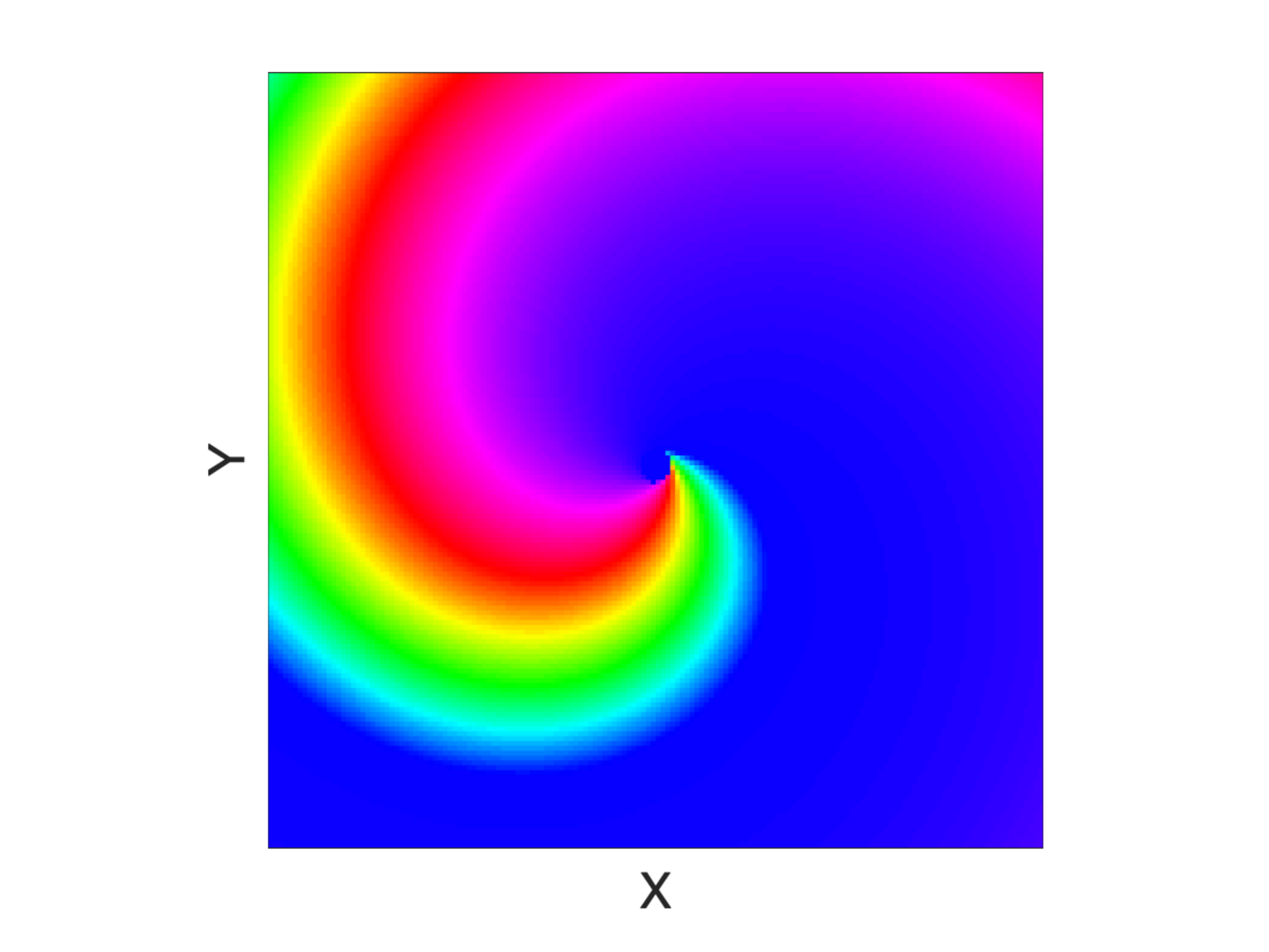}&
\includegraphics[width=\fs\textwidth]{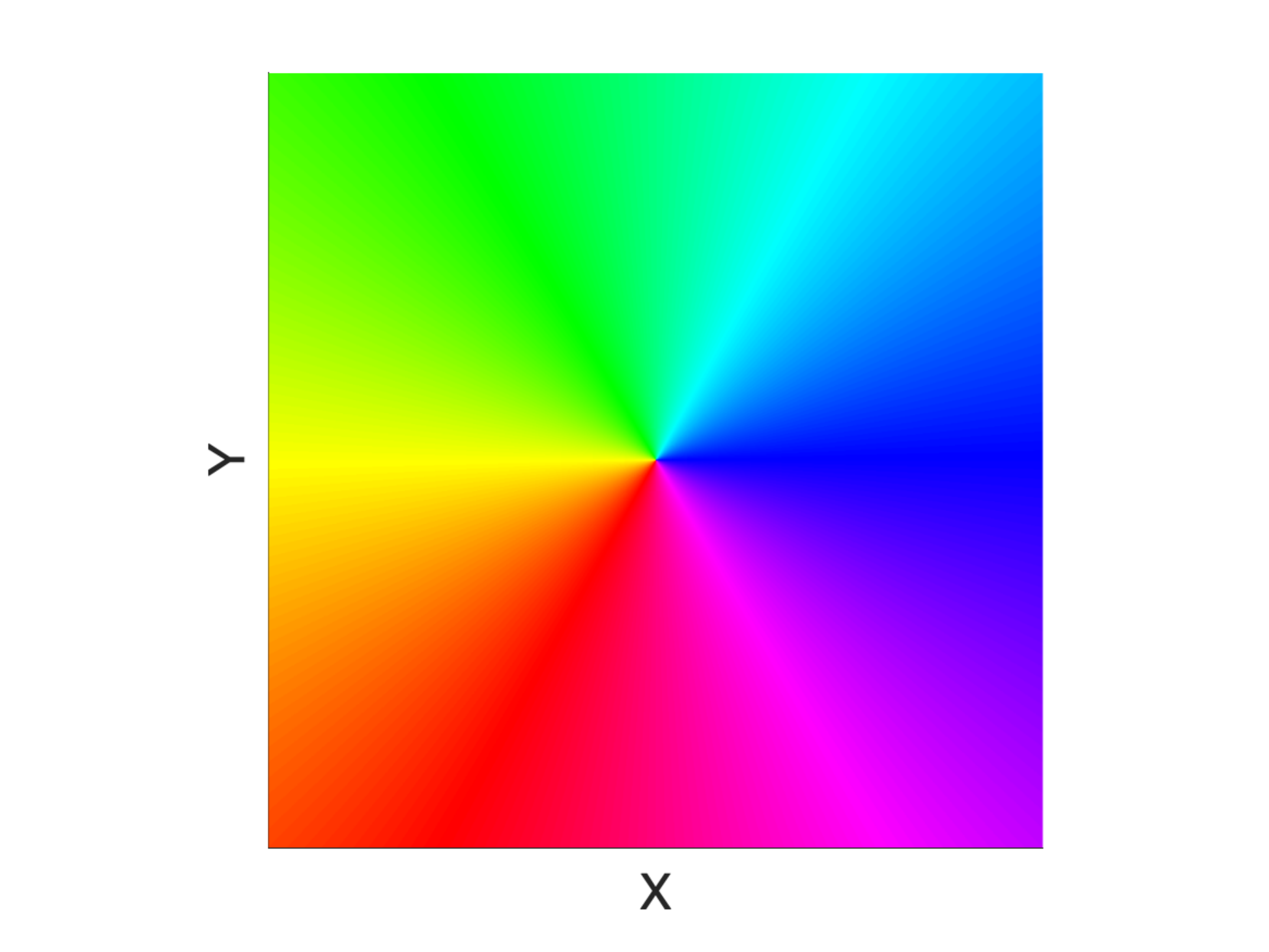}&
\includegraphics[width=\fs\textwidth]{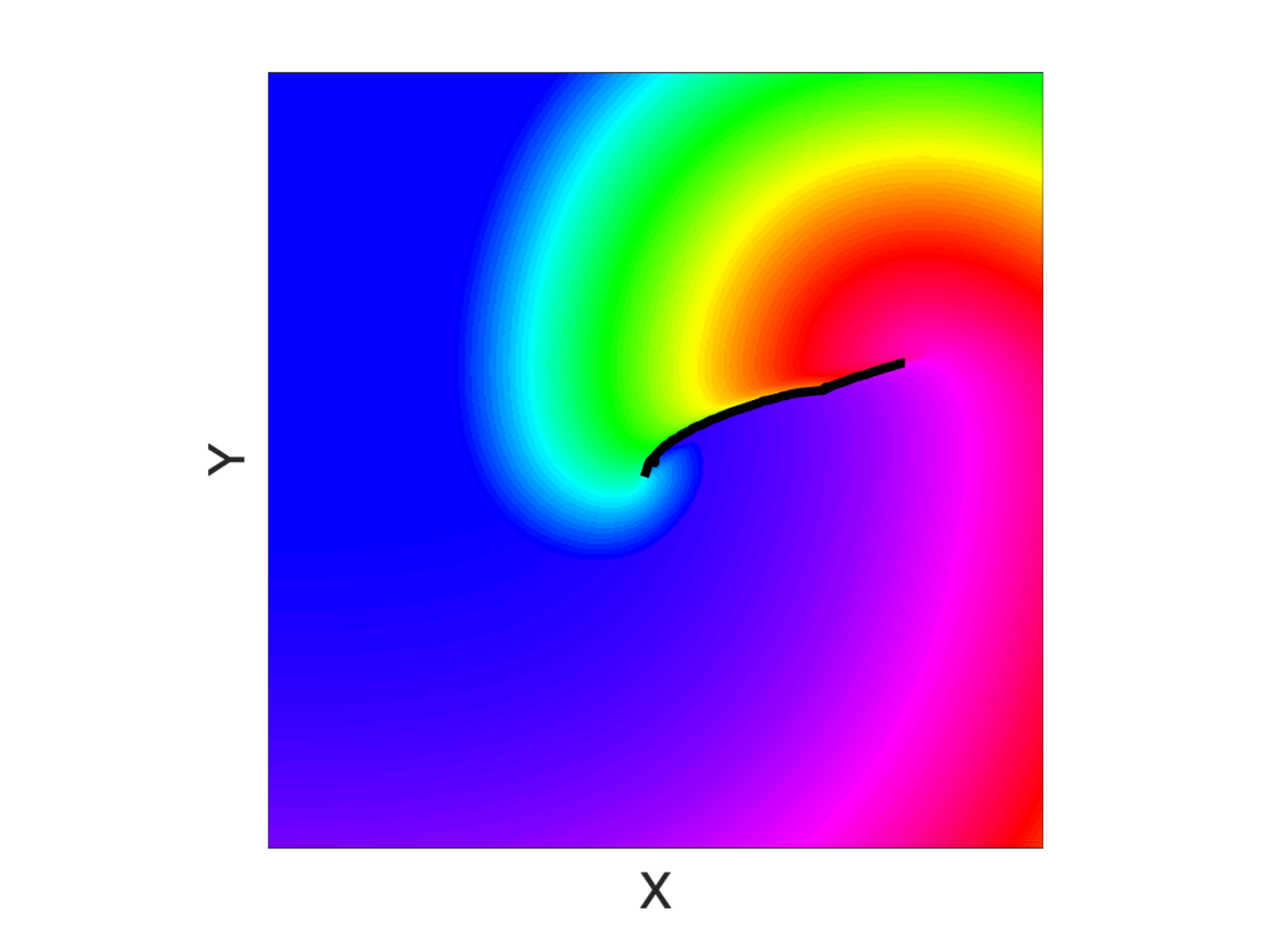} &
\includegraphics[width=\fs\textwidth]{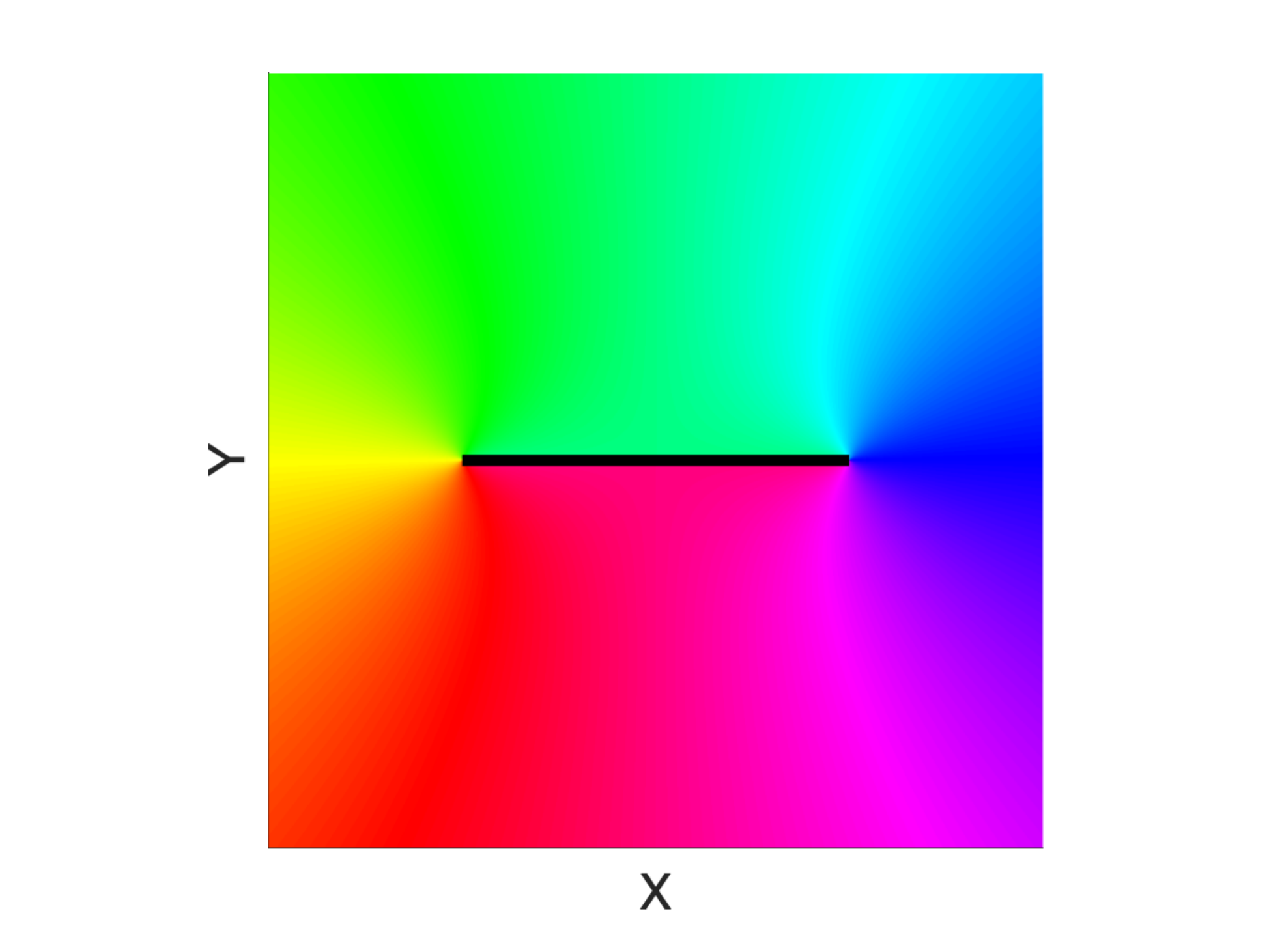} \\
 \includegraphics[width=\fs\textwidth]{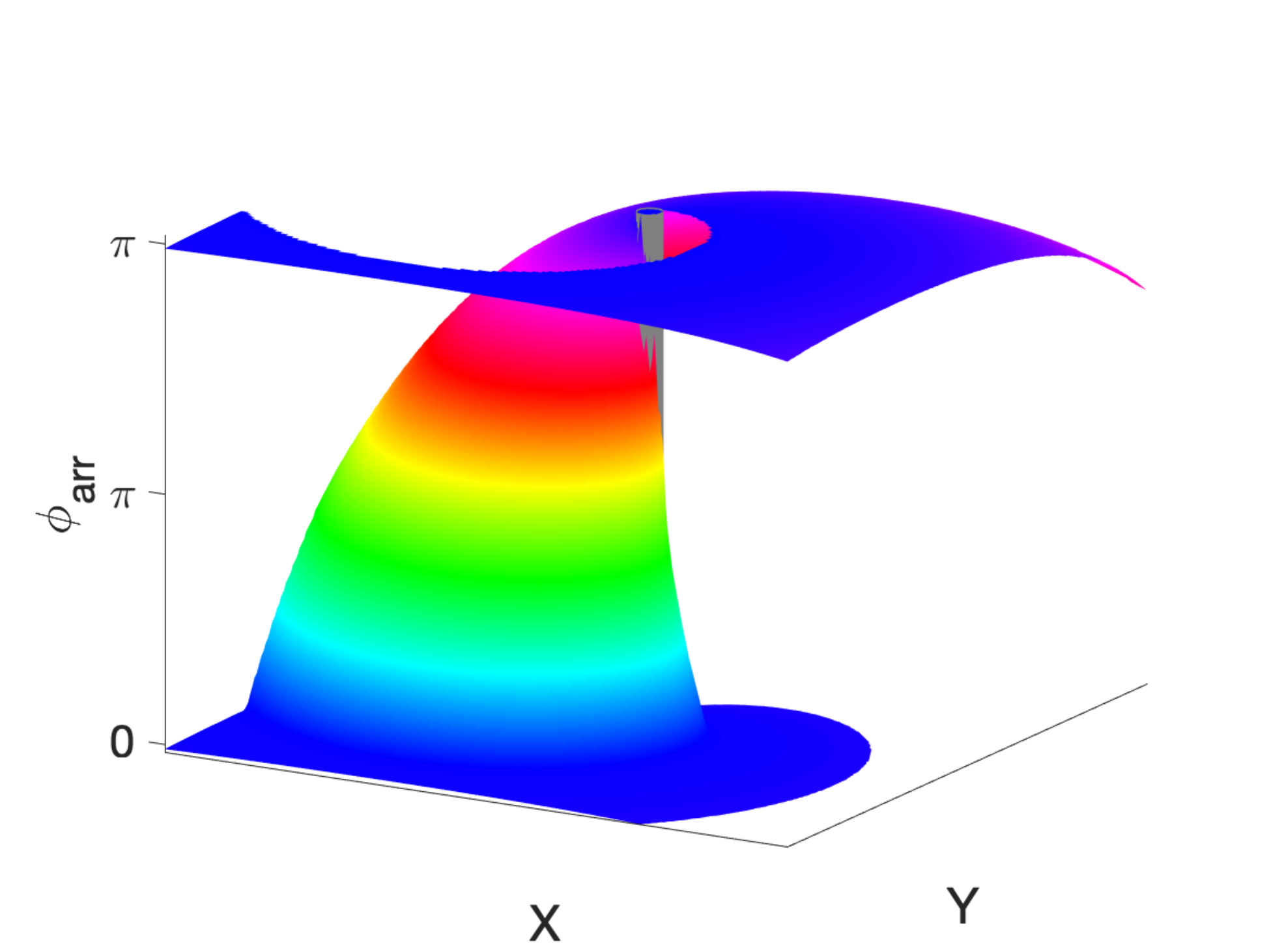}&
  \includegraphics[width=\fs\textwidth]{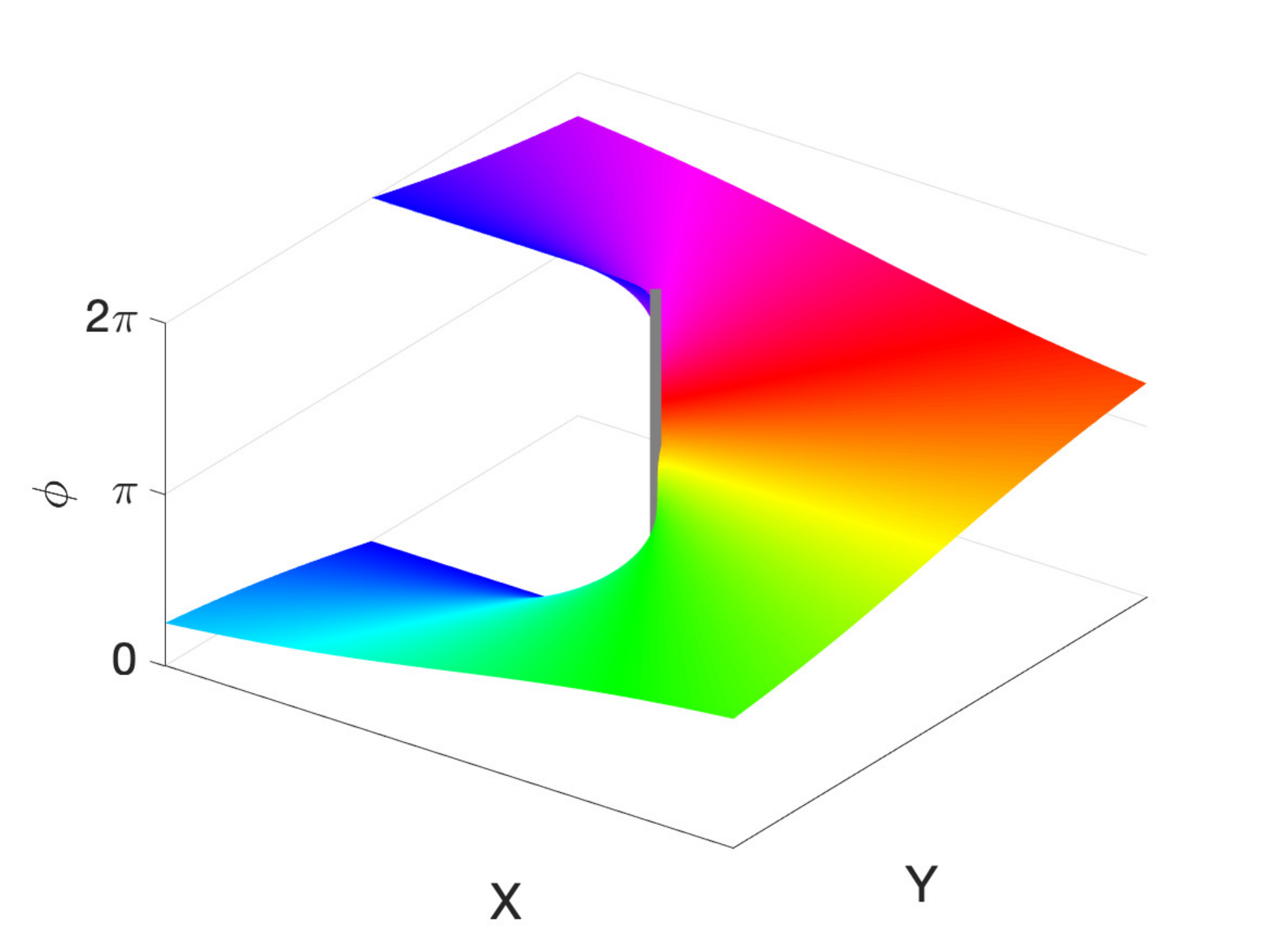}&
  \includegraphics[width=\fs\textwidth]{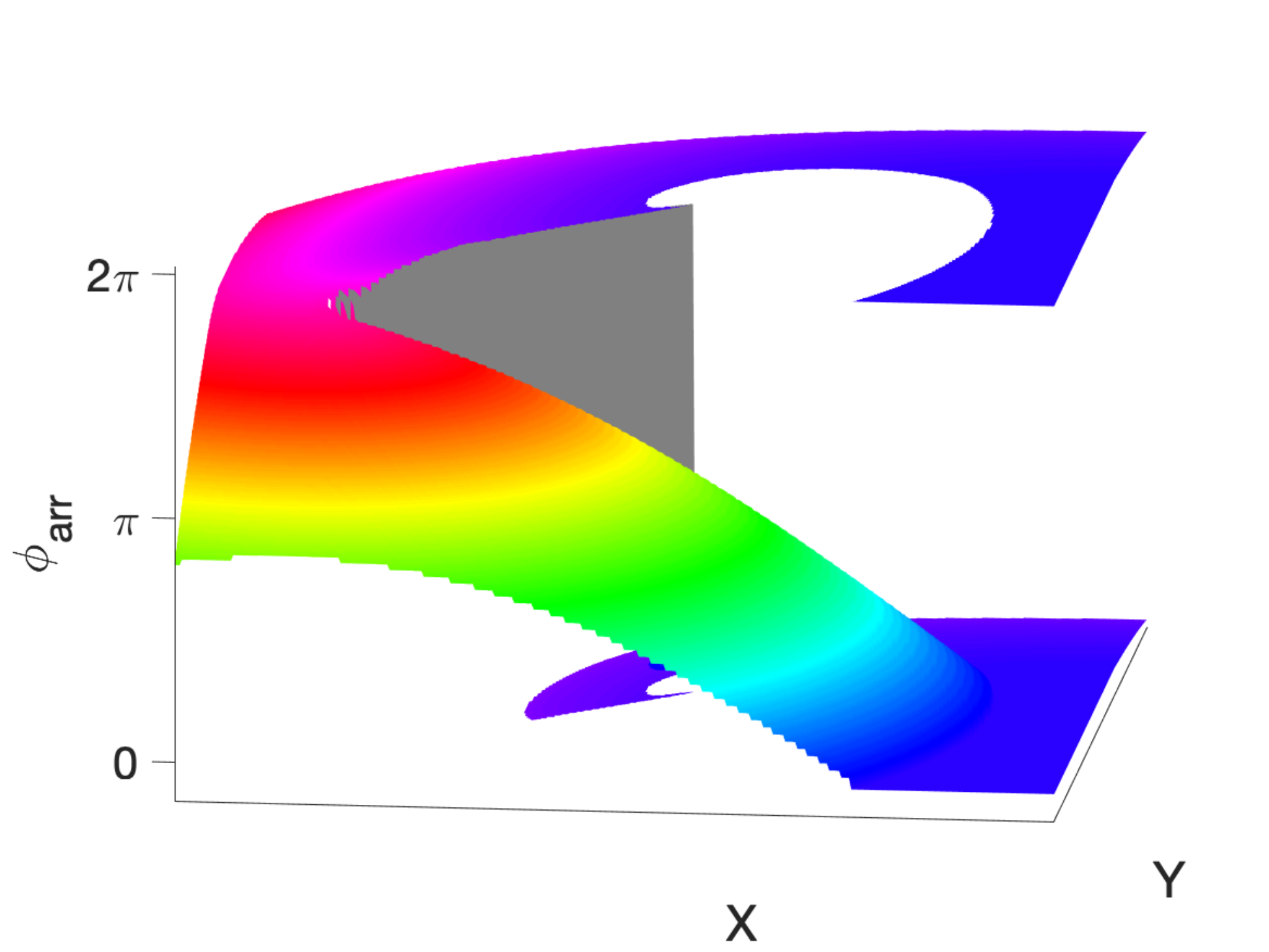} &
  \includegraphics[width=\fs\textwidth]{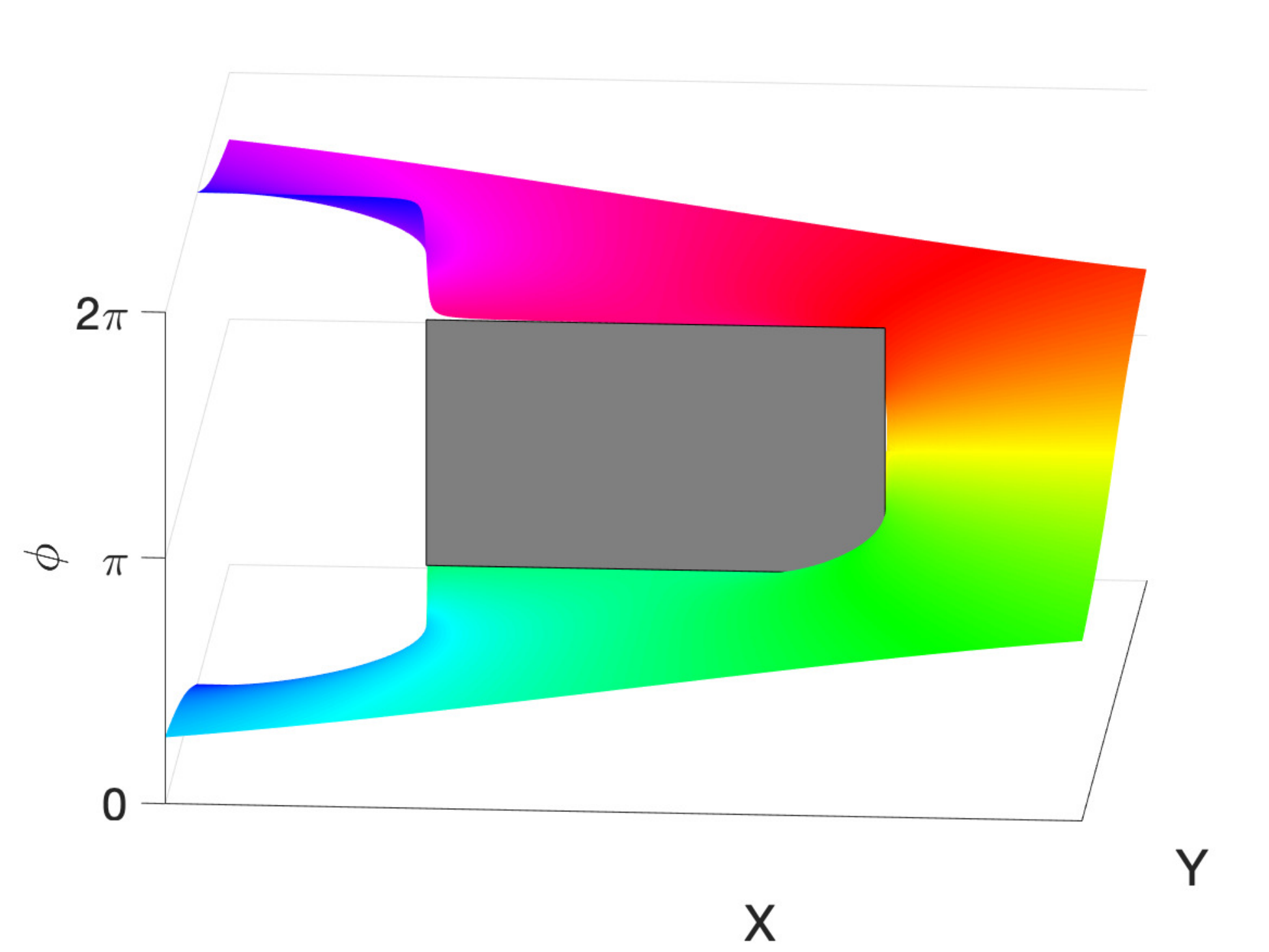} \\
  \multicolumn{4}{c}{
\includegraphics[width=0.45\textwidth]{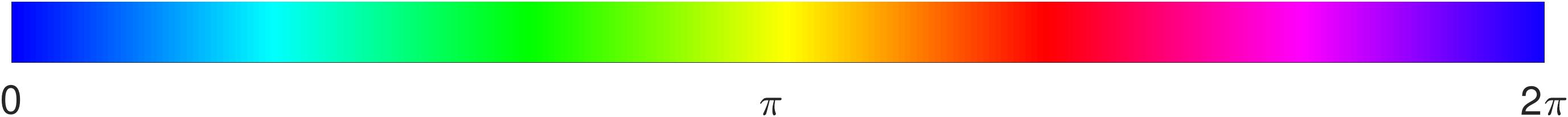}
}
\end{tabular}
\caption{ PSs versus PDLs in cardiac models and complex analysis. Phases are rendered in-plane (top row) and in 3D, as a Riemannian surface (bottom row). (A) Rigidly rotating spirals, as in the Aliev-Panfilov (AP) reaction-diffusion model \citep{Aliev:1996} correspond to a PS (gray), similar to the mathematical function $\phi(x,y) = \arg(z)$ shown to the right of it. (B) Linear-core cardiac models, e.g. \citep{Fenton:1998} exhibit a PDL or branch cut (black/gray), like the mathematical function $w=f(z) = \arg(\sqrt{z^2-1})$ shown to the right of it. Gray areas denote a jump in the phase over a quantity not equal to an integer multiple of $2\pi$, i.e. a PDL (physics) or branch cut (mathematics). 
}\label{fig:complex}
\end{figure}

\subsection{Methods for phase defect detection} \label{sec:PDLdetection}

The numerical identification of PDLs in cardiac excitation patterns can be performed using various methods. Based on a phase (either $\phiact$ or $\phiarr$), one can assign a PDL to the midpoint of edges along which the phase gradient exceeds a predefined value, taking into account that phase differences of $2\pi$ are excluded: 
\begin{align}
| U( \phi(\vec{r}_1) - \phi(\vec{r}_2)) | > \Delta\phi_{\rm crit}     \Rightarrow \frac{\vec{r_1} + \vec{r_2}}{2} \in PDL.  \label{PDLdetection}
\end{align}
However, if the LAT is known, this method approximates the more direct method of identifying a CBL, see Eq. \eqref{CBLdetection}. A comparison between different manners to numerically compute PDLs will be deferred to another publication. 

Below, we also compare the lifetime of PSs and PDLs in experimental recordings. To estimate PS lifetime, we calculated PSs at each time frame between 501 and 1500\,ms using the Hilbert transform of $V(t)$, followed by PS detection using $2\times2+4\times4$-method, which is considered as robust in literature \citep{Kuklik:2017}. If a PS moved between two subsequent frames at most 1 pixel in the horizontal, vertical or diagonal direction, while keeping its chirality, it was considered the same PS. Otherwise the PS was considered a new one. 

A similar approach was taken for PDLs. PDLs in experiments were computed using Eq. \eqref{PDLdetection}, with manually chosen threshold $\Delta\phi_{\rm crit} = 2.22$. If the middle of an edge was on a PDL, both neighboring nodes of the Cartesian grid were considered to be part of the PDL. A PDL was assumed to persist in time if the PDL at the new time and at the old time had points which were not further than 1 pixel away from each other (in horizontal, vertical or diagonal direction). PDLs were allowed to branch and merge, and the lifetime was computed as the earliest and latest time which were connected by this family of PDLs. 

A PS was furthermore considered to lie on a PDL if its distance was less than 0.5 pixels from a point on the PDL. PDLs in Fig. \ref{fig:problems}c and \ref{fig:experiments} were visually rendered using splines. The wave front was added to these figures using Eqs. \eqref{defphi12} - \eqref{WF}, with $\VS = 0.5$.

\section{Results}\label{sec:results}

Here we illustrate the appearance and relation between PSs and PDLs in  simulations and an optical mapping experiment. 

\subsection{Numerical results in two dimensions}

\subsubsection{Rendering of cardiac activation phase around a conduction block line as Riemannian surfaces}
\begin{figure}
    \centering
\raisebox{2.5cm}{\textbf{A}}    \includegraphics[trim={7cm 0cm 9cm 1cm},clip,width=0.2 \textwidth]{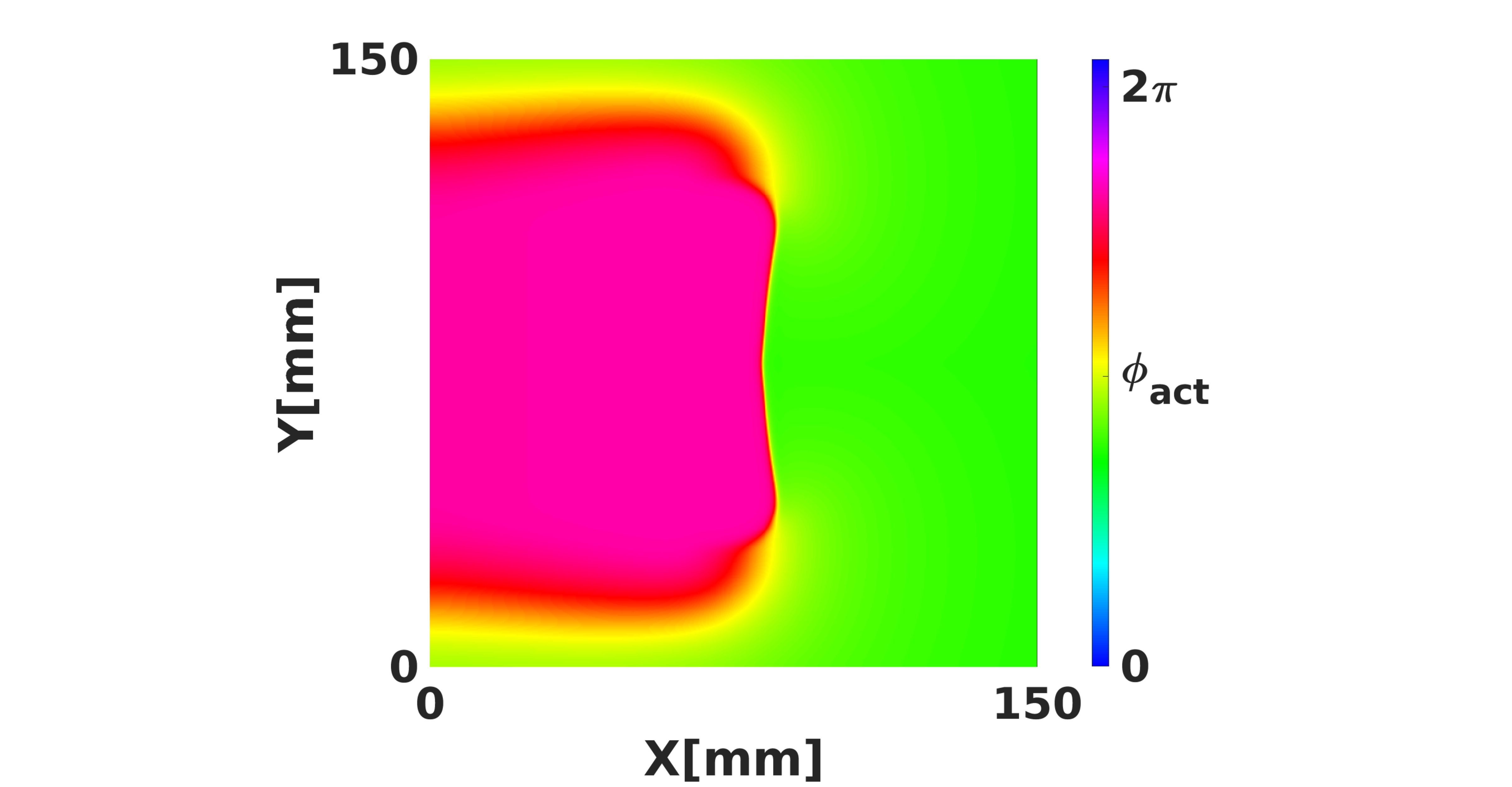}
    \includegraphics[trim={1cm 0cm 0cm 0cm},clip,width=0.23 \textwidth]{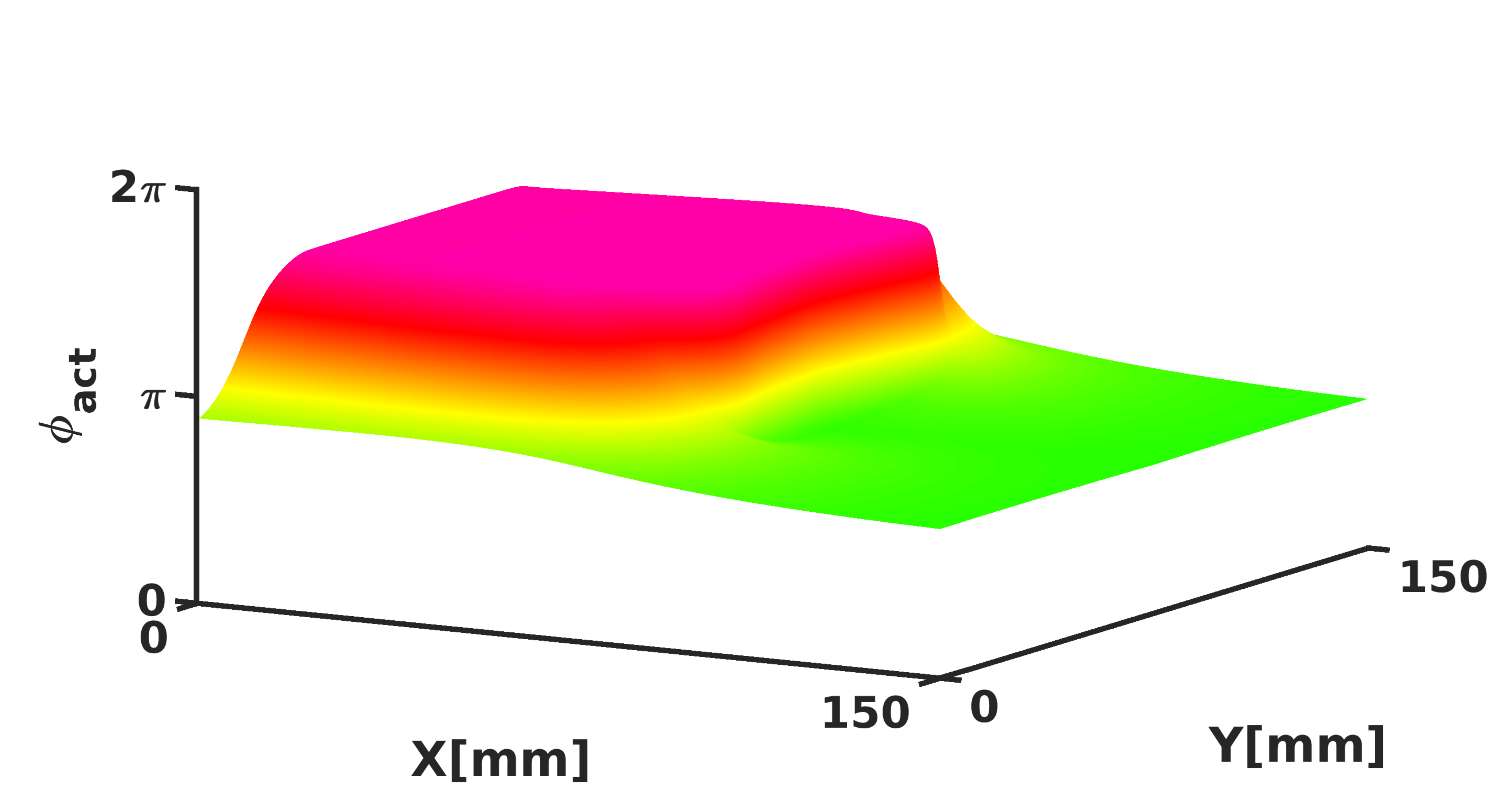}
\raisebox{2.5cm}{\textbf{B}}    \includegraphics[trim={7cm 0cm 9cm 1cm},clip,width=0.2 \textwidth]{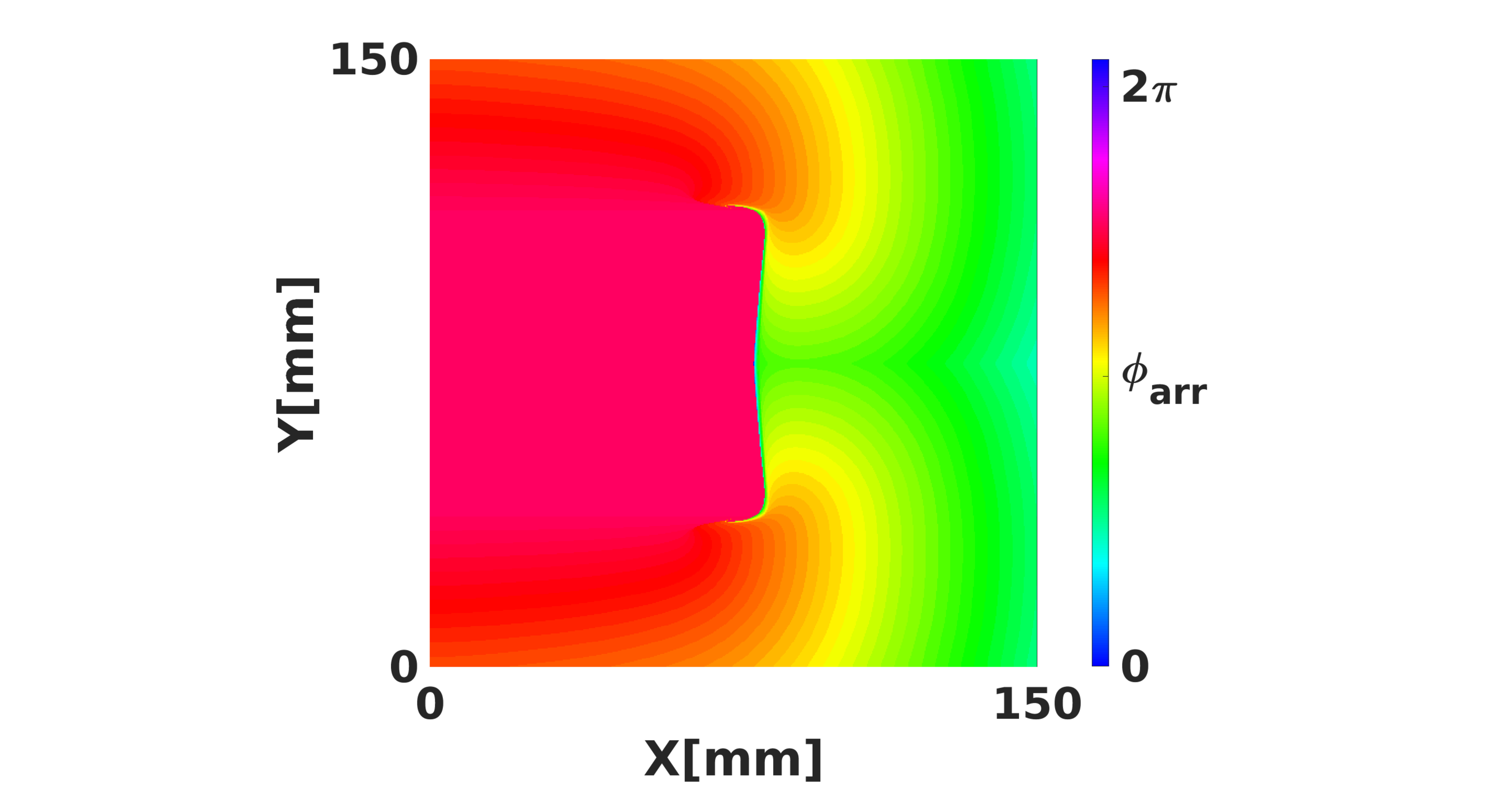}
    \includegraphics[trim={1cm 0cm 0cm 0cm},clip,width=0.23 \textwidth]{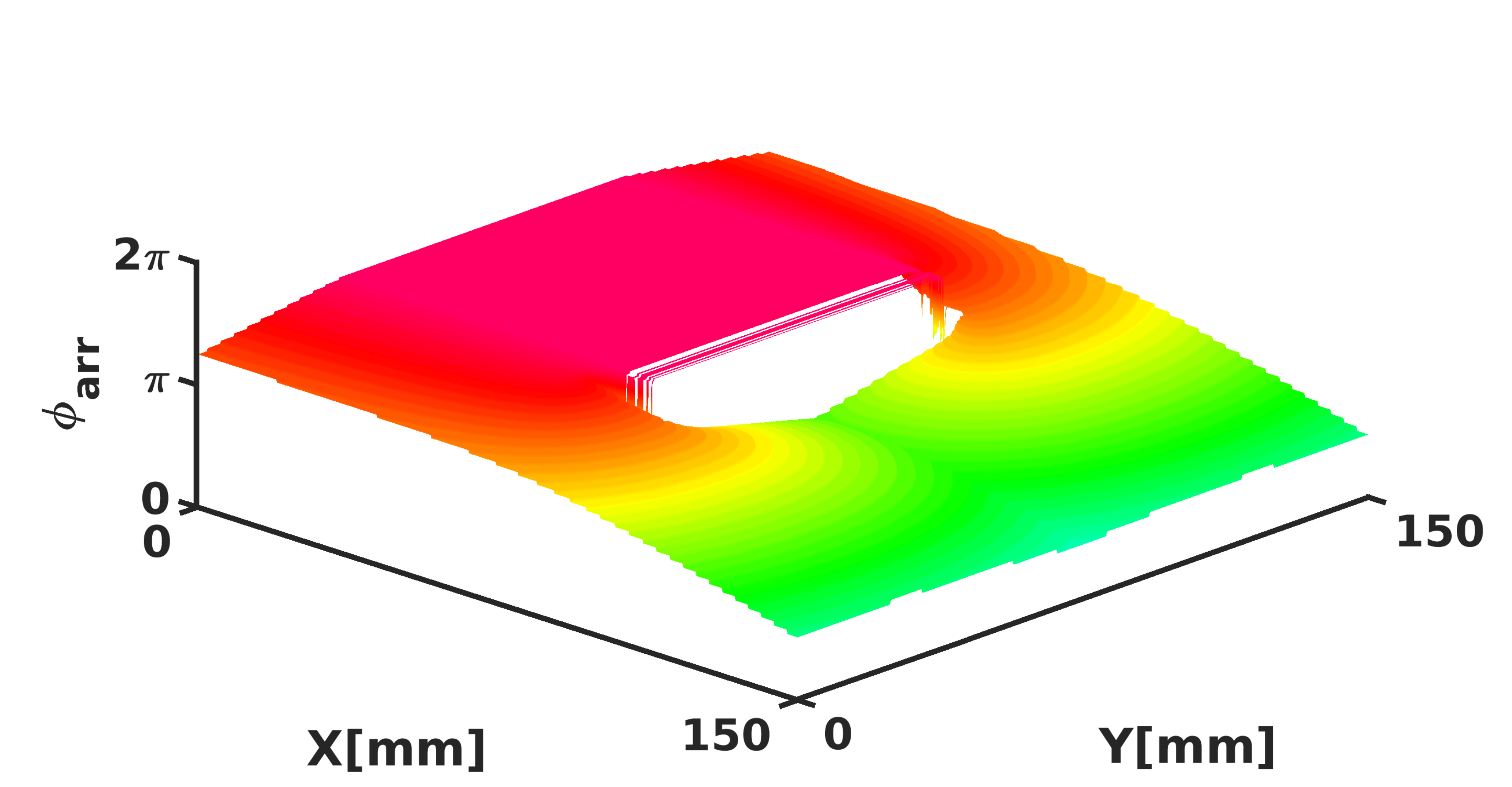} 
    \caption{Interpretation of a CBL as a phase defect clarifies why a PS detection finds PSs on it. (A) $\phiact$, rendered in 2D and 3D as a Riemannian surface. (B) $\phiarr$, rendered in 2D and 3D as a Riemannian surface. At the CBL/PDL, the phase surface has a cliff-like appearance. }
    \label{fig:CBL_Riemann}
\end{figure}

The CBLs presented in Fig. \ref{fig:problems} are examples of PDLs: to this purpose, we redraw the rightmost panel of Fig. \ref{fig:problems}A here using $\phiact$ and $\phiarr$, rendered in 2D and 3D in Fig. \ref{fig:CBL_Riemann}. Here, it can be seen that the suspected PSs in fact lie on the line where phase is nearly discontinuous. 

We distinguish two cases here. If no rotor is attached to the phase defect, the Riemann surface looks like a sheet of paper with a cut in it and both edges shifted relative to each other, see Fig. \ref{fig:CBL_Riemann}. If a rotor is attached to the phase defect, the Riemann surface looks like two rising slopes connecting two floors in a parking lot, see lower right panel in Fig. \ref{fig:complex}B. 

\subsubsection{Coexistence of PS and PDL in the same system}

\begin{figure}
    \centering
\raisebox{2.5cm}{\textbf{A}} 
\includegraphics[trim={8cm 0.3cm 10cm 1cm},clip,height= 2.9cm]{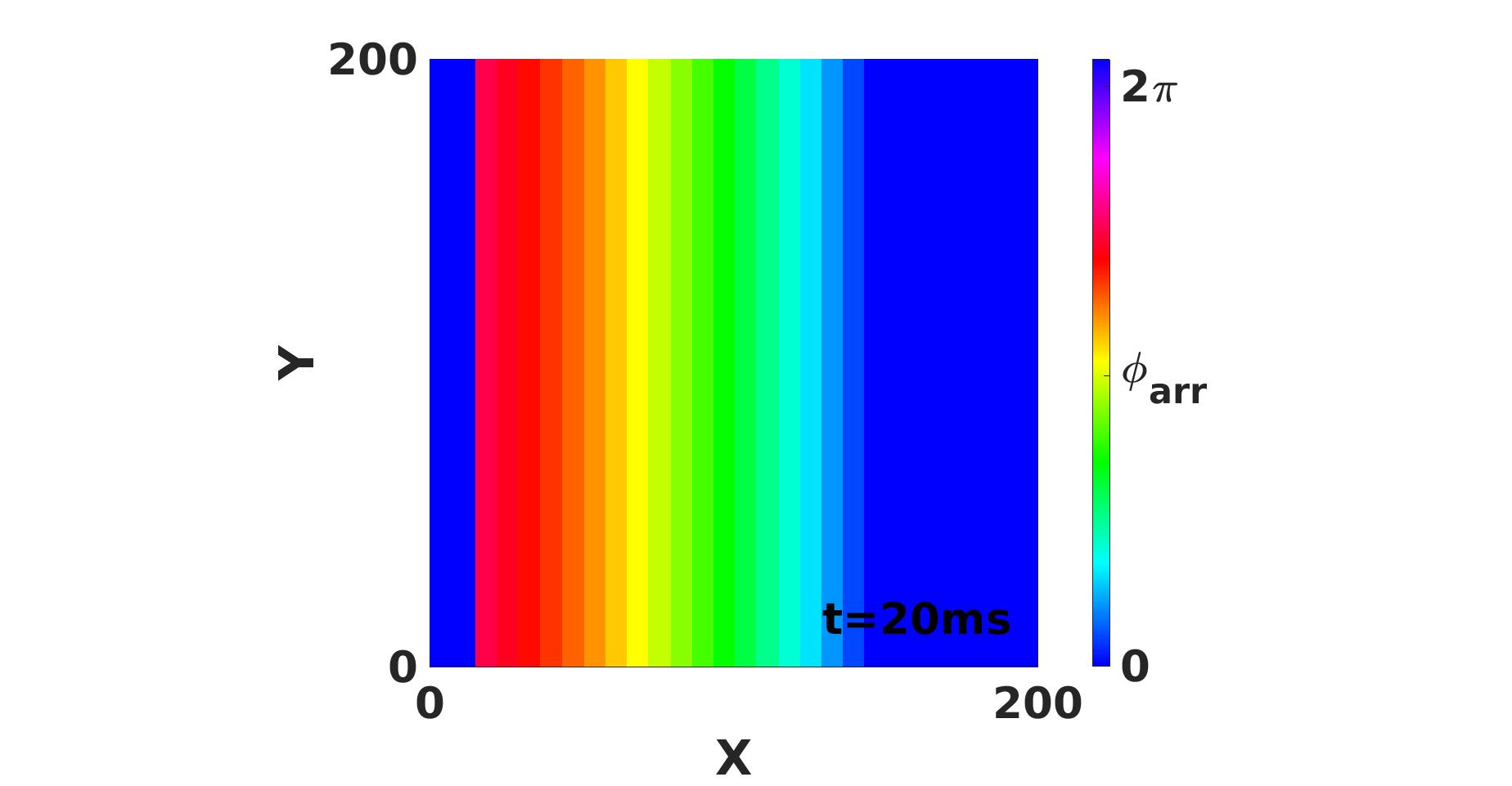}
\raisebox{2.5cm}{\textbf{B}} 
\includegraphics[trim={8cm 0.3cm 10cm 1cm},clip, height= 2.9cm]{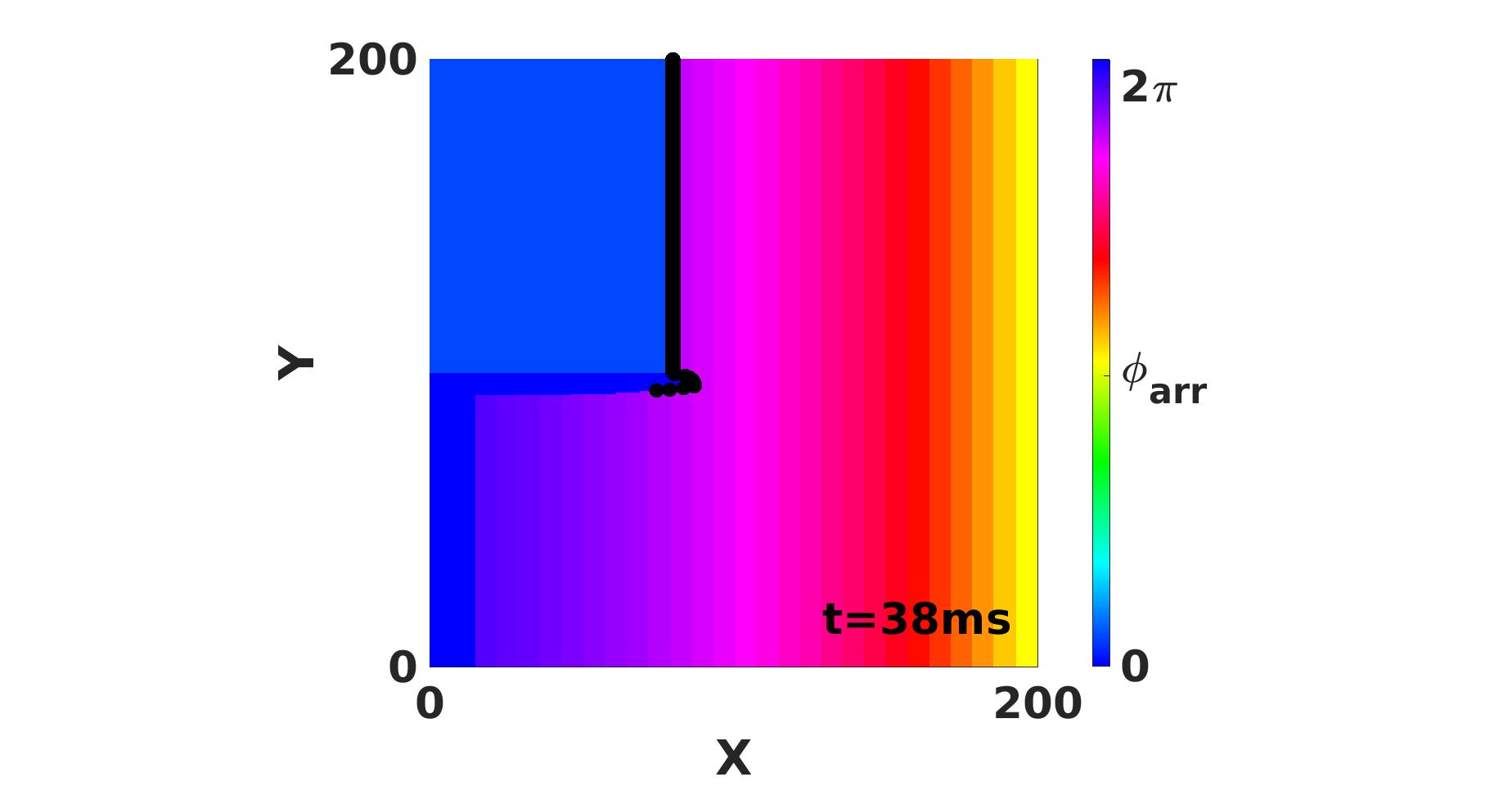}\\
\raisebox{2.5cm}{\textbf{C}} 
\includegraphics[trim={8cm 0.3cm 10cm 1cm},clip, height= 2.9cm]{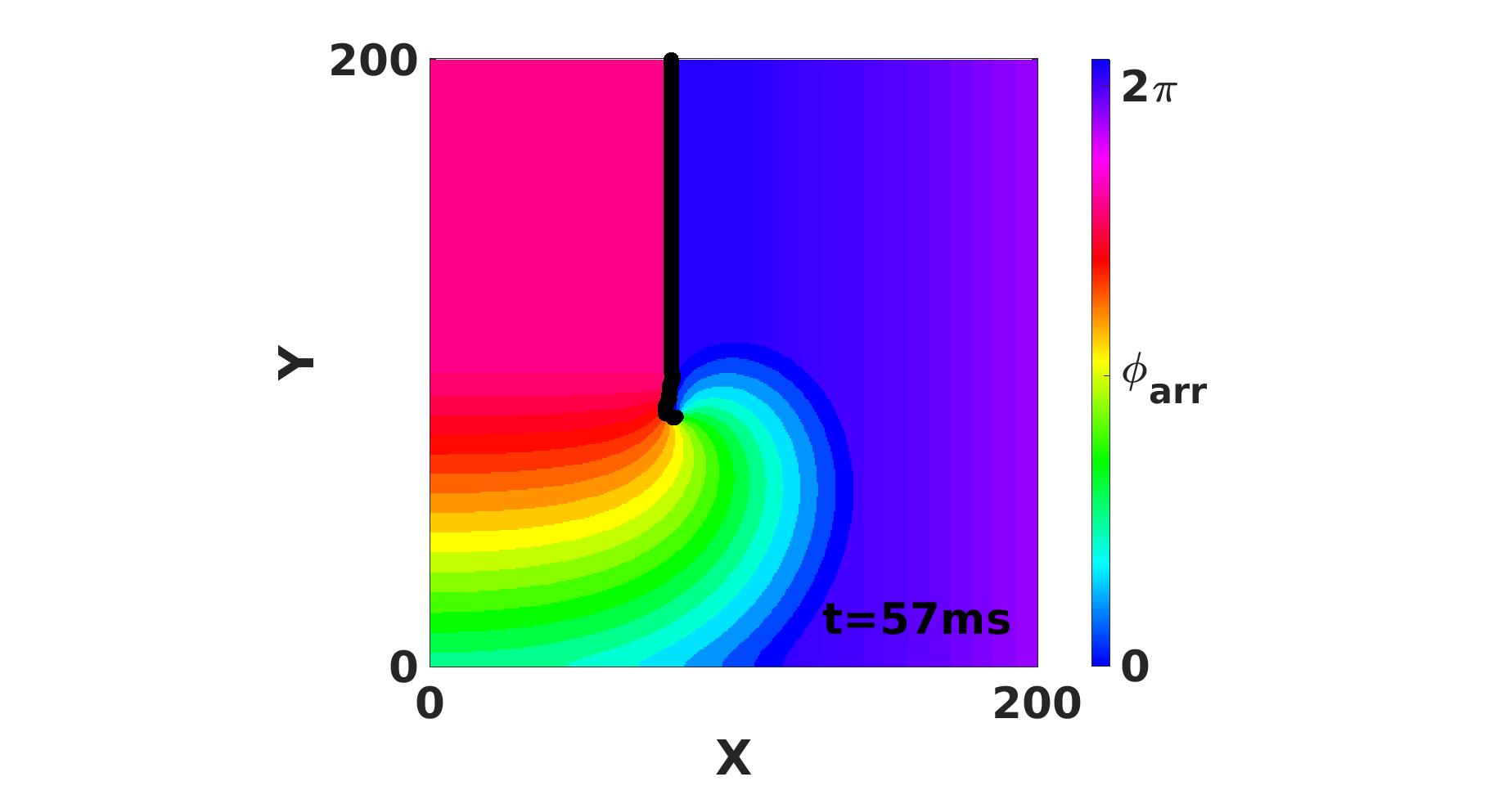}
\raisebox{2.5cm}{\textbf{D}} 
\includegraphics[trim={8cm 0.3cm 10cm 1cm},clip, height= 2.9cm]{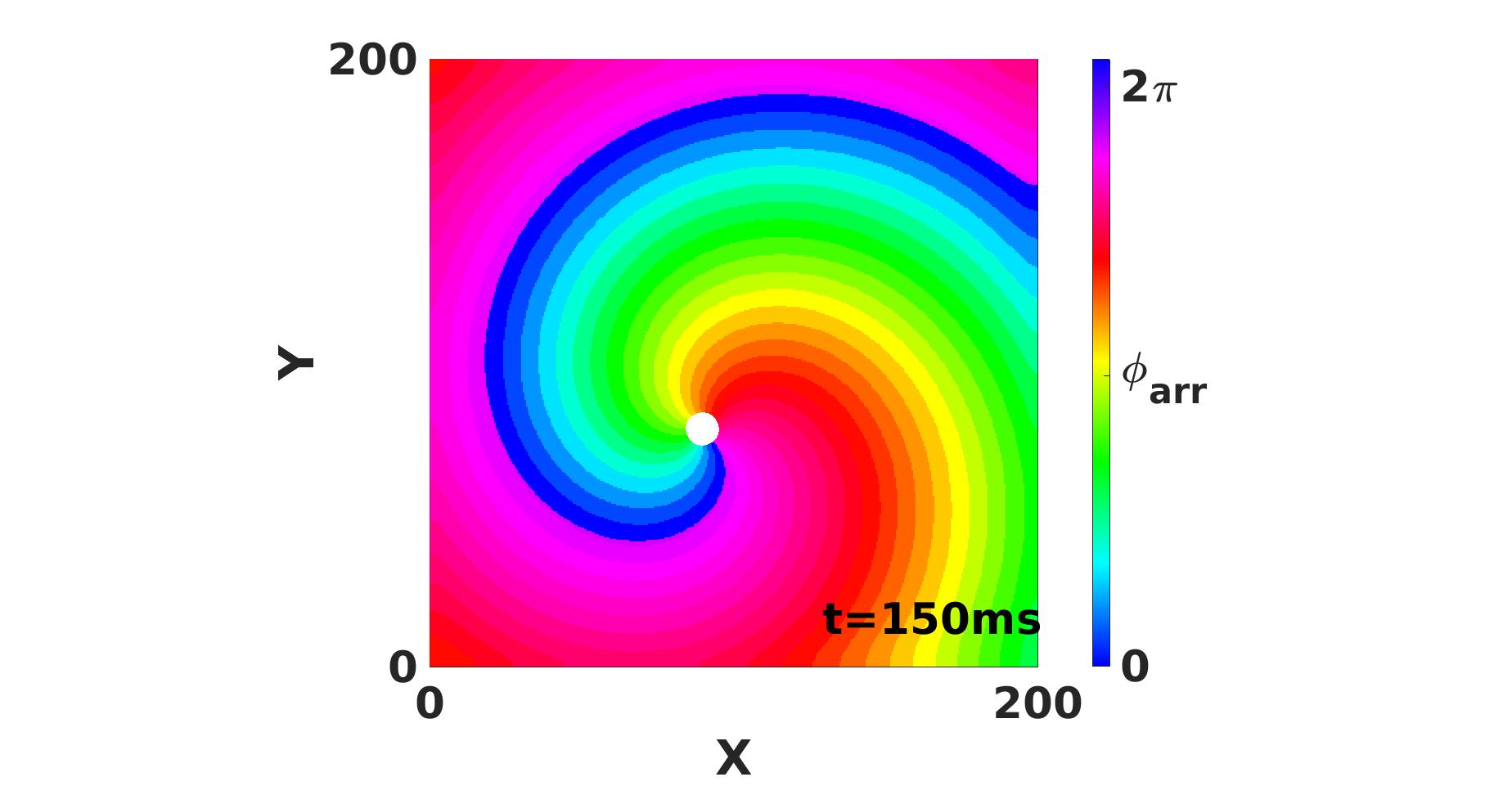}
    \caption{Creation of a PS from a PDL using a S1-S2 protocol with AP kinetics, showing coexistence of PDL and PS in models that generate circular-core spirals. Snapshots A-D shown at times t=20ms, t=38ms, t=57ms and t=150ms.  }
    \label{fig:AP_spiral}
\end{figure}

Here we demonstrate that PDLs can also exist in media that sustain spiral waves with PSs. This is illustrated in simulations where application of S1S2 pacing induces a PDL at the WB of the first stimulus, see Fig. \ref{fig:AP_spiral}. Since AP reaction kinetics were used \citep{Aliev:1996}, the PDL will disappear after both sides have reached full recovery, resulting in a PS at the core of the spiral wave, see Fig. \ref{fig:AP_spiral}D. Hence, we conclude that PDLs and PSs can coexist in media supporting circular-core spiral waves, since the PDL is found at conduction block sites and PSs are found in the spiral wave core.  


\subsubsection{Absence of a PS at the edge of a PDL}

Although Fig. \ref{fig:AP_spiral} shows a PS situated near the end of a PDL, this is certainly not always the case. Counterexamples of PSs located away from end points of a PDL can be seen in Fig. \ref{fig:problems} and \ref{fig:S1S2revisited}. 

The underlying mathematical reason is the following. At the edge of a PDL, the phase surface starts to be `torn', see Fig. \ref{fig:complex}B. Still, at this point itself, the phase is well-defined. This situation is different from a PS, see Fig. \ref{fig:complex}A, where the Riemann surface locally resembles a staircase surface and the phase itself cannot be determined.  

\subsubsection{Phase defect analysis of scroll wave turbulence}

\begin{figure}
    \centering
\raisebox{2cm}{\textbf{A}} 
\includegraphics[trim={1cm 0.3cm 0cm 1cm},clip, height=\figbreakup]{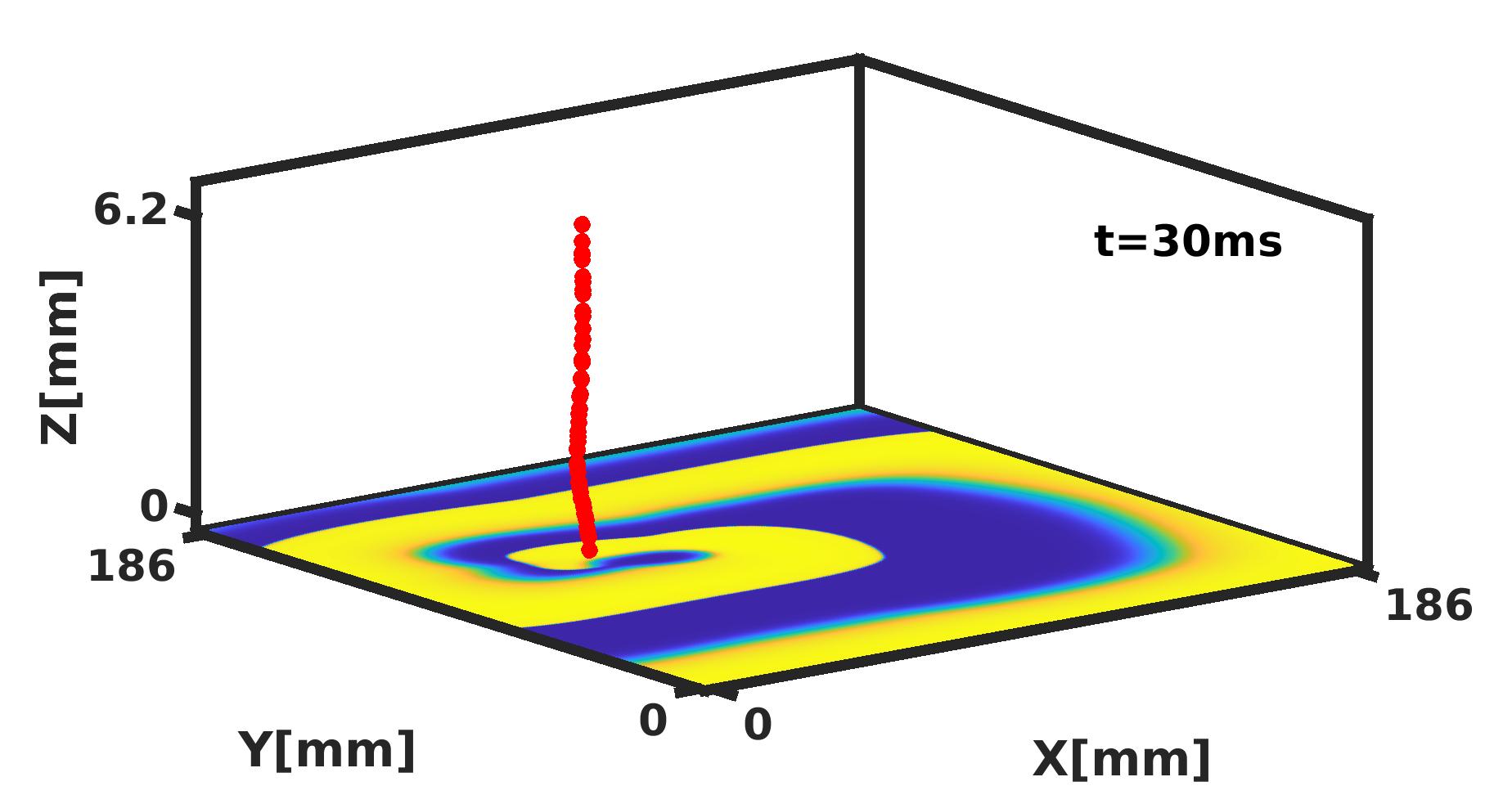}
\raisebox{2cm}{\textbf{B}} 
\includegraphics[trim={1cm 0.3cm 0cm 1cm},clip, height=\figbreakup]{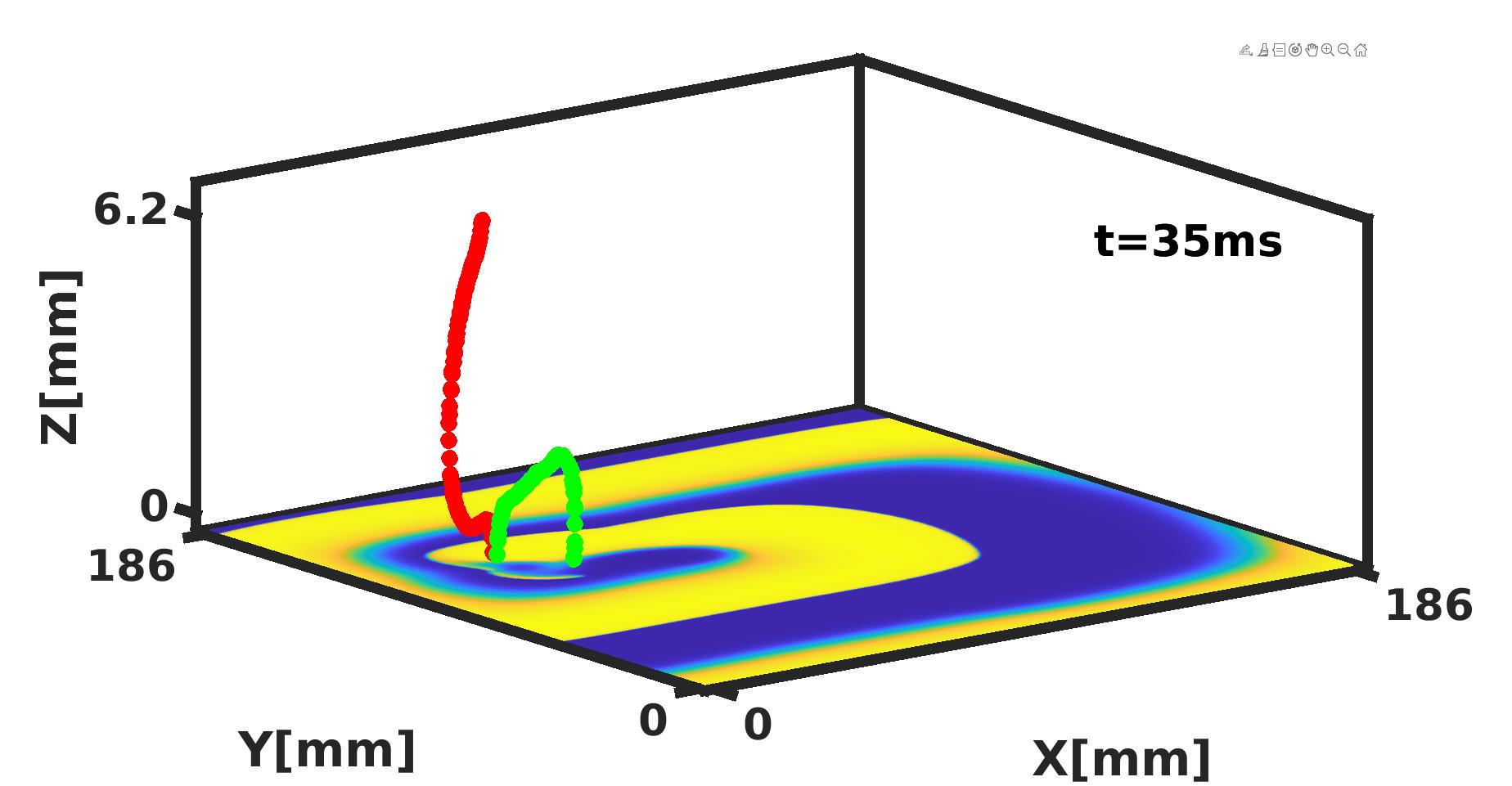}
\raisebox{2cm}{\textbf{C}} 
\includegraphics[trim={1cm 0.3cm 0cm 1cm},clip, height=\figbreakup]{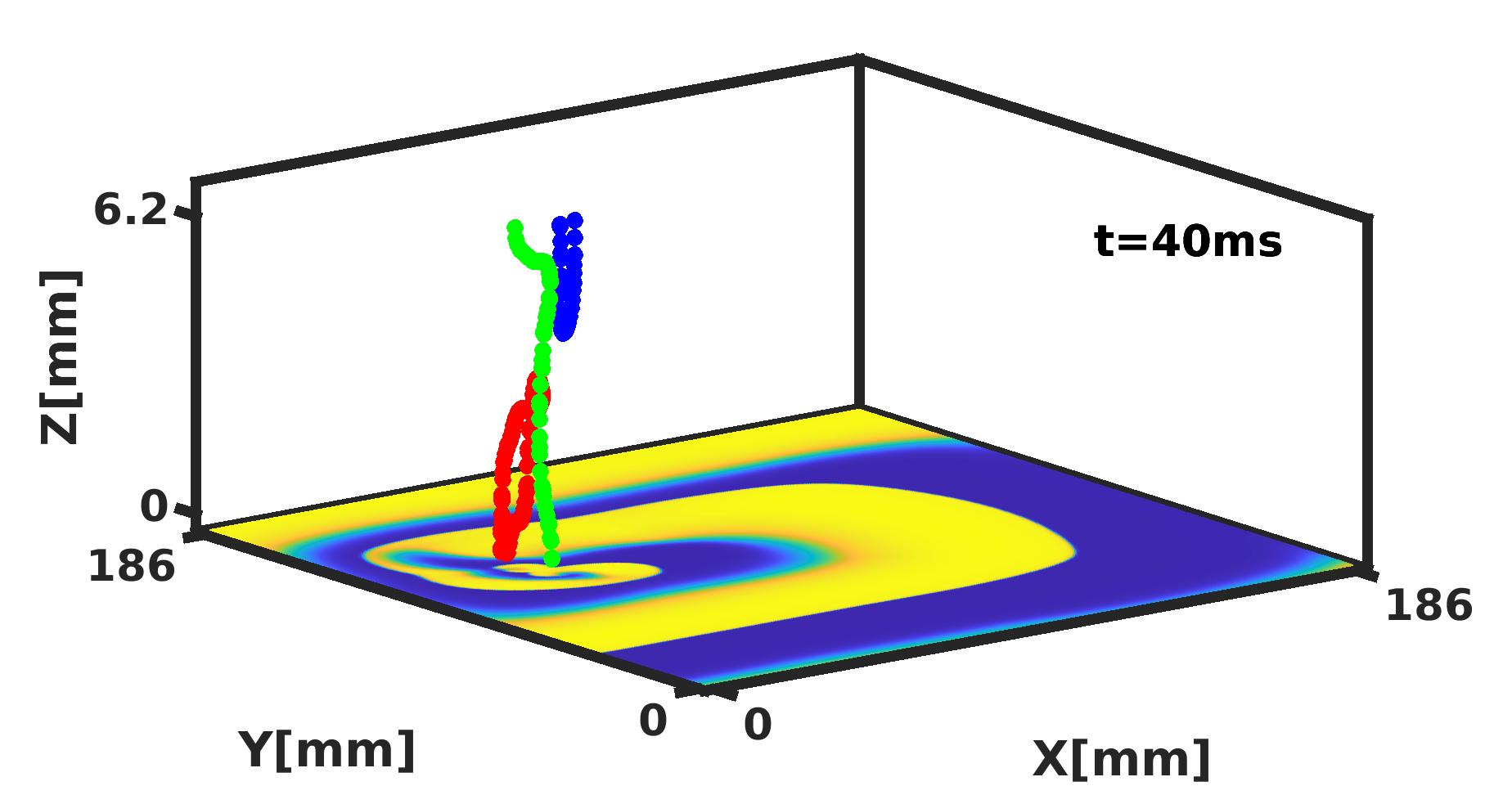}
\\ 
\includegraphics[trim={8cm 0.3cm 9cm 1cm},clip,height = 3.5cm]{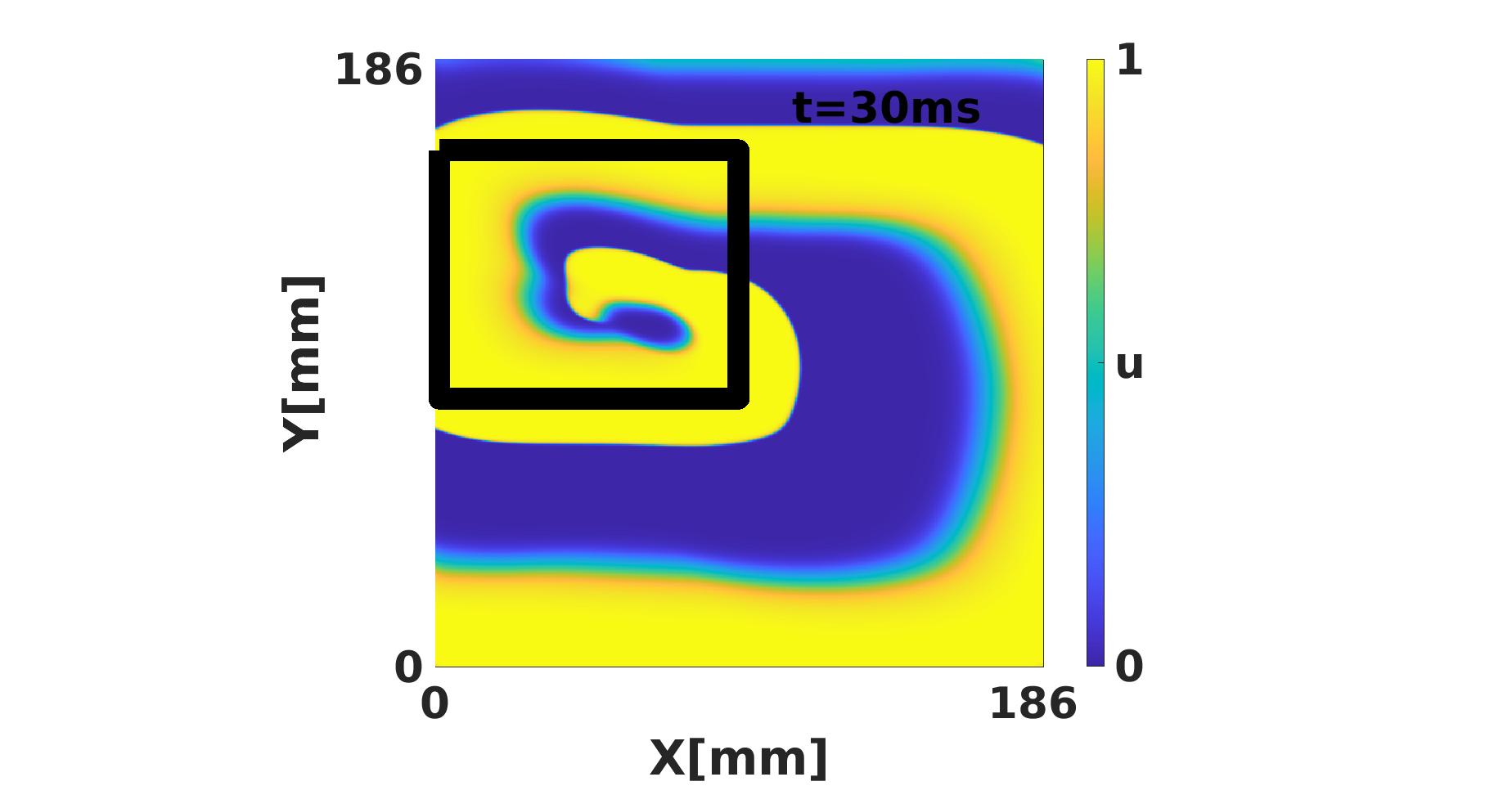}
\includegraphics[trim={8cm 0.3cm 9cm 1cm},clip, height = 3.5cm]{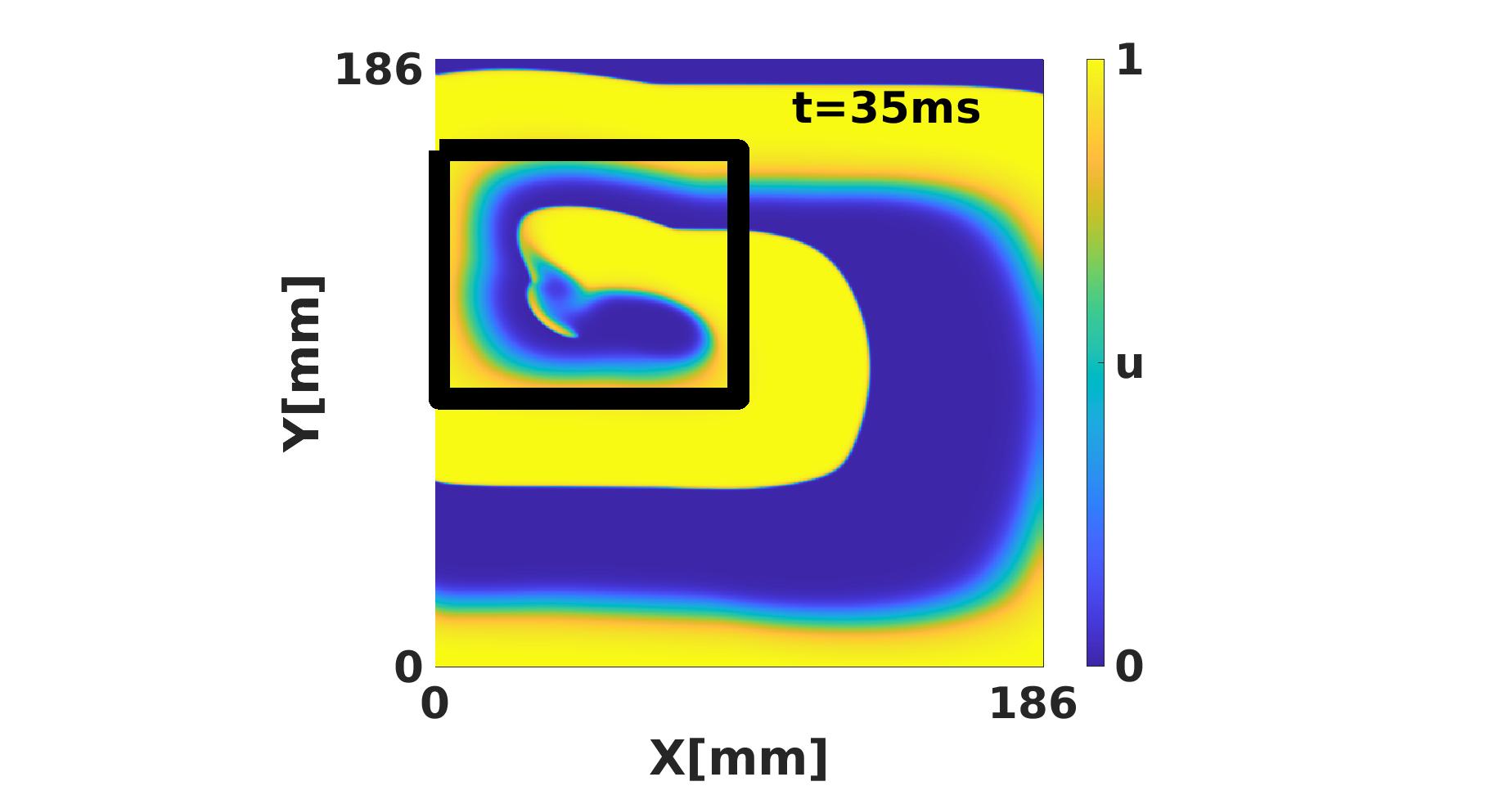}
\includegraphics[trim={8cm 0.3cm 9cm 1cm},clip, height = 3.5cm]{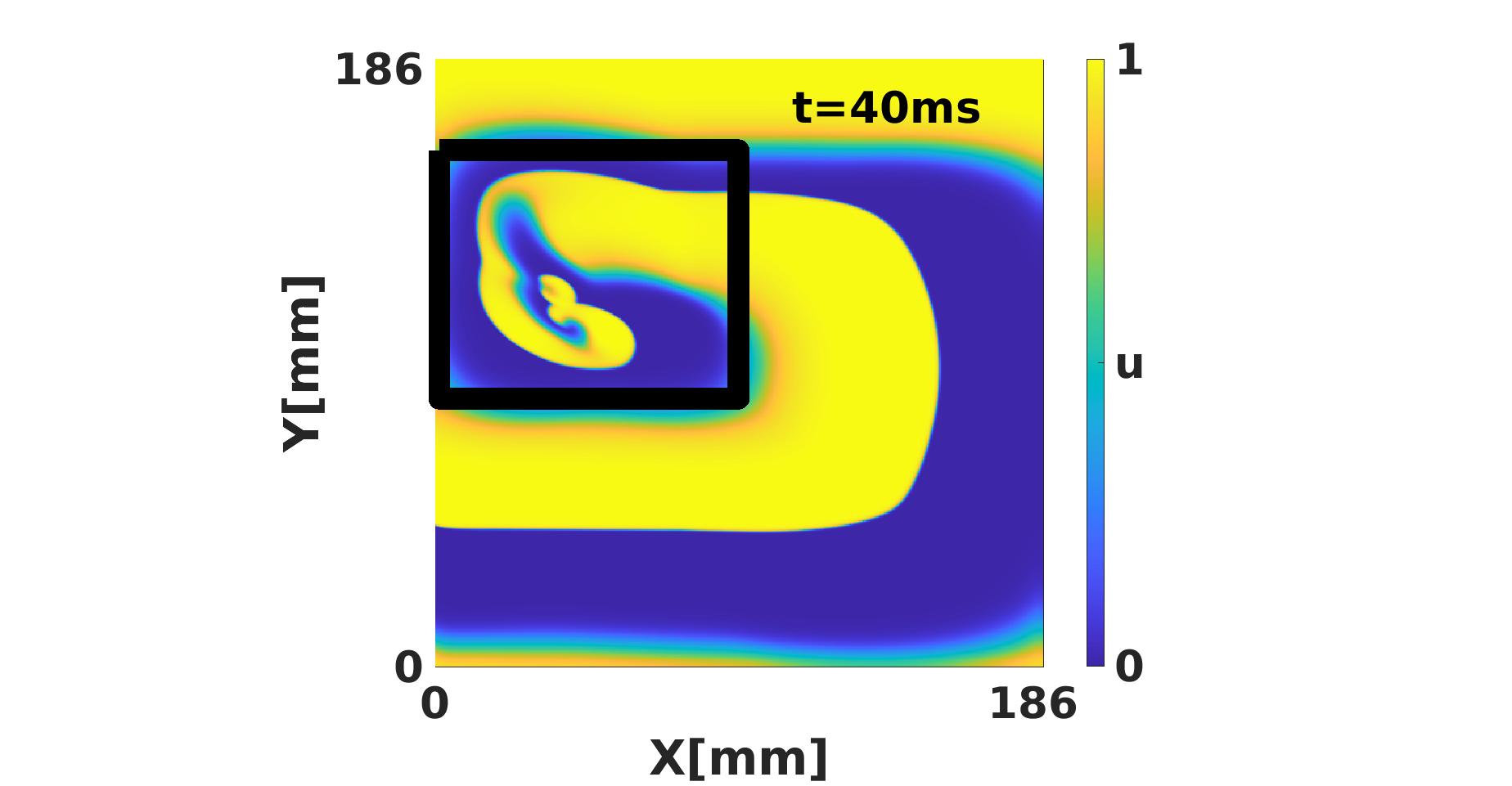}
\\
\includegraphics[trim={8cm 0.3cm 10cm 1cm},clip, height=  3.5cm]{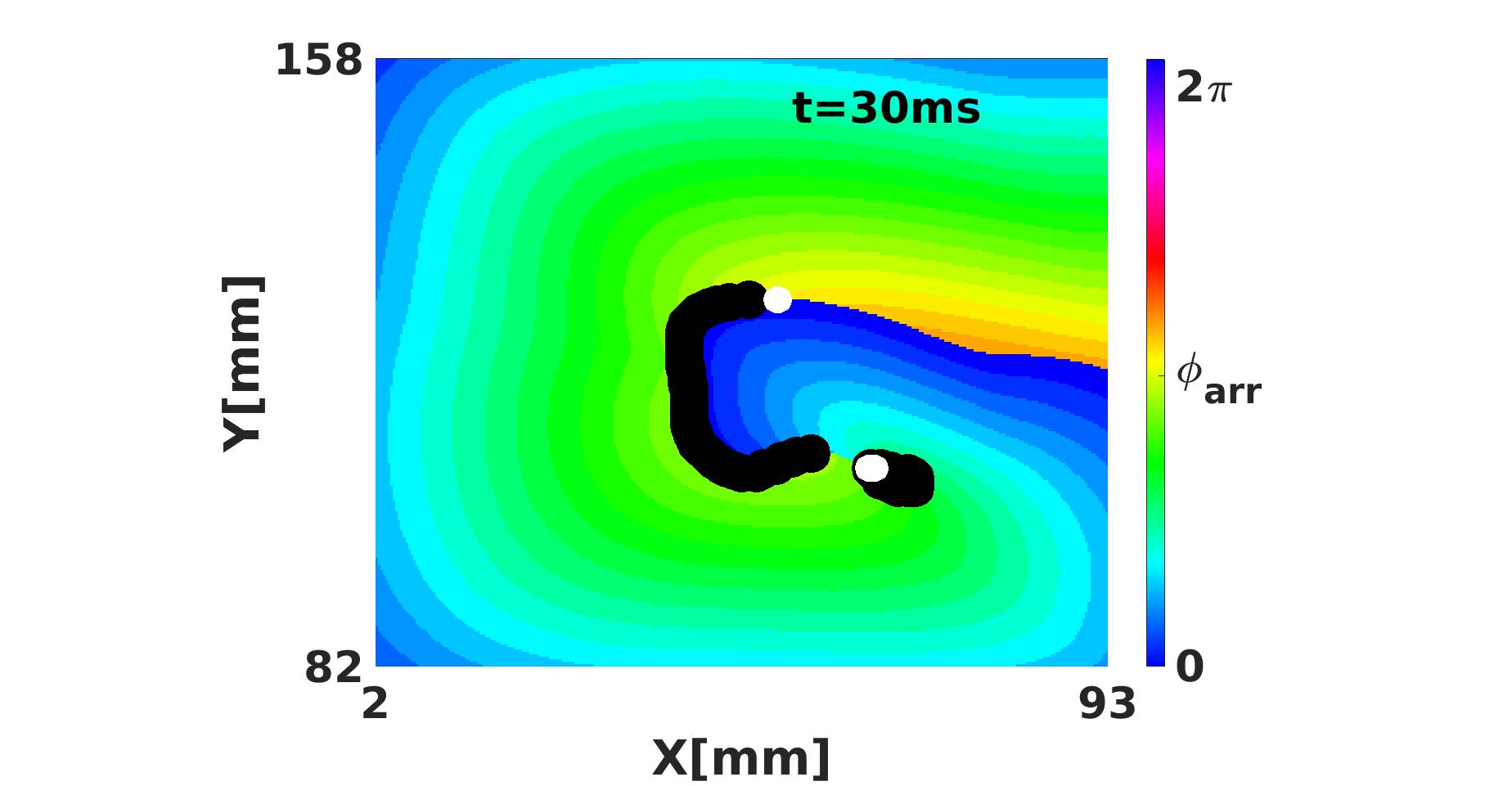}
\includegraphics[trim={8cm 0.3cm 10cm 1cm},clip, height=  3.5cm]{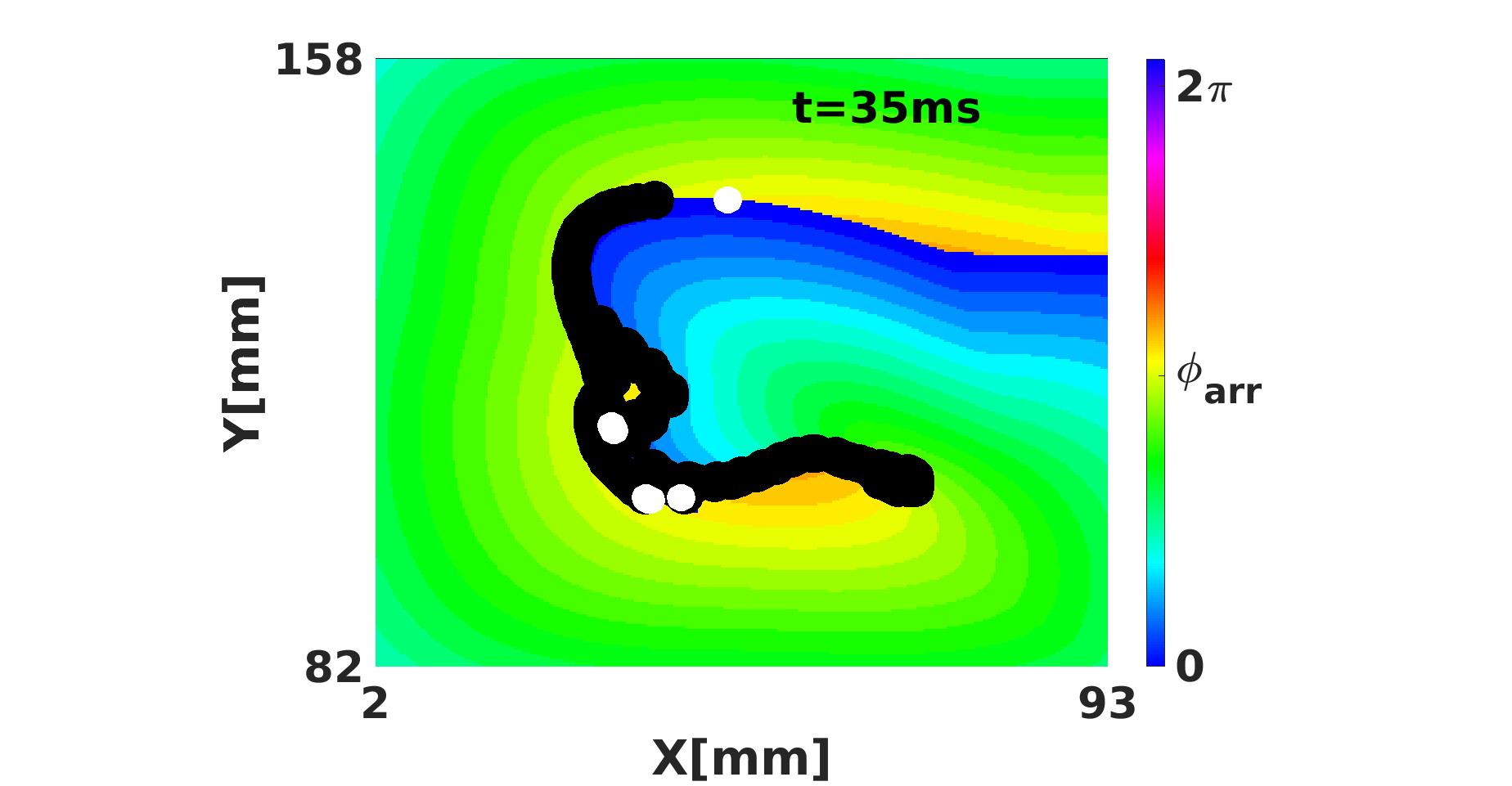}
\includegraphics[trim={8cm 0.3cm 10cm 1cm},clip, height=  3.5cm]{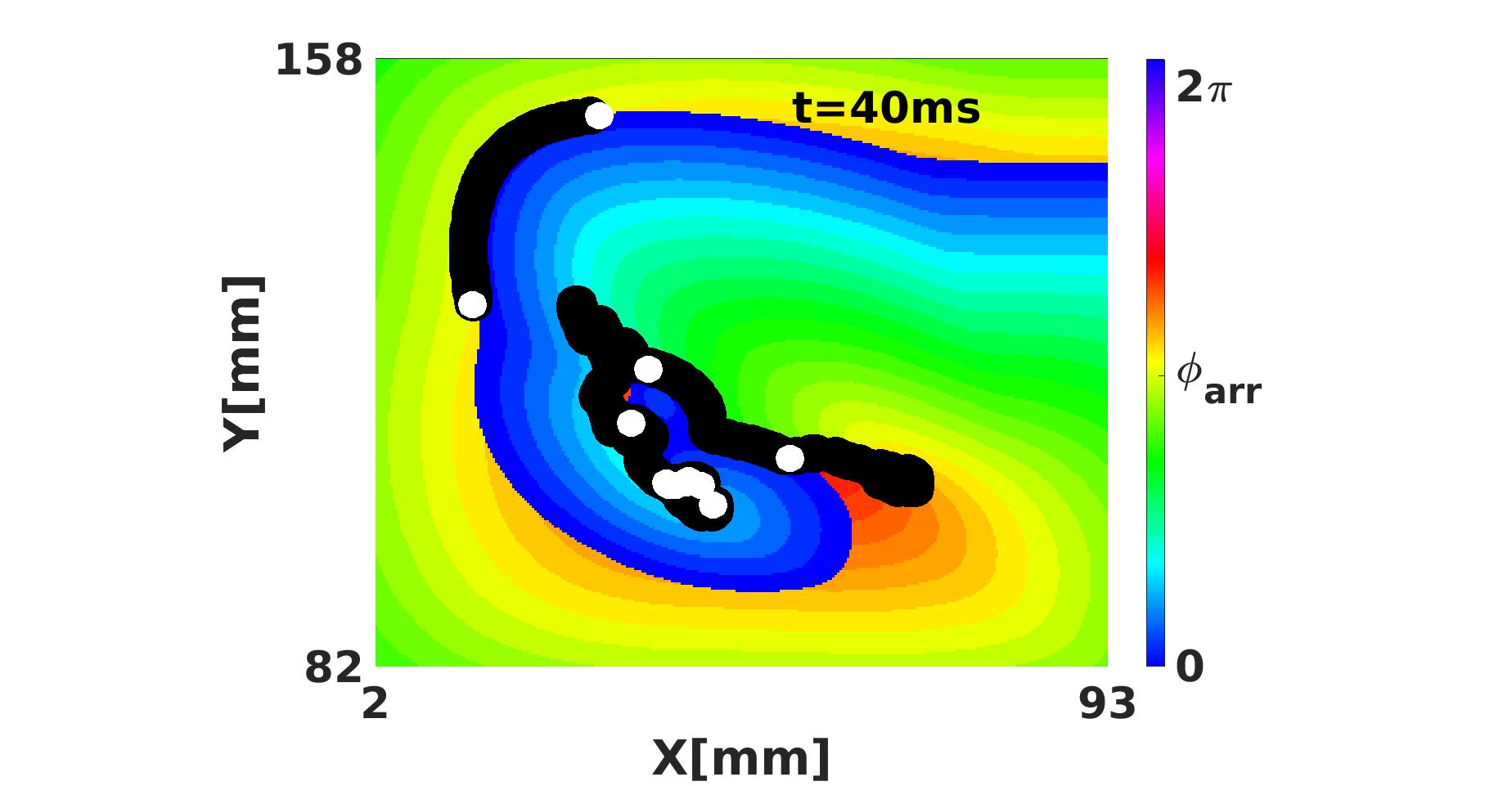}
\\ 
\includegraphics[trim={8cm 0.3cm 10cm 1cm},clip, height= 3.5cm]{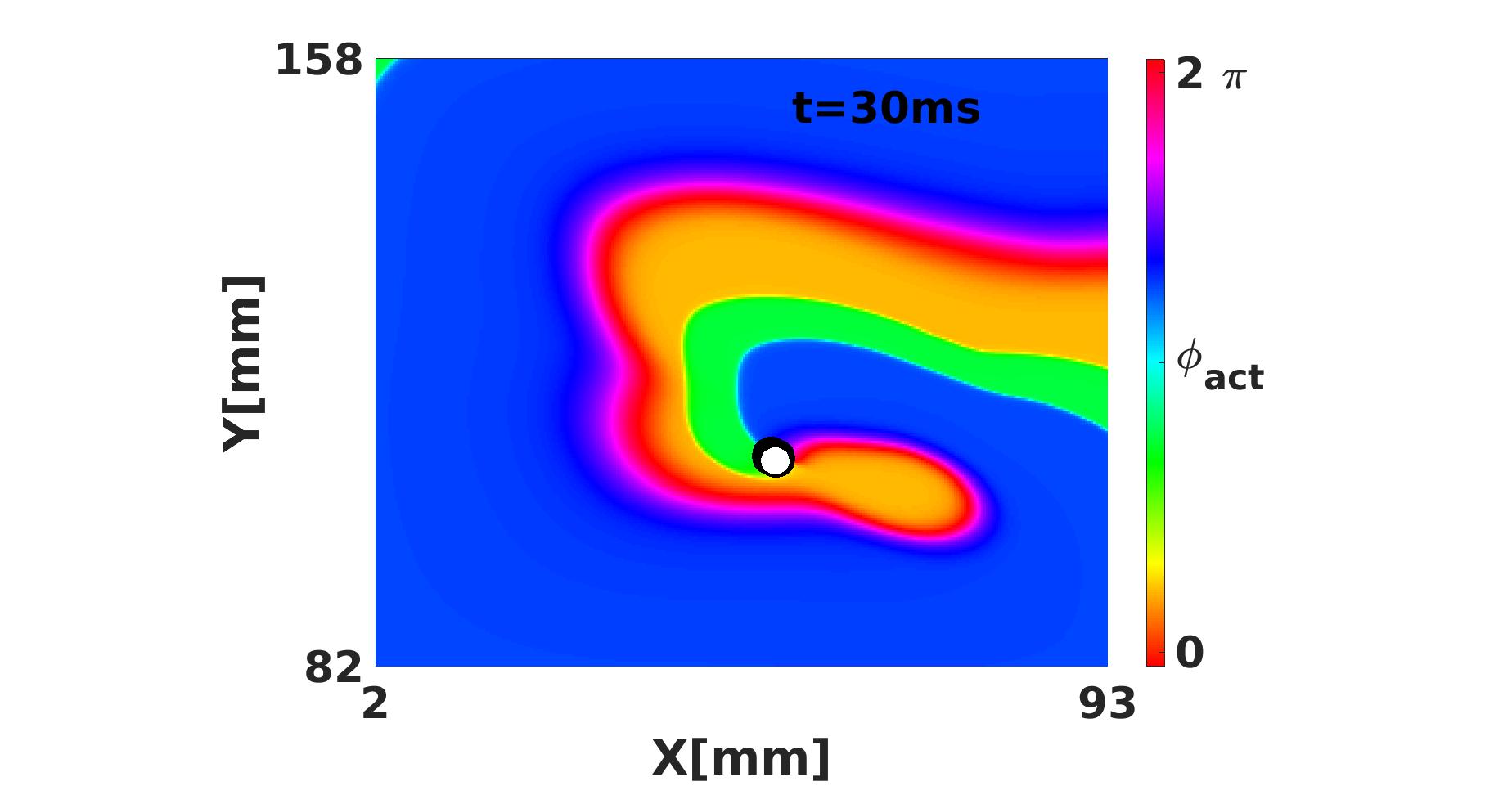}
\includegraphics[trim={8cm 0.3cm 10cm 1cm},clip, height= 3.5cm]{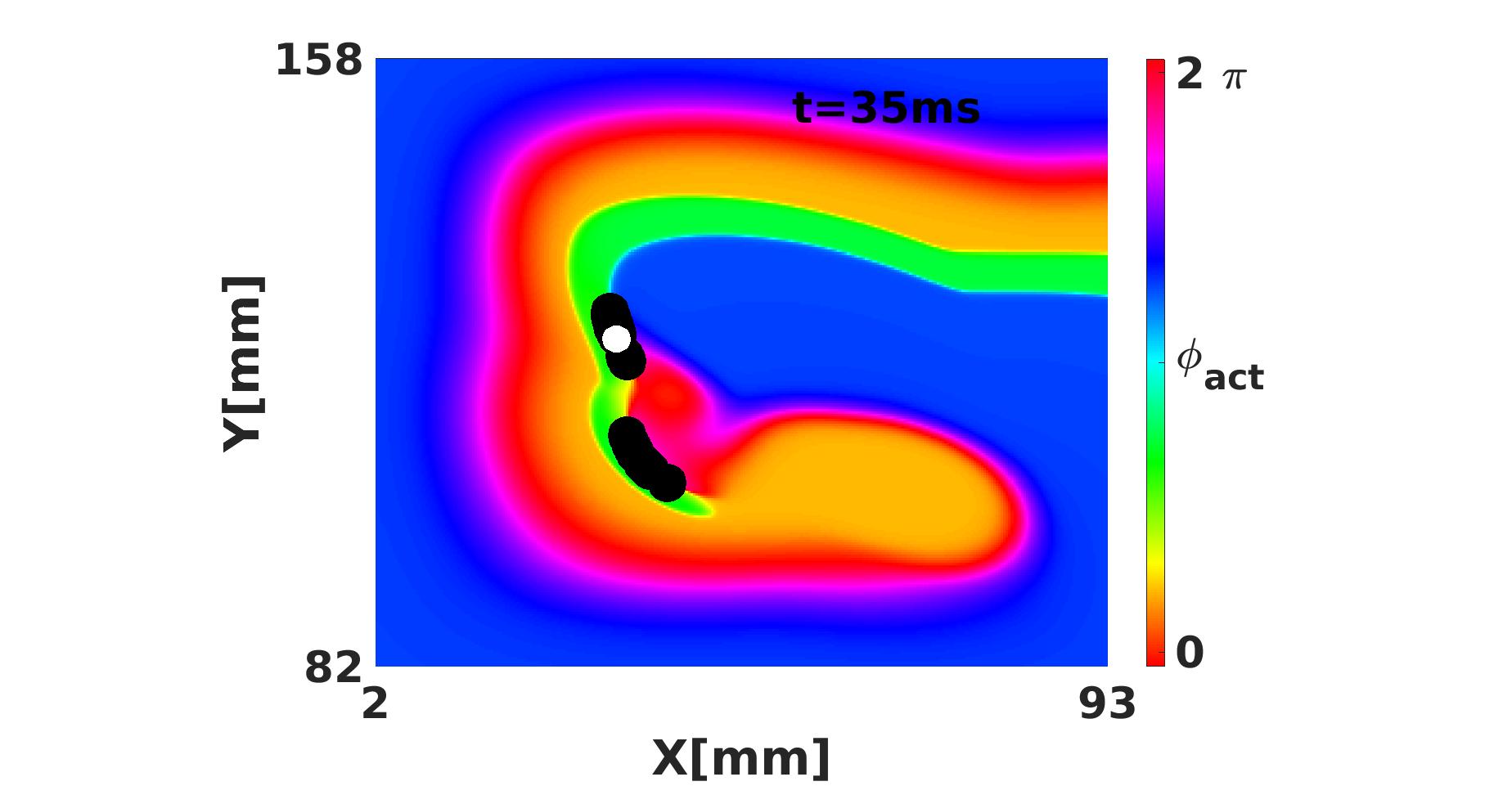}
\includegraphics[trim={8cm 0.3cm 10cm 1cm},clip, height= 3.5cm]{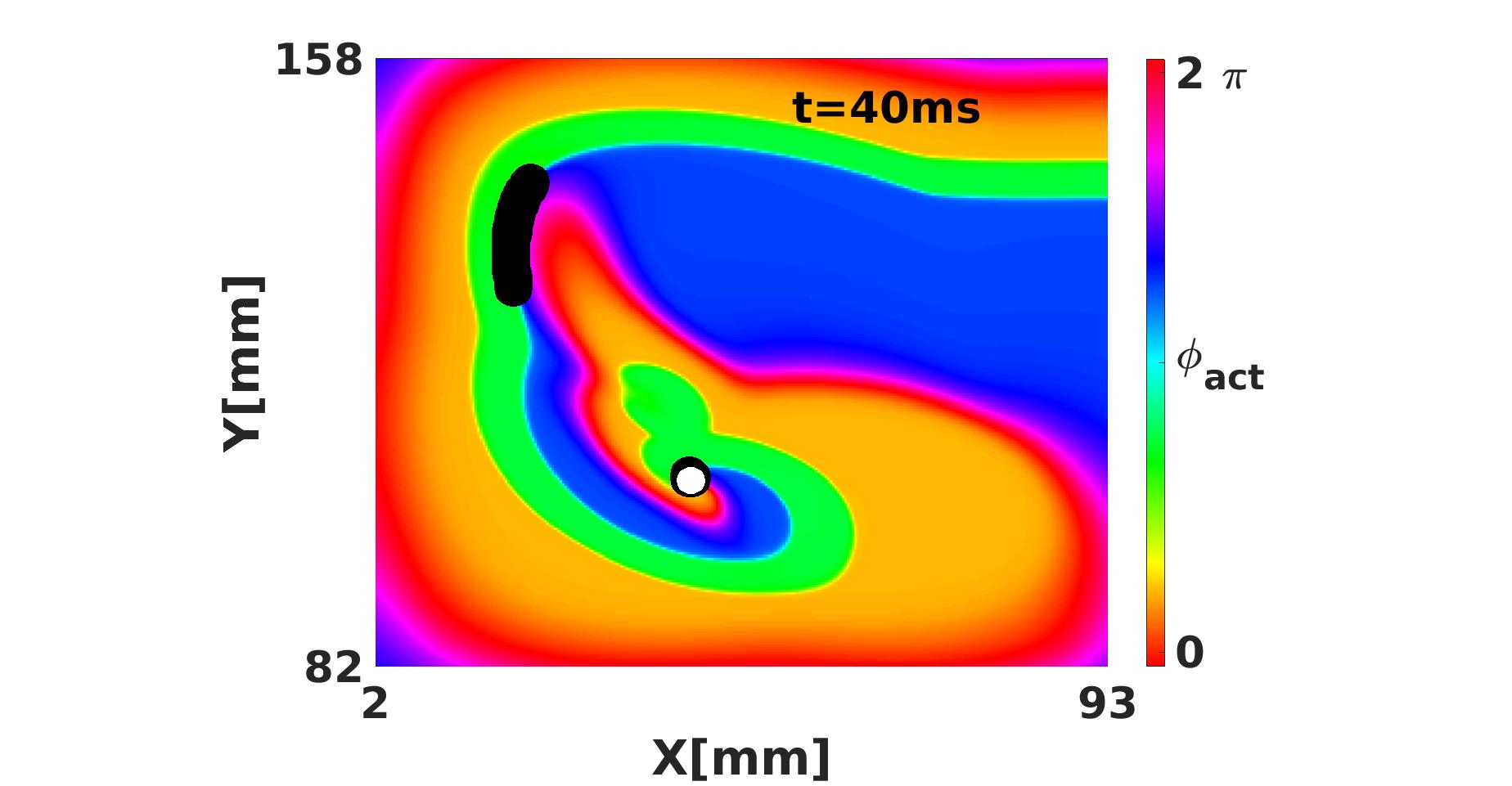}
\\
\includegraphics[trim={1cm 0.3cm 0cm 1cm},clip, height=\figbreakup]{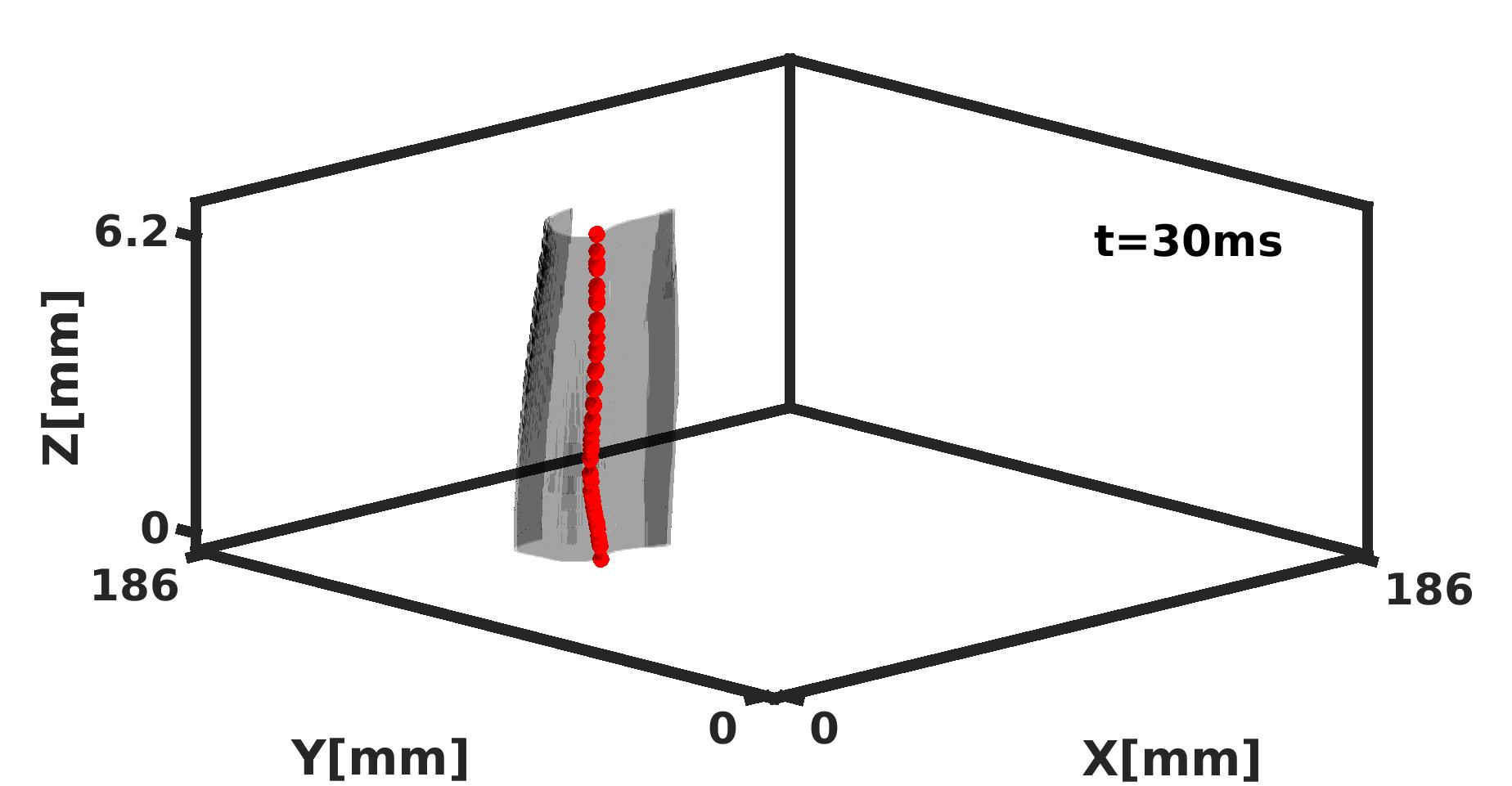}
\hspace{0.3cm}
\includegraphics[trim={1cm 0.3cm 0cm 1cm},clip, height=\figbreakup]{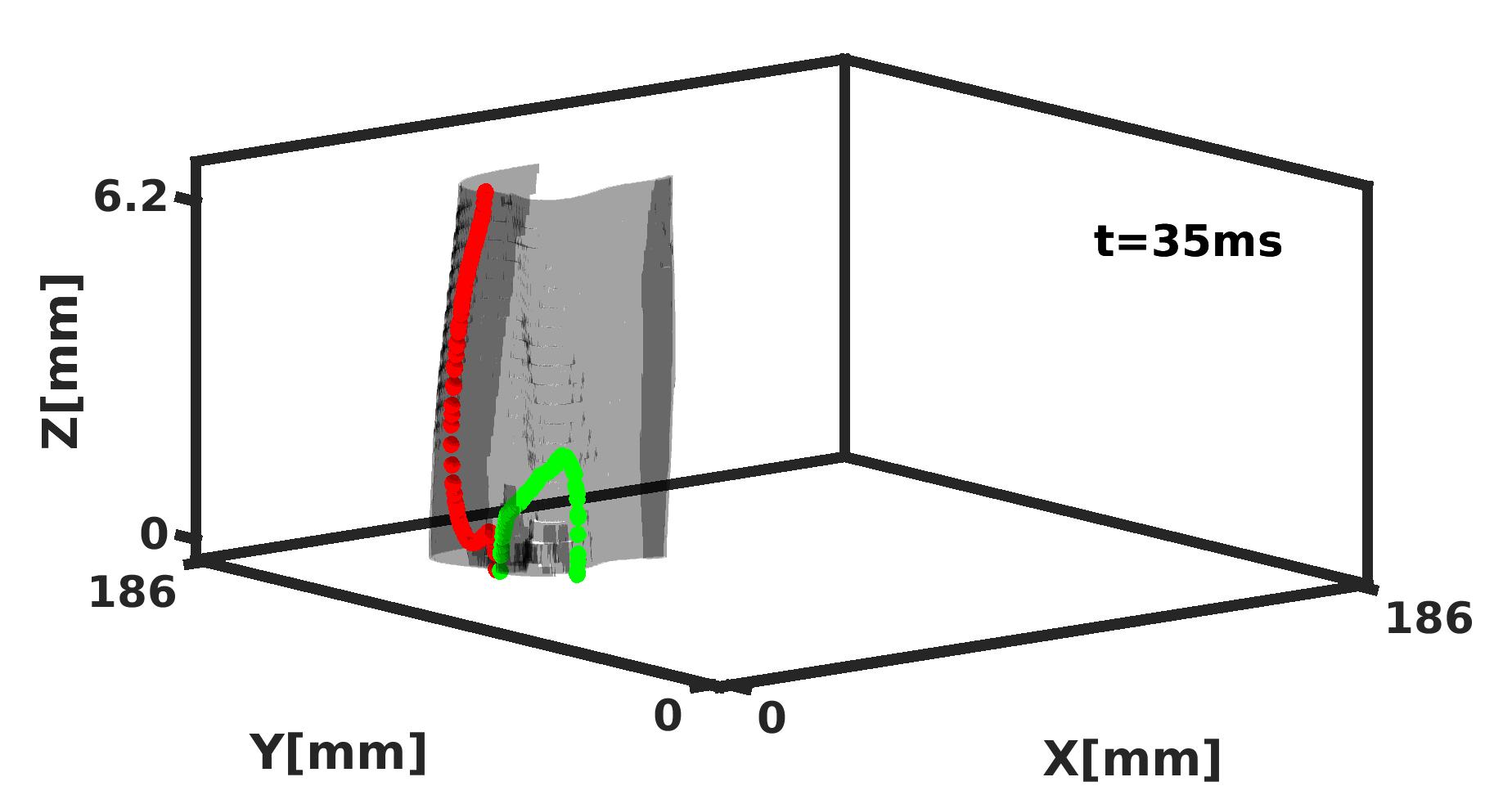}
\hspace{0.3cm}
\includegraphics[trim={1cm 0.3cm 0cm 1cm},clip, height=\figbreakup]{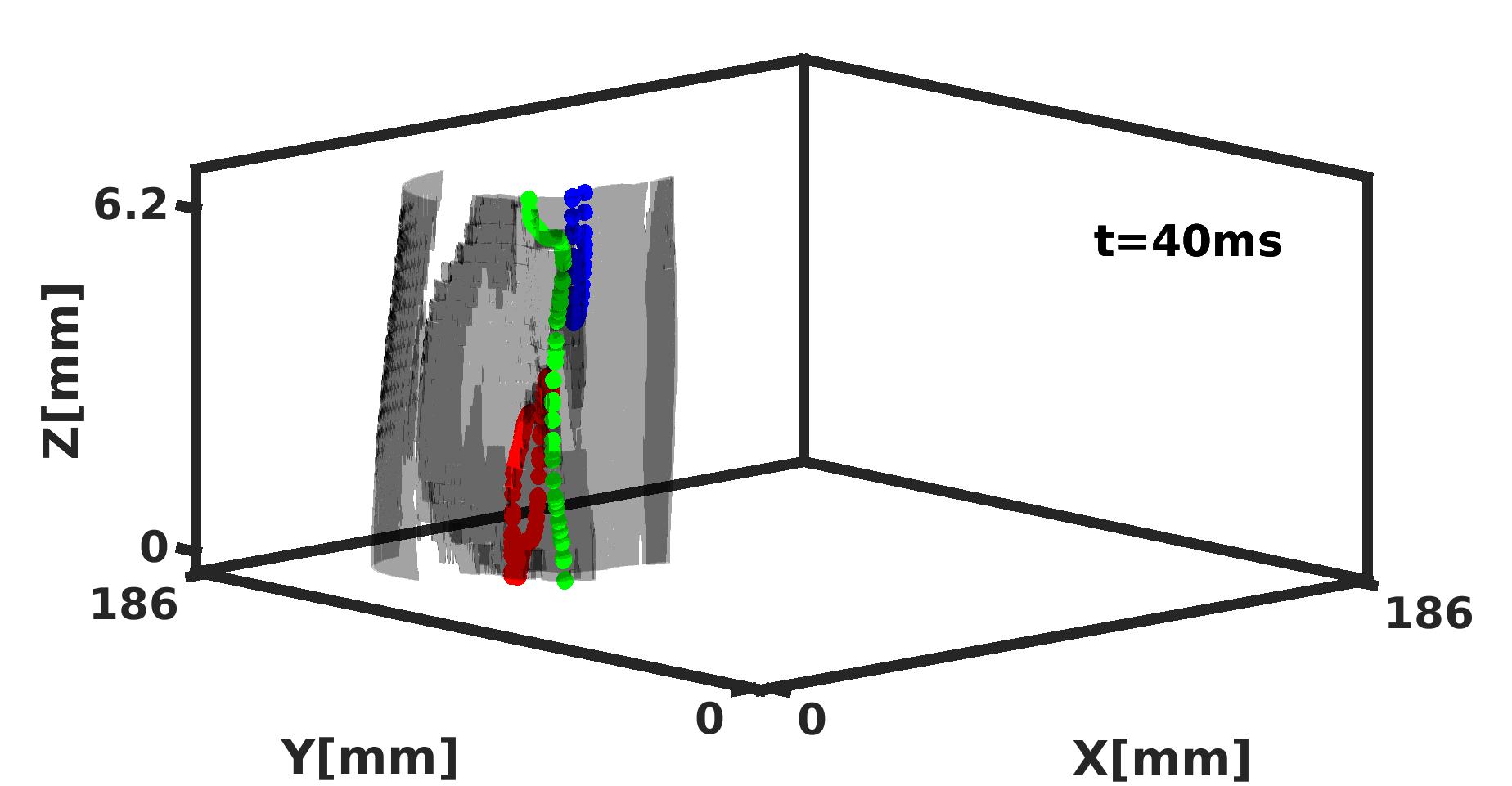}
       
  
    \caption{Analysis of a break-up pattern in a 3D slab geometry with linear-core rotors and rotational anistropy in the FK model. Panels A-C show activity at times t=30, 35 and 40\,ms, respectively. The rows show respectively 3D activity with filaments (colors indicate different filaments),  normalized transmembrane voltage on the bottom surface, and analysis of the surface patterns in terms of $\phiarr$ and $\phiact$ and 3D activity with filaments colored as above and PDSs indicated in grey. Note that several phase defects are observed, indicating conduction block near the spiral wave core.}
    \label{fig:breakup}
\end{figure}

The phase defect framework was illustrated above on simple examples, but is designed to analyse much more (complex) dynamics in excitation patterns. We conducted a 3D simulation in the FK model with rotational anisotropy-induced break-up, as detailed in \citep{Fenton:1998}. Fig. \ref{fig:breakup} shows snapshots of the turbulent pattern on the bottom surface of the slab (representing endo- or epicardial view in this geometrical model). Here, it can be seen that multiple PSs are detected on CBLs, that correspond to the core of cardiac rotors. 

\subsection{Numerical results in three dimensions}

In classical 3D theory, the PS extends to a dynamically varying curve called the rotor filament. With linear cores in the phase defect framework, the PDL will also extend in three dimensions, to form a phase defect surface (PDS). The bottom row of Fig. \ref{fig:breakup} shows the PDSs for the full 3D break-up patterns in the FK model. Here, it is seen that the PDLs that we identified on the myocardial surface extend to surfaces in three dimenions. From the surface recording (2D) already, it can be seen that these surfaces can branch and merge in time. 

\begin{figure}[b]
\centering
\raisebox{4cm}{\textbf{A}} 
\includegraphics[trim=1.5cm 1cm 0.5cm 0.5cm, clip,height=5cm]{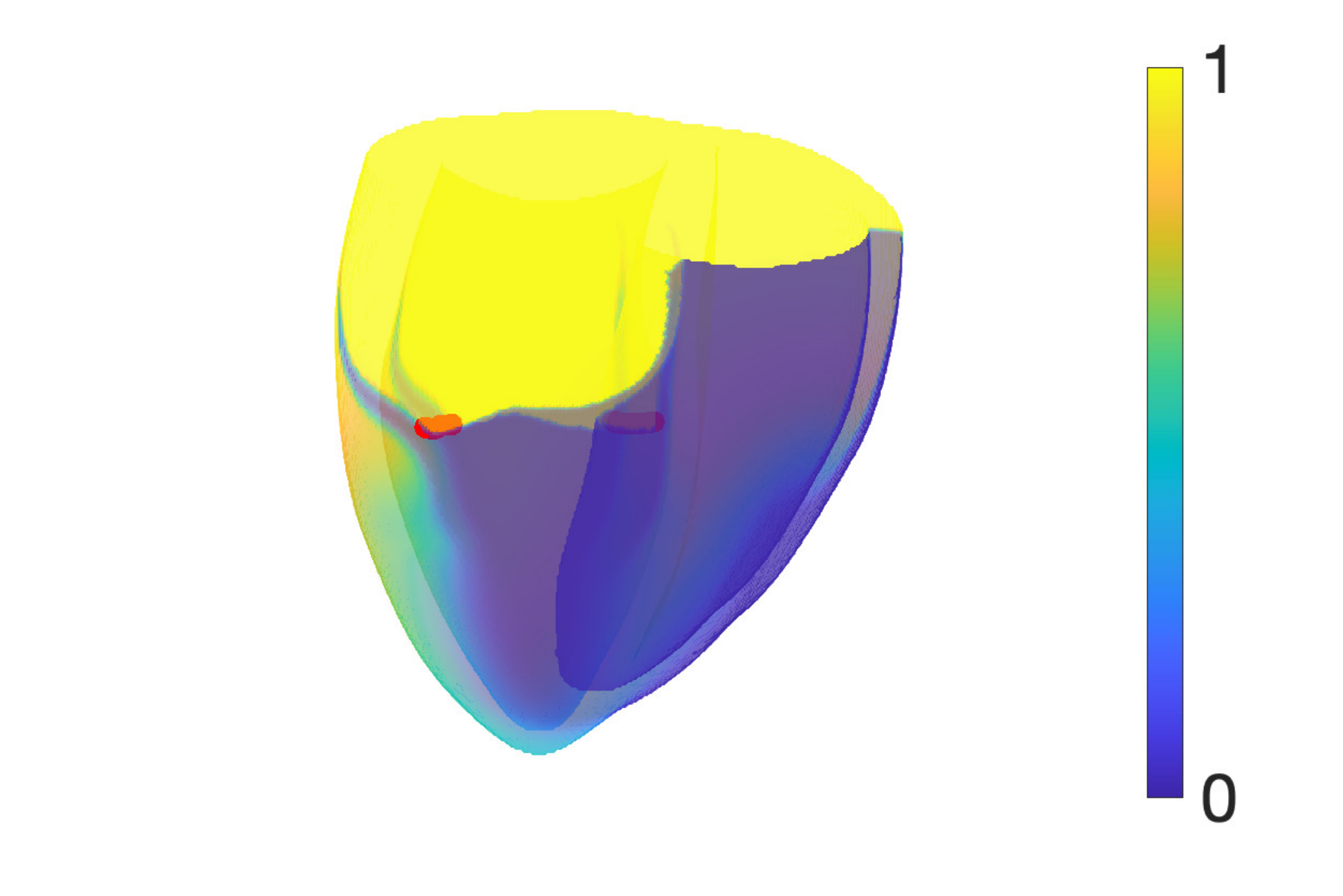}
\raisebox{4cm}{\textbf{B}} 
\includegraphics[trim=1.5cm 1cm 2.5cm 1.5cm, clip,height=5cm]{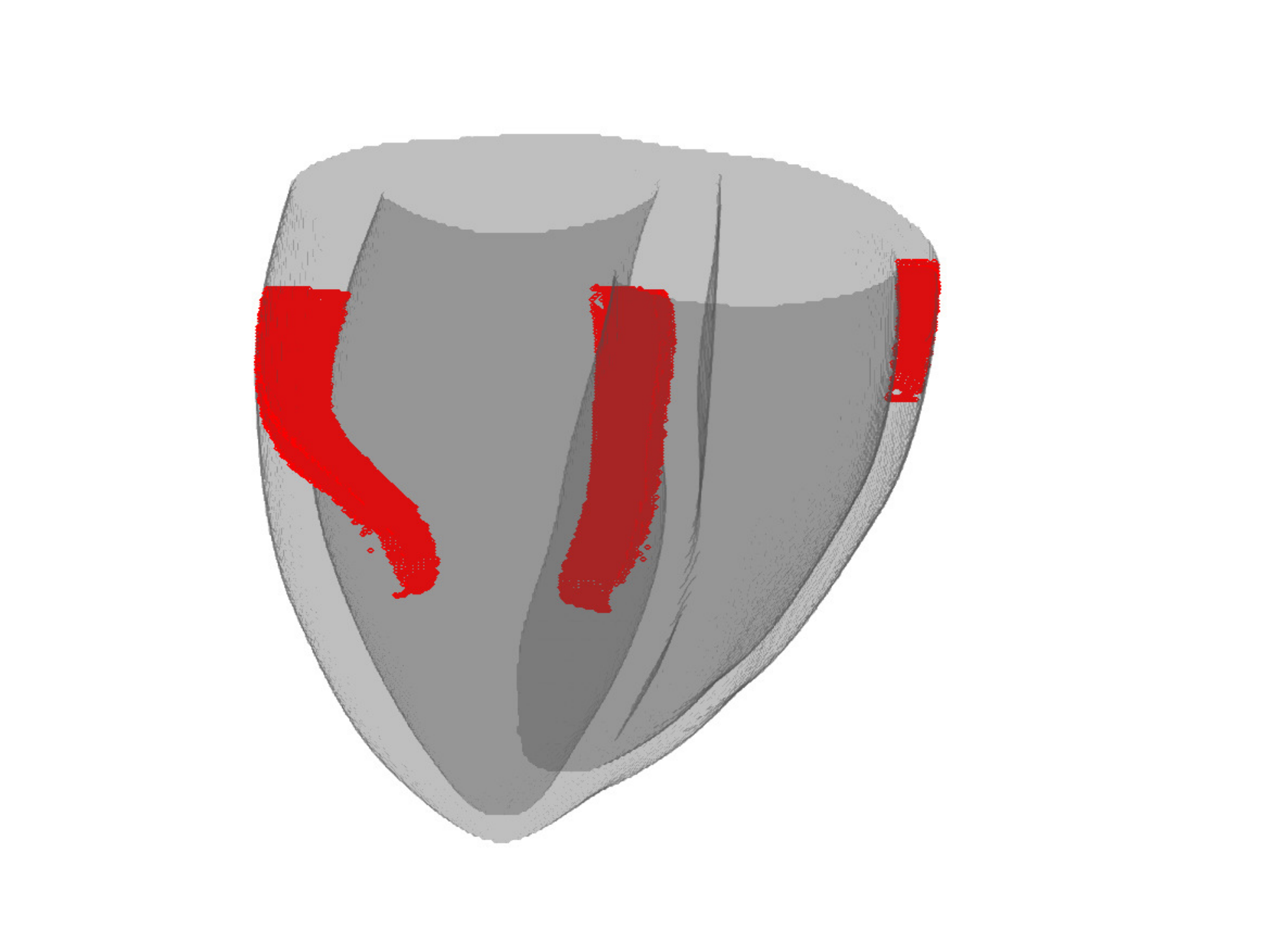}
\caption{
Numerical simulation of rotors in the BOCF model in a biventricular human geometry. (A) Snapshot colored according to normalized transmembrane voltage, showing 2 classical rotor filaments in red. (B) Phase defect surfaces (PDSs) for the same snapshot. The two leftmost PDS bear a classical filament in them, the rightmost one (in the right-ventricular free wall) is a site of conduction block. 
\label{fig:3D} }
\end{figure}

A second example of PDSs is shown in Fig. \ref{fig:3D}A, where a scroll wave with linear core was initiated by setting initial conditions with unidirectional block in the left-ventricular and right-ventricular free wall and the intraventricular septum. When viewed in terms of phase defects, 3 PDSs are seen. The two PDSs on the left correspond to filaments, and the rightmost PDS is a CBL; no matching filament is seen in Fig. \ref{fig:3D}A. 

\subsection{Analysis of rotors from isolated rabbit hearts}
\label{sec:experiment}


\begin{figure*}[t]
\centering
\includegraphics[width=0.8\textwidth]{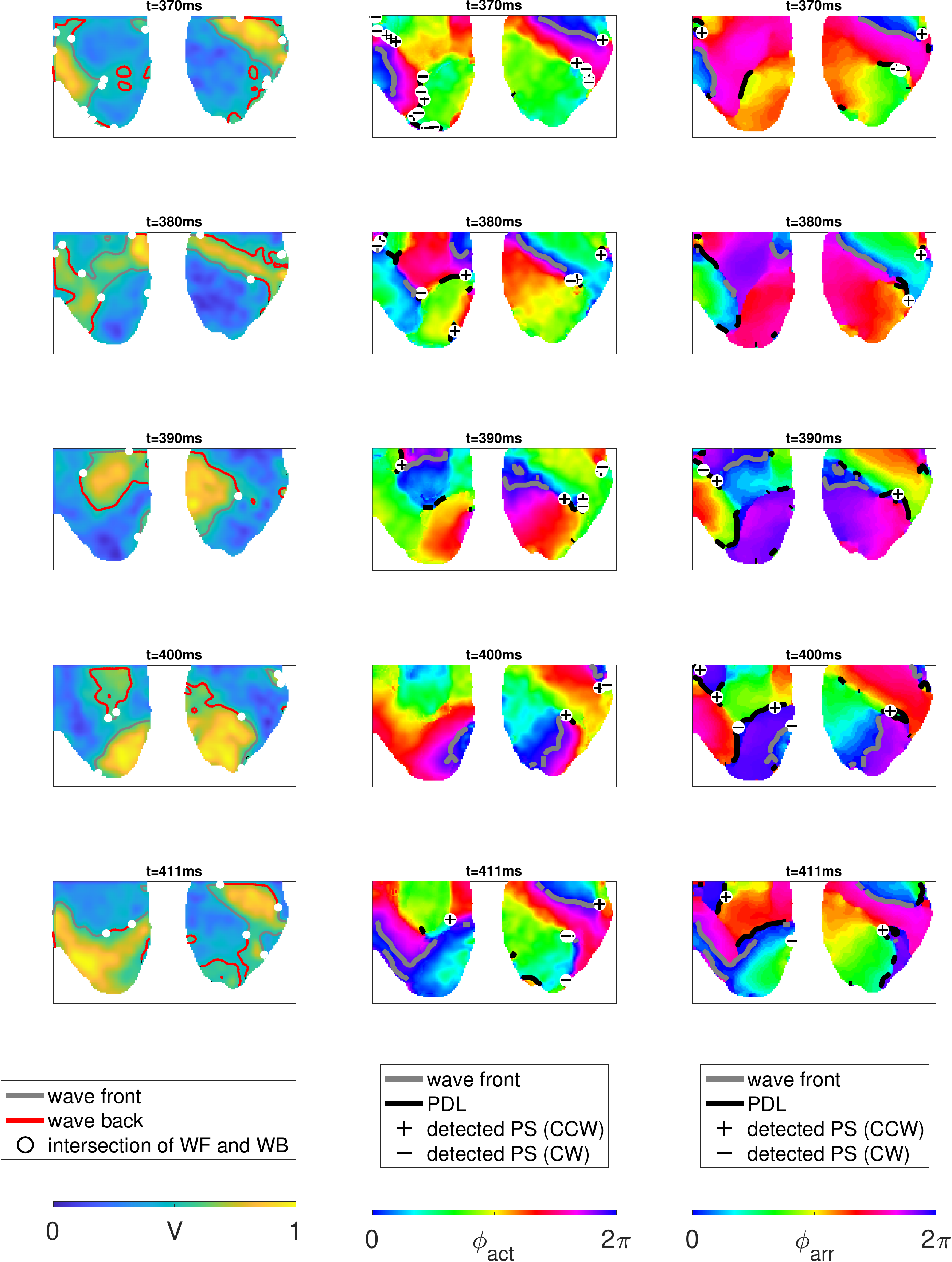}
\caption{Analysis of two-sided optical mapping data in rabbit hearts during ventricular tachycardia. Left column: normalized optical intensity (transmembrane voltage) $V$, together with WF ($V=0.5, \dot{V}>0$) and wave back ($V=0.5, \dot{V} < 0$). Middle column: same data series, $\phiact$ computed with $R$ the Hilbert transform of $V$, with PSs and PDLs computed from $\phiact$. Right column: colormap indicates $\phiarr$, computed with $\tau = 99$\,ms, equal to the inverse dominant frequency. PSs and PDLs computed from $\phiarr$ are also shown. PS detection was done using the $2\times 2 + 4\times 4$ method of \citep{Kuklik:2017}. 
\label{fig:experiments}
}
\end{figure*}

Electrical activity of a Langendorff-perfused rabbit heart was visualized during ventricular tachycardia via optical mapping experiments \citep{Kulkarni:2018} as detailed in Sec. \ref{sec:opticalmappingmethods}.
The processed movies of the different experiments performed are accessible in the Supplementary Material, and various characteristics of rotors in these movies are summarized in Table \ref{tab:stats}.

A first important observation in Fig. \ref{fig:experiments} and Table \ref{tab:stats} is that in more than 99\% of the cases where a PS was found, it was situated on a PDL. A typical sequence of the activation is presented in Fig. \ref{fig:experiments}. Here, a classical rotor can be recognized from the progression in time of a wave front (gray line) that starts in the top-right corner and circles counterclockwise around a line of conduction block that comprises the center of this rotor. Note that the PS detection algorithm finds a variable number of PSs on a PDL over time and therefore, the lifetime of the PS are smaller than the mean PDL lifetime, as is shown in Table \ref{tab:stats}. In contrast, the mean lifetime of PSs was about 3 frames. 

Second, two types of PDLs were observed: approximately two thirds of them were not associated with any PS during their lifetime and we recognize these as conduction block lines without a rotor attached to them. About one third of PDLs/CBLs had at least one PS on them during their lifetime, and these are therefore, at the rotor core. 

A third observation is that no rotors in the observed ventricular tachycardia completed a full turn on the epicardial surface, confirming other results in literature; a representative example is shown in the rightmost panel of Fig. \ref{fig:experiments}. Prior to completing a full rotation, excitation is surfacing from deeper layers, leading to a full conduction block. After that, the excitation loops in the same sequence, leading to a nearly periodic ventricular tachycardia. Note that the fact that the rotor is not performing a full rotation on the epicardial surface is not reflected by the PS analysis, but becomes visible only in the PDL framework. We verified visually that this scenario was true for most re-entrant activity in the data. 

\begin{figure}[htp]
\centering
\includegraphics[height=5cm]{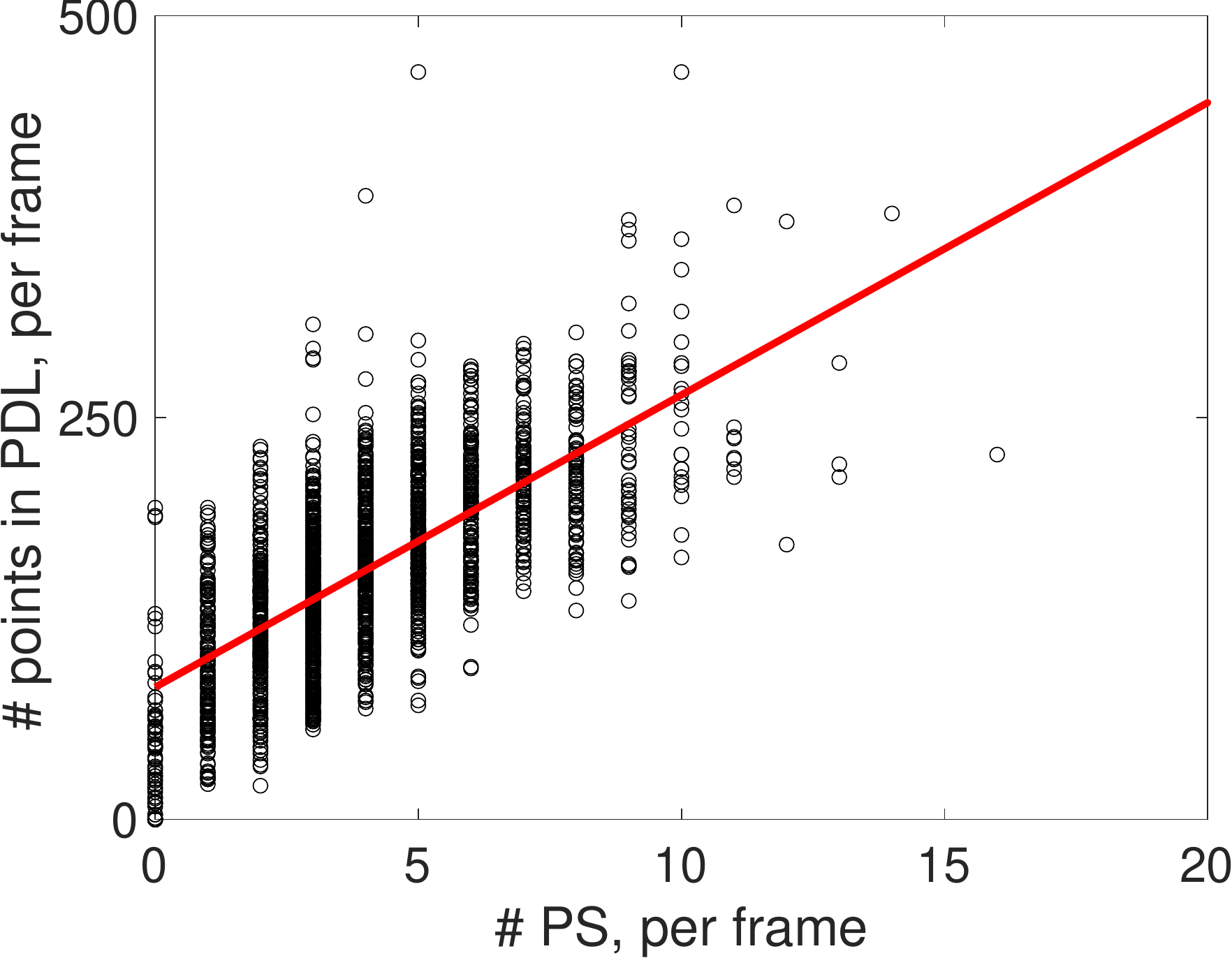}
\caption{Post-processing of rotors observed via optical mapping experiments from n=6 rabbit hearts. 
Positive correlation (red line) between number of detected PS and number of points in the PDL for the experiment shown in Fig. \ref{fig:experiments}. \label{fig:post}
}
\end{figure}

Fig. \ref{fig:post} shows that in every recorded frame the number of PSs correlates with the total length of PDLs in that frame (data from one experiment is shown, counting PSs observed with two cameras on both sides of the heart). The intersection of the trend line with the vertical axis at finite distance shows that even without PSs, PDLs are found, which we attribute to CBLs without a rotor connected to it. 

\begin{table*}[htp]
    \centering
 \begin{tabular}{l | c c c c c c| c }
 exp ID & 1& 2& 3&4 &5 &6 & average \\ \hline
 \# PS / frame & 6.55 & 7.38 & 7.29 & 5.96 & 4.85 & 9.00 & 6.84 \\ 
 \# PDL / frame& 11.82 & 13.14 & 13.20 & 13.23 & 11.54 & 17.59 & 13.42 \\
\# PS not on a PDL / \# PS & 0.05 & 0.07 & 0.03 & 0.10 & 0.04 & 0.10 & 0.06 \\ 
 \# PDL without a PS /frame  & 7.29 & 8.13 & 8.35 & 9.04 & 8.38 & 11.50 & 8.78 \\ 
\# new PS/frame & 1.45 & 1.51 & 1.65 & 1.47 & 1.23 & 2.02 & 1.55 \\ 
\# new PDL/frame & 0.41 & 0.46 & 0.42 & 0.43 & 0.36 & 0.66 & 0.46 \\ \hline
fraction of PS never on PDL & 0.35\% & 0.07\% & 0.24\% & 0.61\% & 0.41\% & 0.40\% & 0.34\% \\ 
fraction of PDL never hosting PS & 61\% & 60\% & 63\% & 64\% & 66\% & 64\% & 63\%\\ \hline
dominant frequency (Hz) & 20.57 & 20.45 & 17.38 & 18.59 & 19.31 & 20.14 & 19.40\\ 
dominant period (ms) & 48.63 & 48.89 & 57.55 & 53.80 & 51.80 & 49.66 & 51.72 \\ 
mean PS lifespan (ms) & 3.42 & 3.75 & 3.31 & 2.94 & 2.83 & 3.33 & 3.26 \\ 
mean PDL lifespan (ms) & 15.22 & 15.14 & 16.39 & 15.25 & 13.96 & 12.50 & 14.74 \\
 \end{tabular}
     \caption{PS and PDL properties observed via optical mapping experiments from n=6 rabbit hearts using two cameras, such that the entire epicardial wall is imaged. Here, PSs were computed using the integral method using the `$2\times2+4\times4$' ring of points \citep{Kuklik:2017}. Phase was computed for both PSs and PDLs with $V(t)$ the normalized optical intensity and $R(t)$ its Hilbert transform.}
    \label{tab:stats}
\end{table*}

\section{Discussion}\label{sec:discussion}

In this manuscript, we proposed a topological framework for excitable systems that feature conduction block lines. We brought terminology from complex analysis (Riemann surfaces, branch cuts) to a cardiac electrophysiology context and demonstrated that the phase defect concept can describe more structures than the classical PS. Moreover, we introduced the arrival time phase, which allows to convert activation time measurements to phase, and thereby unifies the isochrone and phase description of cardiac tissue. 

Our main findings are the following (1) the distribution of phase in our cardiac simulations and experiments is organized in regions with slow spatial variations of phase, separated from each other by localized interfaces, which we called phase defects; (2) that when arrival time phase is used, the only aberrant interfaces occur at CBLs; (3) At these PDLs, classical detection methods tend to localize PSs. 

\subsection{PDLs versus PSs}

The concept of extended PDLs can in our opinion explain some limitations associated with classical PS analysis that we presented in Fig. \ref{fig:problems}. First, near the central region of a PDL, the phase changes abruptly, and if this phase difference exceeds $\pi$, these points on a PDL can be classified by traditional methods as PSs, see Fig. \ref{fig:problems}A. One option is to filter the resulting PSs and remove closely spaced PSs, but another option suggested here is to recognize the structures as line defects rather than point defects of phase.  

The observation that points on a PDL can be detected as PSs at once explains why PSs are often found on CBLs, as was seen in \ref{fig:problems}C. Here, we conclude that in systems that form linear-core rotors (even short-lived) do not possess PSs, but only PDLs. Conversely, systems that sustain circular-core rotors can have PSs at the spiral core, and PDLs at conduction block lines. 

In Fig. \ref{fig:problems}B, we argued that a PS cannot capture the shape of a linear core. In contrast, a PDL can do so, as it is a line spanning the entire conduction block region. 

Geometrically speaking, PDLs and PSs are related to each other: a PS can be seen as the limit of a PDL having length zero. Conversely, a PDL can also be regarded as some kind of extended PS. Hence, the PDL concept can be seen as a refinement of a PS. Therefore, we make the suggestion to adapt current phase analysis methods to accommodate for PDLs, which can potentially make analysis more robust and accurate.  

\subsection{Physiological interpretation of phase defects}

In analogy to the term `phase singularity' referring to spiral wave tips, we here suggested the term `phase defect line' (PDL) to refer to CBLs when their phase structure needs to be emphasized. PDLs are ubiquitous in other branches of physics, e.g. magnetism \citep{Landau:1935}, liquid crystals \citep{Williams:1963} and string theory \citep{Vilenkin:1985}.

Our justification as to why PDLs exist in these systems is based on biological rather than mathematical arguments and essentially traces back to the arguments of \cite{Winfree:1974}. Consider the activation cycle of an excitable cell, i.e. its action potential, as shown in Fig. \ref{fig:classic}B. Now, if one expects a PS in the center of a vortex, one assumes that the cells can also be in a state that lies somewhere in the middle of the cycle in state space. However, this situation may be biologically impossible: electrophysiology processes during activation will push the cell along its activation cycle, not necessarily allowing it to occupy the middle state. In case of a PDL, the `forbidden' state does not need to be realized. Going back from state space to real space, the region with forbidden states is mapped to the PDL. 

If one, however, models cardiac tissue in the continuum approximation using a diffusion term, the transmembrane potential becomes continuous, and if this potential is used as one of the observables $(V,R)$ to infer phase, points in the `forbidden zone' of state space are actually realized by the smoothing effect, leading to a finite thickness of border between regions of different phase. Since this diffusive effect acts during one action potential with duration $\tau$, we estimate the effective thickness of the phase defect line to be
\begin{align}
    d = \sqrt{D \tau}
\end{align}
with $D$ the diffusion coefficient for transmembrane potential in the direction perpendicular to the PDL. For the BOCF model for human ventricle, one has $D=1.171\,$cm$^2/$s and $\tau = 269\,$ms, leading to a PDL thickness of $d=5.6$\,mm if the PDL is perpendicular to the myofiber direction. With conduction velocity anisotropy ratio $c_1/c_2 \approx 2.5$, one finds $d=14\,$mm if the PDL is oriented parallel to the myofiber direction. These are rough estimates in a minimal monodomain model. 

Thus, as with other interfaces between phases in nature, a boundary layer forms with a thickness related to fundamental constants of the problem. This boundary layer effect explains why the common PS detection methods usually do return a result, even in the region of conduction block. When the new framework is applied to a linear-core rotor, the classical spiral wave tip or PS is expected to be the point where a wave front ends on a PDL. However, this may depend on the precise tip structure, spatial resolution and PS detection method, see e.g. Fig. \ref{fig:S1S2revisited}. 


\subsection{Relation to other works}

Between the submission of our preprint \cite{Arno:2021arxiv} and acceptance of this paper, a paper was published by \cite{Tomii:2021} et al., who also identify a phase discontinuity at the center of cardiac vortices, and compares it to a branch cut in complex analysis. This effect was demonstrated by Tomii et al. in numerical simulations only. They detected PDLs using
\begin{align}
\cos ( \phi(\vec{r}_1) - \phi(\vec{r}_2)) > A     \Rightarrow \frac{\vec{r_1} + \vec{r_2}}{2} \in PDL.  \label{PDLdetection_Tomii}
\end{align}
This method is very similar to our Eq. \eqref{PDLdetection_Tomii}. However, our study goes further than their result, as we introduce arrival time phase to link the phase picture with LAT, we demonstrate the presence of PDLs in optical mapping experiments, and report about PDSs in 3D as well. 

We are also aware that the concept itself of a CBL is not new at all, as conduction blocks are mentioned for more than a century in electrophysiology reports \citep{Mines:1913}. In 3D, filaments have also been reported as being ribbon-like before \citep{Efimov:1999}. However, these observations have not been thoroughly integrated in pattern analysis methods (which is usually based on PSs) or the dynamical systems approach to better understand arrhythmias (which has long considered circular-cores only). 

Meanwhile, CBLs are routinely localized and visualized during clinical procedures, e.g. as the result of an ablation. A prime insight advocated in this manuscript, and by \cite{Tomii:2021}, is that the CBL itself is an extended line of abrupt change between two regions with different activation phase, and that this resembles more a mathematical branch cut than a PS. Further studies are recommended to see if the phase defect concept can help to solve some basic questions in the field, such as why rotors do not complete full turns in tissue, and which characteristics of the reaction kinetics determine the core shape. 

\subsection{Relevance of phase defects for theory of arrhythmias}

The recognition that a PS may not be the best option to describe linear core rotors, is in our opinion opening up several paths for further analysis and insight in cardiac arrhythmia patterns. 

First, by introducing the arrival time phase $\phiarr$, we have unified the LAT description with the phase description, which allows computing $\phiact$. Which phase to use in practice will depend on the availability of data and their quality. Fig. \ref{fig:phiarr} furthermore shows that both phases can be converted into each other. Hence, activation phase can also be estimated from LAT, and arrival time phase from a single snapshot of $V$ and $R$ at a given time.  

Second, this work was inspired by localizing the regions where an external stimulus can affect rotor dynamics. The relevant sensitivity function is here known as a `response function' (RF) \citep{Biktasheva:2003}, and these have been computed recently also for meandering and linear-core rotors \citep{Marcotte:2015, Dierckx:2017}. 
For circular-core rotors, it has been numerically demonstrated that the RF is located near the spiral wave tip, or its PS. So, rather than acting as a point particle, the rotor has a finite extent given by the spatial decay width of its RFs, with sensitivity localized near the spiral wave tip or PS, not at its instantaneous rotation center. 
For linear-core spirals, the PDL at their core indicates that the sensitivity is located around this zone of rapidly varying phase. More specifically, their RF was shown to be localized near the end points of the PDL \citep{Dierckx:2017} (`turning points'), suggesting that linear-core rotors resemble a localized particle that hops from turning point to turning point. That the sensitivity is lower in the middle part of the PDL can be understood from the conduction block interpretation: this block line cannot move much, as the tissue on one side is inexcitable. Hence, the PDL concept confirms insights from RF theory, and it can be hoped that combining both may open up ways to deepen the understanding of how point dynamics (local electrophysiology) affect the emerging patterns. 

Third, it is yet unclear how essential structures such as WBs and CBLs move under the influence of external stimuli such as impeding waves or electrotonic currents from boundaries or 3D effects. To answer this question is a key step towards designing better methods to control wave patterns, e.g. as in defibrillation. The identification of the phase defect, with boundary layer, could allow in the near future to design a quantitative theory of their dynamics, in analogy to earlier successes on rotor filaments with circular cores \citep{Keener:1988} and excitation fronts \citep{Kuramoto:1980}, which have lead to instability criteria for pattern formation in the heart \citep{Biktashev:1994, Dierckx:2012}. 

Fourth, a consequence of this investigation is that describing fibrillation patterns using PSs only \citep{Gray:1995,Gurevich:2019} may not be sufficient to fully understand the dynamics if conduction block occurs. Therefore, these studies could benefit from additional identification of PDLs, of which we made first stpdf in Figs. \ref{fig:breakup} and \ref{fig:experiments}. 

Finally, our preliminary results on 3D dynamics in Fig. \ref{fig:breakup}, and \ref{fig:3D} are a first step to extending the filament concept to the PDSs, like was done here to go from PSs to  PDLs. In 2D, we here demonstrated that circular-core spiral waves are centered around a PS and linear-core meandering spirals rotate around a PDL. Since in 3D, the collection of classical PSs forms a filament curve, linear-core scroll waves extend to PDSs. Therefore, the study of the dynamics of meandering scroll waves can benefit from its description in terms of phase defect surfaces. These surfaces, as shown in Fig. \ref{fig:3D} have a ribbon shape \citep{Efimov:1999}, but the meaning of the ribbon is different from the ribbon model for filaments \citep{Echebarria:2006}, where the surface normal of the ribbon was purely related to scroll wave twist. In our framework, the width of the PDS (ribbon) is approximately equal to the length of the PDL in 2D. How dynamical concepts such as filament tension \citep{Biktashev:1994} and rigidity \citep{Dierckx:2012} generalize to PDSs is still an open problem. For the intermediate case of flower-like meander \citep{Barkley:1990b}, between the extremes of circular and linear cores, it still needs to be determined whether the phase structure  is a PS, PDL or a different structure. Further analysis of complex 3D dynamics in the phase defect framework will be the topic of further investigation.

\subsection{Interpretation of experimental results}

Our analysis of rotors observed in isolated rabbit hearts during experiments shows in short two observations. First, that PSs on the epicardium were in this system always located on conduction block lines (PDLs). This observation is in line with older and more recent literature \citep{Efimov:1999,Podziemski:2018}. Second, the rotors were not performing more than one full rotation. To our knowledge, the question on why rotors are not performing multiple rotations in ventricular tissue is unresolved. The phase defect framework may be used to answer this question in a comprehensive manner, rather than resorting to numerical simulations only. 

Another relevant question is whether the PS or PDL is the structure that governs the organisation of cardiac arrhythmias. Here, we would answer that both are intricately related to each other: in systems with linear-core rotors, we think the PS only arises due to smoothing of the phase field and is therefore located at a specific spot on the PDL, often where it meets the wave front. Whether one should localize PSs or PDLs in experiments or clinical situation, depends on the application and desired accuracy.

\subsection{Limitations and outlook}

The concept of PDLs was illustrated here in a very simple setting: we mostly considered single rotors or rotor creation in 2D. Our initial results in 3D (see Fig. \ref{fig:breakup}) and with break-up patterns (see Fig. \ref{fig:3D}) show that there is rich dynamics present that can be further analysed using the phase defect framework. We also did not consider alternans here, as in that case the inertial manifold in state space will be not a simple closed curve, and may require two or more phase parameters to describe those states. This effort will be undertaken in future work. 

Also, we have described the framework in a general setting, without explicit attention to the reason why rotors form in the system. Future work could be directed to elucidate differences between rotors formed by dynamical breakup (e.g. in atrial fibrillation models) or by interaction with inhomogeneities of the medium. 

In this paper, the phase defect concept was used only to describe the observed patterns, without yet looking into how the topology determines the evolution of the structures such as WFs, WBs and PDLs. Still, we developed the phase defect framework with the aim of providing a comprehensive quantitative analysis of excitation patterns, in the line of previous works on circular-core rotors and filaments \citep{Keener:1988,Wellner:2002,Verschelde:2007,Dierckx:2012}. The PDL concept provides not only a terminology for it, but also a way forward: solutions will need to be stitched together at the PDL interface, to elucidate their dynamics, much like was done before for wave front dynamics \citep{Keener:1986}. 

In the optical mapping experiments, the mechanical uncoupler blebbastatin was used, which however also affects the physiology of the cardiac cells. Therefore, the phase defect structures that we observed could be different from real cardiac tissue where no such uncoupler is present. 

Furthermore, we used simple methods to find PDLs and WFs in the experimental data. As a result of representing the PDL as a spline curve, it is seen in Fig. \ref{fig:experiments} that the wave fronts are not exactly touching the PDLs, while this is expected from theory. We will continue refining our numerical processing methods to make this image more consistent. 

This study was also set in a fundamental science setting. Yet, it is inspired by the clinically relevant length and timescales: how do local dynamics self-organize in complex patterns with rotors and conduction block lines? In our opinion, contributions to the answer can be found from scrutinizing the patterns themselves, and building up a comprehensive understanding after the correct building blocks have been identified. 

Finally, we suggest to try our methods to clinically relevant datasets, such as LAT maps derived from local electrograms. Existing algorithms could be tuned or extended, given that the classical PS was found in our simulations and experiments on a line of nearly discontinous phase. 

\subsection{Conclusion}
In summary, the key ideas presented in this manuscript are:
\begin{enumerate}
\item near conduction block lines, the phase of cardiac cells is close to discontinuous along a line, rather than along a point, as assumed in classical phase analysis; 
\item therefore, in excitable systems that have a forbidden zone in the inner region of their typical excursion (action potential) in state space, detected PSs are located at PDLs;
    \item there is more than one definition of phase, and these different phases can be translated into each other depending on the goal and available data. 
\end{enumerate}

As a historical note, we find that the name of mathematician Bernhard Riemann is inscribed twice in our hearts: the heart is not only a Riemannian manifold due to anisotropy of wave propagation \citep{Wellner:2002,Verschelde:2007,Young:2010}, but also features phase defect lines showing non-trivial Riemannian surfaces that organize the electrical patterns during arrhythmias.

\section*{Author Contributions}

H.D. and E.G.T. conceived the study. E.G.T. conducted optical mapping experiments. H.D. and L.A. ran the numerical simulations. H.D and L.A. performed experimental data analysis together with J.Q., N.N. and M.V. All authors contributed to manuscript writing and internal reviewing. 

\section*{Funding}
E.G.T. was supported by National Science Foundation DCSD grant 1662250. H.D. received mobility funding from the FWO-Flanders, grant K145019N. L. Arno was supported by a KU Leuven FLOF grant.

\section*{Acknowledgments}
We thank Alexander V. Panfilov, A. M. Pertsov and D.A. Pijnappels for helpful comments on the manuscript and L. Leenknegt and D. Kabus for proofreading. 

\section*{Data Availability Statement}
The optical mapping movies, analyzed and generated datasets and processing methods are available from the authors upon reasonable request.

\clearpage

\bibliographystyle{unsrtnat}

\bibliography{PDarxiv}

\end{document}